# Agentic Architecture of Information Systems

The AAM/5C Model: A Reference Model for Distributed Autonomy

Laurent Goujon

Conceptual volume

**Author:** Laurent Goujon — IBM France

**ORCID:** 0009-0000-8127-5732 (https://orcid.org/0009-0000-8127-5732)

**Contact:** laurent_goujon@fr.ibm.com

## Trademarks and Legal Notices

The **AAM/5C** model (Autonomic Architecture Model / 5 Capabilities) is proposed by the author, Laurent Goujon; what the book claims as its own is delimited in the state of the art of the Introduction.

The trademarks and projects cited belong to their respective holders: the Model Context Protocol (MCP), introduced by Anthropic, PBC in November 2024, has been a project of the Agentic AI Foundation (LF Projects, LLC) since December 2025; Agent2Agent (A2A) is placed under the governance of the Linux Foundation; TOGAF and ArchiMate are registered trademarks of The Open Group. The other trademarks cited belong to their respective owners; citing them serves identification and suggests no affiliation.

# Executive Summary

The technical leadership of large organizations faces the cognitive complexity wall: at the scale of hundreds of interacting components within an Information System, maintaining the dependency map and reacting at the pace of incidents increasingly exceed what a human team can deliver [RH]. This book argues that the answer is not more tooling but a change of architecture: entrusting the components of the Information System with a share of perception, decision and coordination, without giving up control.

To this end, the book proposes and develops the **AAM/5C** model — Autonomic Architecture Model / 5 Capabilities — built around five Capabilities (Capture, Comprehend, Choose, Coordinate, Control) instantiated in seven logical layers, L1 to L7. The guiding principle is that intelligence proposes but never acts without validation: every act executed autonomously must be explicitly authorized, continuously audited and reversible, through a separation of powers built into the architecture.

The contribution of the book is a prescriptive reference model that articulates, under a single authorization rule, known mechanisms — separation of decision, issuance and enforcement, bounded authorization, the autonomic loop — and adds three constructs, all to be validated: the Validation step of the Distributed Autonomy Loop, the ephemeral authority surface and the cognitive dependency plane.

The targeted benefits are quantified as design targets [DT] to be validated and organized into ten falsifiable hypotheses (H1 to H10):

- Remediation time under one minute;
- Proactive incident avoidance above 85%;
- Peer consensus within 500 ms at the 95$^{th}$ percentile;
- No agentic act outside the log or outside the authorization path;
- Toil reduction above 75%;
- Ratio of net savings to total cost of ownership above 2.5.

The model distinguishes what its architecture guarantees under explicit assumptions — every admitted act carries a verified authorization — from what remains to be measured, and it targets large-scale Information Systems. The book has a **conceptual** status: it sets out and defends the model, without claiming to deploy it as is. Its research core is located as follows: the authorization rule R-AUT (§2.2.3) and the guarantee contract (§7.4.3), under which only H7a and H8a hold by construction [TG]; the refutation convention for the hypotheses (Appendix 1); a partial TLA+ specification of peer consensus (Appendix 2); the limits of the model (§9.2). Its adoption in a heavily regulated organization (for example, large European banking and insurance groups) calls for complementary work, explicitly deferred to future work whose conditions are laid out in Chapter 10:

- Compliance: the DORA regulation on digital operational resilience, the EU AI Act, model risk governance;
- Governance of high-stakes decisions;
- Supply chain security;
- Operational resilience;

- Cloud data security (ISO/IEC 27017 and 27018) and data sovereignty;
- Architecture standards;
- Empirical evidence.

Recommendation: use this book as a shared conceptual reference to structure thinking on the agentification of Information Systems, and begin the empirical validation of its hypotheses on measured pilots and prototypes before any scale-up.

**Keywords**: agentic architecture; autonomous Information Systems; multi-agent systems; AI governance; large language models; governance by design; enterprise architecture; distributed autonomy.

# Contents

# List of Figures

# List of Tables

## List of Listings

# Abbreviations and Acronyms

English-origin acronyms are expanded at their first occurrence in the text; this list collects them with their expansions, together with the acronyms of French origin, given with their English expansions. Identifiers (DOI, ISBN, CVE…), standard designations, names of companies and commercial products, the role codes of Table 14 and the notations specific to the model (L1 to L7, R-AUT, H1 to H10…), defined in the Definitions, are not listed.

**A2A** — Agent2Agent
**ACPR** — French Prudential Supervision and Resolution Authority
**AI** — Artificial Intelligence
**AIOps** — Artificial Intelligence for IT Operations
**AOM** — Agentic Operating Model
**API** — Application Programming Interface
**ATLAS** — Adversarial Threat Landscape for Artificial-Intelligence Systems
**BDI** — Belief-Desire-Intention
**C4** — Context, Containers, Components, Code
**CISO** — Chief Information Security Officer
**CMDB** — Configuration Management Database
**CoT** — Chain-of-Thought
**CSF** — Cybersecurity Framework
**CTO** — Chief Technology Officer
**CVSS** — Common Vulnerability Scoring System
**DAL** — Distributed Autonomy Loop
**DEVS** — Discrete Event System Specification
**DORA** — Digital Operational Resilience Act
**EBA** — European Banking Authority
**ECB** — European Central Bank
**ERP** — Enterprise Resource Planning
**EU** — European Union
**FPR** — False Positive Rate
**FTE** — Full-Time Equivalent
**GDPR** — General Data Protection Regulation
**HTTP** — Hypertext Transfer Protocol
**HTTPS** — Hypertext Transfer Protocol Secure
**IoT** — Internet of Things
**IP** — Internet Protocol
**IS** — Information System
**IT** — Information Technology
**JSON-RPC** — JavaScript Object Notation – Remote Procedure Call
**LLM** — Large Language Model (in the body text, the acronym appears only in names: OWASP LLM Top 10, LLM node)
**MAPE-K** — Monitor, Analyze, Plan, Execute – Knowledge
**MCP** — Model Context Protocol
**MTTR** — Mean Time to Recovery
**NASA-TLX** — Task Load Index
**NIS2** — Network and Information Systems
**OIDC** — OpenID Connect

**OODA** — Observe, Orient, Decide, Act
**OPA** — Open Policy Agent
**OSI** — Open Systems Interconnection
**PAR** — Proactive Avoidance Rate
**PDCA** — Plan–Do–Check–Act
**PEP** — Policy Enforcement Point
**RACI** — Responsible, Accountable, Consulted, Informed
**ROI** — Return on Investment (here, for convenience, the ratio of H10: G.93)
**RPO** — Recovery Point Objective
**RTO** — Recovery Time Objective
**SaaS** — Software as a Service
**SABSA** — Sherwood Applied Business Security Architecture
**SBOM** — Software Bill of Materials
**SLSA** — Supply-chain Levels for Software Artifacts
**SRE** — Site Reliability Engineering
**TCR** — Task Completion Rate
**TIBER-EU** — Threat Intelligence-Based Ethical Red Teaming
**TLA+** — Temporal Logic of Actions
**TLS** — Transport Layer Security
**TPR** — True Positive Rate
**TRIM** — Targeted Review of Internal Models

# Preface

## The Author

This book is the work of Laurent Goujon, Distinguished Engineer and Executive Architect, CTO (Chief Technology Officer) of IBM Consulting France. Over nearly thirty years at IBM, he has devoted his practice to evolving the Information Systems of large organizations by introducing emerging technologies — from distributed systems to generative and agentic AI, by way of the emergence of the Internet, the cloud, big data, cognitive computing and advanced analytics — and, above all, by scaling them up in extremely regulated environments.

Among other roles, he has been IBM CTO for a large banking group, then for the Cognitive & Analytics practice of IBM Consulting in Europe. He has had the opportunity to lead the technical teams of major transformation programs across multiple sectors: banking, insurance, aerospace, transport, telecommunications. He is a member of IBM's Open Innovation Community (formerly the Academy of Technology).

The conviction underlying this book was born of that practice: the distributed autonomy of Information Systems is an operational necessity in the face of the scale and entanglement of architectures, provided that it is governed — which is the challenge of deploying it in a regulated environment.

## Intended Audience

The book is addressed to architects (enterprise, solution, infrastructure, security), to IT (Information Technology) and SRE (Site Reliability Engineering) engineers, to heads of operational platforms, to the leadership of Information Systems and AI governance functions, as well as to researchers and students.

## Acknowledgements

### Sources of Intellectual Inspiration

This book owes much to the work of Michael Wooldridge, W. Ross Ashby, Edwin Hutchins, Jeffrey O. Kephart and David M. Chess, Artur d'Avila Garcez and Luís C. Lamb, and Enid Mumford, whose contributions are discussed in more detail in §2.1.

### Gratitude

The author extends particular thanks to the first readers who agreed to submit the manuscript to their critical eye, and whose feedback advanced the argument decisively.

# Introduction

**IN BRIEF**

Can an Information System perceive, decide and act without permanent human intervention, without the enterprise losing control of its decisions? The book answers with an architecture model, AAM/5C, which it presents, justifies and submits to criticism. This introduction does not yet enter the model. It sets out its objective, its three-part outline, the scope of a conceptual volume and the reading conventions, foremost among them the epistemic grid, which distinguishes throughout what is guaranteed, argued, targeted and measured. Finally, it situates the model within the state of the art. The “In Brief” boxes, read in sequence, give a short, continuous version of the argument; Chapter 1 opens it with the imperative of autonomy.

### Objective

This book is devoted to the distributed autonomy of Information Systems. It proposes, formalizes and submits to critical examination the **AAM/5C** model — Autonomic Architecture Model / 5 Capabilities. The five Capabilities — Capture, Comprehend, Choose, Coordinate, Control — are instantiated in seven logical layers, L1 to L7. The objective is to equip the architect with a shared reference for conducting the progressive agentification of the Information System under governance and security constraints.

For the research reader, the book is a design contribution: it proposes an artifact (a reference model and its authorization rule), derives from it analytical statements that hold under explicit assumptions, and fixes, before any measurement, the protocol that could refute its hypotheses; it reports no empirical result.

### Outline and Reading Paths

The book follows a ternary rhythm:

- **Part I — Why** (Chapters 1 to 3): the stakes of distributed autonomy, the AAM/5C model, then the problem statement, the hypotheses and the evaluation framework;
- **Part II — Where and How** (Chapters 4 to 8): the conceptual model, the overview, the implementation of the upper layers (L1 to L3) and lower layers (L4 to L7), the governance and ethics of autonomy;
- **Part III — Assessment and Outlook** (Chapters 9 to 11): the assessment of the model, the conditions for adoption in a regulated environment, then the general conclusion.

The glossary (Definitions) and the Reference Documents are placed at the end of the book, before the appendices. The five appendices bring together the experimental protocols identified to date (Appendix 1), the mathematical formalization (Appendix 2), the supporting material (Appendix 3), the objections (Appendix 4) and the gallery of the personas who animate the chapter vignettes (Appendix 5).

Four reading paths are possible. **Sequential reading** follows the argumentative sequence intended by the author. **Reading the Why alone** (Chapters 1 to 3) gives the stakes and the hypotheses. **Reading the architecture** (Chapters 4 to 8) serves the practitioner who wants to enter the model directly. **Critical reading** starts with Appendix 4 (objections) and is completed by the Reference Documents.

### Scope and Status of the Book

The book has a **conceptual** status, with no value as a compliance standard. The empirical validation of hypotheses H1 to H10 is a matter for future work, according to the protocols of Chapter 3 and Appendix 1; the conditions for adoption in a heavily regulated environment are laid out in Chapter 10.

### Reading Conventions

**Epistemic grid.** The structuring statements — targets of hypotheses H1 to H10, governance thresholds, parameters of the business model, properties of the architecture — carry one of the following four statuses, in a grid that is the only one in the book:

- **[TG] — Theoretically Guaranteed by construction**: a property that the architecture holds from its very structure, under its explicit assumptions — a correct enforcement point interposed on every execution path of an agentic act, intact keys, executed act identical to the verified act (guarantee contract, §7.4.3); it is demonstrated on the model, without measurement. Only the analytical statements H7a and H8a (§3.2.4) fall under this status; H8a restates the invariant of rule R-AUT (§2.2.3): every act admitted by a Policy Enforcement Point (PEP) carries a verified act token or capability envelope.
- **[RH] — Reasoned Heuristic**: a proposition argued by analogy or by reasoning, without formal proof. Empirical validation may confirm or revoke it without calling into question the structure of the model.
- **[DT] — Design Target**: a quantified target, to be validated by future work.
- **[EM] — Empirically Measured**: a value derived from a documented pilot.

**Default status.** Any unmarked numerical value, other than a value drawn from a cited source or given as illustrative, has [DT] status: an order of magnitude or a design target, never a measurement result. No value in this book has [EM] status. An unmarked figure can therefore under no circumstances be invoked as empirical evidence.

**Analogies.** An illustrative analogy carries no status and proves nothing; an analogy that transposes an established result from a neighboring field falls under [RH]. Thus the correspondence between System 1 and System 2 [Kahneman 2011] and the L4/L6 pair illustrates the duality between probabilistic reasoning and deterministic decision. It does not imply that large language models reproduce human cognition. The architectural guarantee rests on the determinism of the policy engine, not on the analogy.

**Defensibility.** Distinct from the grid, a five-level defensibility scale qualifies the maturity of the six sub-problems of Chapter 3 (§3.3.1). It does not qualify statements, which fall under the grid alone. Each level has its criterion:

- Conjectural: a statement without argument;
- Plausible: a qualitative argument, without established result;
- Supported by analogy: a result established in a neighboring field, whose transposition to the model is not demonstrated;
- Partially verified: part of the statements is established on the model, at [TG] status;
- Demonstrated: a complete formal proof, or a measurement at [EM] status.

**Language conventions.** “Must” expresses a design requirement; “should,” a recommendation; “may,” a possibility; “aims at” introduces a [DT] target. The terms cognitive agent, Capability, layer and primitive action are understood in the sense of the Definitions. “Capability,” capitalized, denotes one of the five Capabilities of the model; the capability envelope (§7.4.3) is a bounded authorization, unrelated to the adjustment of resource capacity. Key concepts are coded by the rank letter of the chapter: B.C1 is the first concept of Chapter 2.

**Book and model.** “The book” denotes the present volume, from the Executive Summary to the Index; “the model,” used alone, denotes AAM/5C, which the book sets out.

**Layers, blocks and agent groups.** Layers L1 to L3 are called “upper” and form Block I (operational anchoring); layers L4 to L7 are called “lower” and form Block II (cognition, governance and execution). This convention deliberately inverts the numbering of the Open Systems Interconnection (OSI) model. Agents are divided into three groups — Perception (L1–L2), Cognition (L3–L4–L5), Governance (L6–L7) — described in §2.3. The stack admits no other partition. Two layers are cited before their presentation in Chapter 2: L4, the orchestration layer where the agent reasons and proposes, and L6, the data and governance layer that carries the signed policies.

**Typographic convention.** Apart from apparatus elements (subtitles of the book and of chapters, table titles, headings, variables, cross-references and reference statuses), italics are reserved for two uses: narrative vignettes, which illustrate a point through a professional situation, and the openers of the three parts. Each vignette is announced by the heading “In Practice”; the personas who animate nine of them are presented in Appendix 5.

## State of the Art and Positioning

The AAM/5C model recomposes, in the service of governed autonomy, six theoretical foundations whose contributions are discussed in §2.1. Four streams of work delimit what it borrows and what it claims; each is examined from the same angle.

### Autonomic Computing and Self-Adaptive Architectures

*What this stream contributes.* [Kephart and Chess 2003] set out the principle of systems that self-configure, self-heal, self-optimize and self-protect, around the MAPE-K loop (Monitor, Analyze, Plan, Execute – Knowledge): monitoring, analyzing, planning and executing on a shared knowledge base. [Garlan et al. 2004] give it its software form: the loop is externalized in a reusable infrastructure that maintains a model of the architecture, its invariants and its adaptation strategies. [Zhang et al. 2024] take up this program with a multi-agent system based on large language models, applied to microservice management.

*How AAM/5C differs.* The Distributed Autonomy Loop (DAL, §4.4) inserts a Validation phase between deliberation and execution, and entrusts it to a deterministic layer distinct from the one that plans, according to rule R-AUT (§2.2.3). MAPE-K, by contrast, leaves planning and execution in the same plane of responsibility. The difference lies in the separation of powers: as long as the planner can execute, no governance property can be established against it, and a probabilistic planner makes this separation necessary. From the autonomic principle, the model retains the closed loop and observability as a prerequisite for autonomy.

### Multi-Agent Systems

*What this stream contributes.* [Wooldridge 2009] and [Ferber 1999] provide the vocabulary of agency — autonomy, goals, interaction — agent architectures and interaction protocols. The contract net of [Smith 1980] provides decentralized task allocation by announcement, bidding and contract award, whose principle the negotiation between agents of the model takes up. This stream also contributes the results on the emergence and instability of agent collectives, the foundation of anti-resonance (Chapter 7).

*How AAM/5C differs.* An agreement between agents obtained through Agent2Agent (A2A) is only a proposal: it produces no effect without an authorization issued in accordance with a signed policy, according to rule R-AUT (§2.2.3). The collective has latitude to coordinate, not the power to execute; this choice places the authorization plane on the path of every agentic act. The compromise of a few agents therefore does not suffice to cause an act to be executed; the rules of admission to the vote among agents and the number of compromised agents tolerated, for their part, protect coordination (Appendix 2, §2.2).

### Enterprise Architecture and Architecture Description

*What this stream contributes.* ISO/IEC/IEEE 42010 [ISO/IEC/IEEE 42010:2022] standardizes architecture description: stakeholders, concerns, viewpoints, views and the traceability among them. TOGAF [TOGAF 2022] provides a development method, ArchiMate [ArchiMate 2026] a modeling language. Together, these frameworks give the form in which an architecture becomes admissible before an architecture board.

*How AAM/5C differs.* These frameworks are frameworks of form; AAM/5C is a prescriptive model. It states which layers must exist, which invariants must hold and who decides what, and it backs its statements with falsifiable hypotheses and an epistemic status. It does not dispense with a standards-compliant architecture description: it fixes its content.

Conformance to 42010 is partial and declarative. Three requirements are satisfied:

- The stakeholders and their concerns are identified — technical leadership, architects, SRE engineers, risk and compliance functions, regulators; autonomy, security, governance, cost, compliance;
- Four viewpoints are declared: conceptual (Chapters 2 to 4), layered implementation (Chapters 6 and 7), governance and ethics (Chapter 8), adoption in a regulated environment (Chapter 10);
- Architecture decisions are stated with their rationale, R-AUT first.

Three remain to be produced: the formal specification of the viewpoints, the correspondence rules between views, the record of unaddressed concerns; in addition, beyond the requirements of the standard, the context view and the deployment topology down to the component level. Their production is a matter for an architecture description (§10.7).

### Security of Agent Protocols and Governance

*What this stream contributes.* The Model Context Protocol (MCP) [Anthropic 2024], for context, and A2A [Google A2A 2025], for interaction among agents, provide the interoperability substrate; [Jeong 2025] shows their complementarity. [Hou et al. 2026] characterize the attack surface of MCP, and [Anbiaee et al. 2026] compare the threat modeling of agentic protocols. Regulation (EU) 2024/1689 [EU AI Act 2024] sets, for high-risk systems, the obligations of risk management, documentation and logging; ISO/IEC 42001 [ISO/

IEC 42001:2023] and the voluntary NIST framework [NIST AI RMF 2023] organize AI risk management, without amounting to compliance with that regulation.

*How AAM/5C differs.* These works describe threats and obligations; they do not say where authorization comes from. The model locates it through rule R-AUT (§2.2.3): decision, issuance of the authorization and verification fall to three distinct components, and local issuance is excluded. A compromised MCP server cannot issue an authorization, but it can distort the context on which L6 decides (§7.7, §8.3); a compromised agent can propose, it cannot authorize itself. It is this localization of the power to authorize, and not the addition of controls, that distinguishes the model from approaches in which governance is a filter applied after the fact.

### What None of These Streams Provides

In the works examined here, no stream provides an Information System architecture model that is at once prescriptive on the location of the power to authorize, compatible with a probabilistic planner and backed by falsifiable hypotheses about its own guarantees. Autonomic computing gives the loop but lets the planner execute; multi-agent systems give coordination but usually let agreement between peers make law; enterprise architecture gives the form without the content; protocol security gives the threats without the architecture that answers them. The model sets out to fill these four gaps together, and submits to refutation, through hypotheses H1 to H10, its targets and the conditions on which its guarantees depend (H7b, H8b); the analytical statements H7a and H8a are not refutable.

The contribution of the book is a prescriptive reference model that articulates, under a single authorization rule, known mechanisms — separation of decision, issuance and enforcement, bounded authorization, the autonomic loop — and adds three constructs, all to be validated: the Validation step of the Distributed Autonomy Loop, the ephemeral authority surface and the cognitive dependency plane. §4.4 realizes the first; §1.1 situates the second, which §7.4.2 realizes; the third (C-Dep, §4.2.3) completes the F-Dep / O-Dep / C-Dep triple graph, whose functional and operational planes belong to the state of the art of dependency graphs. Each is claimed for its contribution, not for its inspiration.

The ephemeral authority surface takes up authorization bounded in time and scope; it adds the binding of each covered act, through the digest of the evaluation proof, to a signed decision of L6, an authority distinct from the token issuer.

The book also draws on notions that do not belong to it: the shift from access control to action control, formulated in industrial practice (§1.1); Zero Trust [NIST SP 800-207], which cognitive Zero Trust extends to the intentions of agents; atomic revocation of an authority, a known mechanism for which the present book proposes a bounded-time target and the articulation with the R-AUT chain.

Three established mechanisms already do part of what R-AUT prescribes (§2.2.3). The NIST Zero Trust architecture separates the policy engine, which decides, the policy administrator, which issues the session credential, and the PEP, which enforces the decision [NIST SP 800-207]. Open Policy Agent, queried by the program that executes an agent’s tool calls, renders a decision on each of them before its execution [OPA 2026]. Agent frameworks suspend execution to submit a critical action to a human, such as the *interrupt* function of LangGraph [LangChain 2024].

These mechanisms have older antecedents: the reference monitor, always invoked, tamper-proof and small enough to be analyzed and tested [Anderson 1972]; capabilities, each of which designates an object and the operations permitted on it [Dennis and Van Horn 1966]; the explicit norms of normative multi-agent systems, which govern autonomous agents while allowing for their violation [Boella et al. 2006].

AAM/5C inherits no guarantee from this lineage; it adds to these mechanisms, as part of the ephemeral authority surface, the signed binding between the evaluation proof of L6 and the act token of L5 and the capability envelope without local issuance, together with an evaluation instrument: an independent measurement of non-bypass (H8b).

The positioning against the market's agent orchestration frameworks is examined in §5.2, where none of the four frameworks studied natively provides an authorization rule comparable to R-AUT; its objective assessment through a comparative benchmark is left to future work (§10.8).

# Part I — Why

*This first part sets out why distributed autonomy is becoming an operational necessity (Chapter 1), lays out the AAM/5C model of governed autonomy (Chapter 2) and translates it into ten hypotheses (Chapter 3).*

## Chapter 1. Agentic Architecture of Information Systems

**IN BRIEF**

Why make an Information System autonomous? Because its complexity exceeds what a team can hold in its head. The obstacle is not the density of the dependency graph — real architectures are sparse — but its concentration, depth and instability. This chapter calls the threshold at which this complexity overwhelms the team the cognitive complexity wall: technical debt becomes cognitive debt. It proposes to respond by delegating perception, decision and coordination to agents, under a governance built into the architecture, and places this ambition on a five-level autonomy scale. It poses the problem, whereas Chapter 2, which follows, will build the model that answers it.

### *Toward Distributed Cognitive Autonomy*

**Key Concepts of the Chapter**

- **A.C1 — Distributed cognitive load** — the share of the operator's mental load that can be transferred to the agent without degrading decision quality; an analogical transposition of cognitive load theory [Sweller 1988] to the human–agent case;
- **A.C2 — Ephemeral authority surface** — the set of acts an agent is authorized to perform within a time × scope window; the act token and the capability envelope bound it for every act admitted by a PEP (§7.4.2, §7.4.3);
- **A.C3 — Atomic revocation** — withdrawal of an agentic authority within a bounded time, independently of the agent's internal state (target [DT]: under 500 ms).

#### In Practice

*Camille, the SRE lead of an online retail site, is responsible for the proper operation and operational readiness of a set of 200 microservices on the eve of a Black Friday–type sales event. A composite incident — a saturated message queue, concurrent automated data collection, a slowdown of the central database — absorbs most of her time in diagnosis and coordination. In the target architecture, the agents qualify the incident, negotiate a response and*

*submit it for authorization. Camille is no longer on the front line and answers for the invariants that the execution policies encode.*

### 1.1. The imperative of autonomy: from technical debt to cognitive debt

The Information Systems of large organizations have become hyperdistributed ecosystems: hybrid cloud, edge, serverless functions, thousands of interdependent microservices. Their technical, operational and regulatory complexity grows faster than the human capacity to master it [RH]. Yet diagnosis, incident resolution, evolution and compliance still rest on teams — operations, security, architecture, data — whose intervention conditions the stability of the whole.

Despite DevOps practices, the emergence of the SRE role and infrastructure as code, critical processes remain manual: diagnosing complex failures, adjusting capacity to unforeseen peaks, arbitrating between competing services, real-time compliance. The engineer serves as "cognitive glue": absorbing the complexity that automation cannot handle. When this load, part of which agents can take over (distributed cognitive load, A.C1), becomes excessive, it slows team velocity and degrades the quality of their output [Skelton and Pais 2019].

The difficulty is not combinatorial. Take the example of 200 microservices: the number of couplings can reach 19,900 pairs. Real architectures can be far from it: Meta's estate, 18,500 services, has fewer than 400,000 directed links, a little over a tenth of a percent of the roughly 342 million directed links of a fully connected topology, and each service calls four others at the median [Huye et al. 2023]. What undoes a team's mental map comes down to three documented traits: the **concentration** of calls on a handful of services, the **depth** of the chains, and the **instability** of a given service's call topologies from one request to the next [Luo et al. 2021].

Automation by scripts is no longer enough [RH]; hence the move to autonomy. Agentic systems open this possibility: components able to observe, reason, cooperate and adapt. From the intentional agent model of the BDI type (Belief-Desire-Intention) [Wooldridge 2009], AAM/5C takes the vocabulary — beliefs, desires, intentions — for its layers L3 and L4; its other lineages are placed in the state of the art (Introduction) and in §2.1.

Autonomy does not erase human expertise: it displaces it. The engineer ceases to be the firefighter of technical debt and becomes the architect of governance policies. He or she no longer treats symptoms; he or she defines the system's invariants, objectives and limits, as a supervisor (human-on-the-loop) guaranteeing coherence between business intentions and agentic execution.

This displacement moves us from access control to action control: we no longer define only who gets in, but which act is authorized, at the moment it is requested, or within a capability envelope decided in advance (§7.4.3). The notion was formulated in industrial practice, notably at Cisco [Patel 2026]. This book proposes an architectural variant of it — ephemeral authority surface (A.C2), atomic act, atomic revocation within a bounded time (A.C3; §7.4) — for which it bears the burden of proof.

Two terms are used in a strict sense:

- **Cognitive**, applied to a component or a relationship of the model — cognitive agent, cognitive dependency, cognitive cache, cognitive Zero Trust — qualifies what pertains to the contextual reasoning of agents, a function carried by layer L4 and bounded by the decision of L6; it does not denote a human mental process. Applied to humans — cognitive load, cognitive debt, distributed cognition — it keeps its usual meaning;

- **Autonomy** denotes an agent's capacity to have an authorized act executed under the R-AUT rule (§2.2.3), without human synchronization, within the limits of its ephemeral authority surface.

## 1.2. Agentic capabilities at the heart of IS

### 1.2.1. Classical architecture facing the cognitive complexity wall

The cognitive complexity wall denotes the threshold beyond which the complexity of the Information System — the concentration, depth and instability of its dependencies (§1.1) — exceeds what a team can hold in its head. Enterprise architectures rest on robust, proven models: functional layers, microservices, integration pipelines, orchestrators, API catalogs (Application Programming Interface), identity management, observability. They nonetheless have structural limits:

- **Insufficient automation**: runbooks — written operating procedures — describe what to do, but do not execute themselves in the face of ambiguity;
- **Illegible dependencies**: concentrated, deep and changing (§1.1), they make overall behavior hard to understand;
- **No coordination between components**: each can be automated, none reasons with the others; a layer that negotiates, arbitrates and explains is missing;
- **Cognitive silos**: components have access neither to the global context nor to the ability to turn a signal into a decision;
- **Observability limited to detection**: signals are captured, rarely analyzed through to diagnosis and action proposal;
- **Burden of recurring operations**: identity governance, schema mapping, certificate management, quota adjustment.

These limits do not stem from poor design: traditional architectures were not designed to host internal decision-making agents, and the complexity described in §1.1 accentuates their effect (Figure F.1).

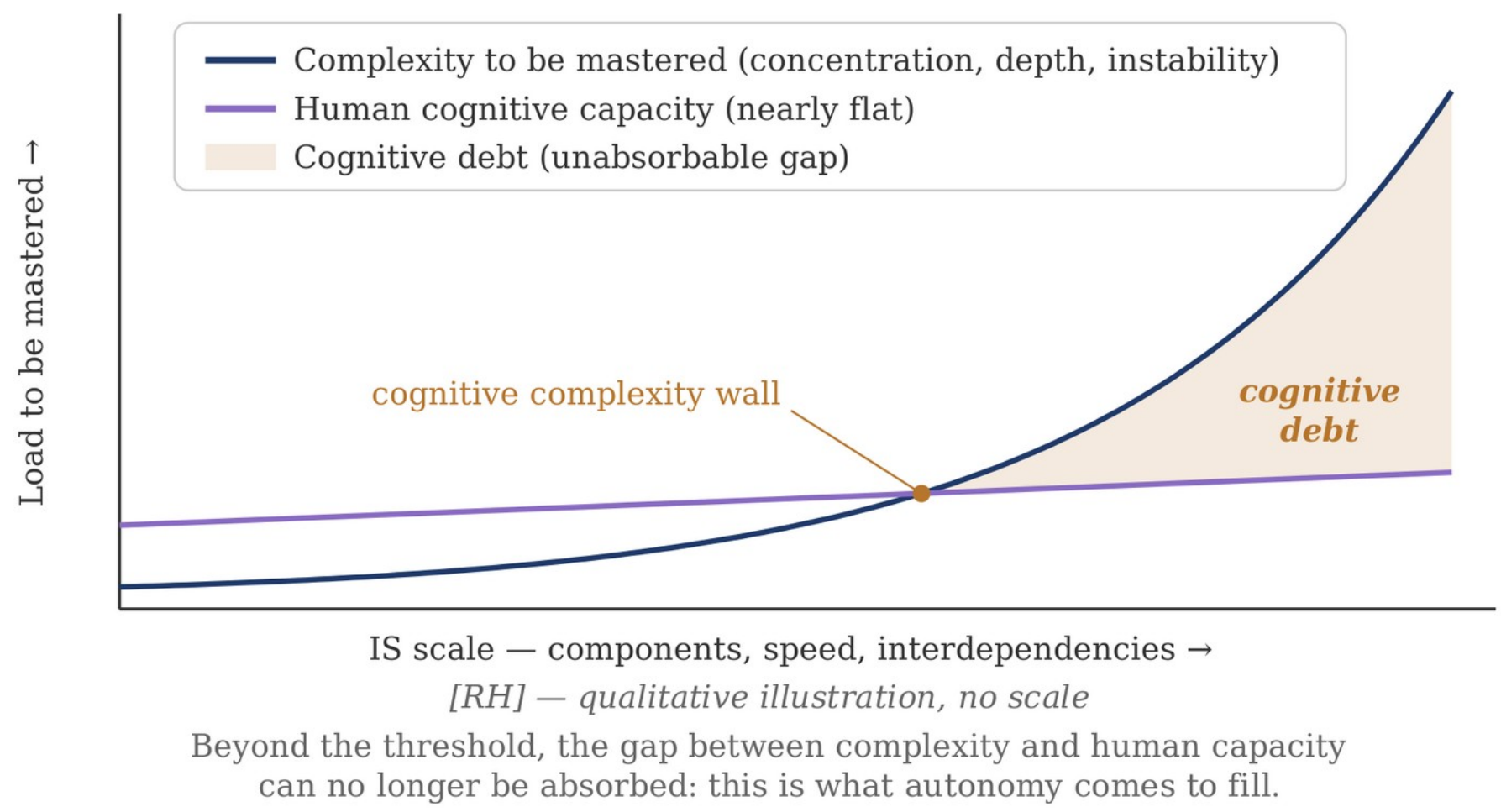

Figure F.1 — Cognitive debt: why human steering saturates [RH].

### 1.2.2. The promise of an agentic IS

An agentic IS is an Information System in which each component — API gateway, microservice, orchestrator, identity module, data pipeline, Kubernetes cluster — is endowed with its own agent. This agent runs through the cycle of the five Capabilities defined in §1.3.1. This layer of agents is superimposed on the existing system without replacing it.

An unexpected load spike on an online retail site gives an order of magnitude, for illustration. In the **classical scenario**, monitoring detects the latency in about 5 minutes; the alert reaches the engineer in 10 minutes; he or she correlates the logs by hand and works out a plan. Mean time to recovery (MTTR) reaches 30 to 45 minutes, with losses that grow with it.

In the **agentic scenario**, the gateway agent detects the latency in 10 seconds. Its reasoning engine consults the context to identify the service objectives at stake, then negotiates through the A2A protocol with the catalog agent and the scaling service. The agreed plan is submitted for authorization (§2.2.3) and executed in 45 seconds. The design target remains H1 — MTTR under one minute (§3.2.1). Table 1 summarizes the contrast between the two approaches.

| | **Classical Information System** | **Agentic architecture (AAM/5C)** |
|---|---|---|
| **Philosophy** | Reactive automation: runbooks and static thresholds | Governed autonomy: reasoning, negotiation, authorization |
| **Decision engine** | Fixed rules; the human interprets | Probabilistic reasoning (L4) bounded by a deterministic decision (L6) |
| **Role of the human** | "Cognitive glue": manual correlation | Architect of governance policies |
| **Remediation** | Detection, then lengthy human intervention | Detection, reasoning and authorized execution, in under one minute [DT] |
| **Control** | Centralized: global supervision, configuration management database (CMDB) | Distributed coordination (A2A); authorization by signed policy (R-AUT, §2.2.3) |

Table 1 — The promise of an agentic IS.

## 1.3. Foundations for an agentic IS architecture model

### 1.3.1. The five expected Capabilities

A natively autonomous architecture model must give a formal framework to the new functions that agents bring to the Information System. These functions boil down to five Capabilities, which give the model its name:

- **Capture** — perceive the state of the component and of its environment, qualify the signal against service commitments, route it to reasoning;
- **Comprehend** — diagnose in context, down to the root cause, by combining probabilistic reasoning and logical invariants [Garcez and Lamb 2023];
- **Choose** — plan an explainable action — remediation, reconfiguration, resource optimization — under cost and risk constraints;
- **Coordinate** — negotiate among peers, through the A2A protocol, access to resources and service commitments, without a central orchestrator;
- **Control** — execute only an act that is authorized under a signed policy, under the R-AUT rule (§2.2.3): this is action control.

Chapter 2 instantiates these Capabilities in the L1–L7 stack (Table 2); Chapter 3 derives ten hypotheses from them, arranged by Capability (Table 4).

### 1.3.2. A2A and MCP: the two interaction protocols

The model relies on two existing protocols, which it does not claim as its own: it fixes their use and subjects them to its authorization rule.

The **A2A** protocol [Google A2A 2025] is the language of cooperation between agents. It conveys intentions — "can you defer your requests?" — rather than function calls. Agent Cards describe each agent's capabilities there, and execution remains opaque: agents collaborate without exposing their memory or their own logic. The model relies on A2A to carry negotiation, conflict resolution and consensus-seeking, which the specification does not define as such. A common protocol also facilitates collaboration among agents from different frameworks and vendors [Jeong 2025].

**MCP** [Anthropic 2024] connects agents to the systems where data resides; the model makes it the channel of their reasoning context: security policies, legal constraints, budget, service objectives, topology, dependencies, data classification. MCP is a context protocol, not a governance system: the signing of the policies it carries is a requirement of the model, not of the specification [MCP 2026]. In short, MCP defines what an agent can know, A2A what it can negotiate with the others; what it can do falls under R-AUT (§2.2.3).

## 1.4. Why now? The emergence of neuro-symbolic AI

The model rests on a requirement that neither classical automation nor large language models meet alone: that an act decided by a machine be verified as compliant with the enterprise's policy before it is executed. A probabilistic engine can propose but cannot prove; a rules engine can verify but cannot propose. Entrusting autonomy to the former alone installs hallucination at the heart of the Information System; entrusting it to the latter alone brings us back to scripts. The neuro-symbolic approach combines the two: L4 proposes through reasoning, L6 decides through a policy engine, and no intention becomes an act without this decision.

Several dynamics converge:

- **An unsustainable scale**: architectures are too distributed to be operated by hand;
- **The maturity of language models**: Chain-of-Thought reasoning makes it possible to go beyond rigid automation;
- **Symbolic verification**: logic solvers can prove that a formal representation of an agent's behavior satisfies formalized constraints [Zhou et al. 2026], in the line of neuro-symbolic AI [Garcez and Lamb 2023]; the model retains the principle, verify before executing, through a policy engine — a guarantee bounded by what the policy can express (L6 expressiveness limit) and by the attestation of its inputs (I-ATT, §7.2);
- **Economic and environmental pressure**: operating cost, notably of inference, and the imperative of frugality.

The model expects a drop in the hallucination rate of L4 when the context supplied by MCP is sufficiently rich [RH] (Figure F.16, §8.6); this behavior remains to be confirmed experimentally.

## 1.5. Taxonomy of autonomy levels: positioning the AAM/5C model

To situate the ambition of the model, the book adopts a five-level scale of Information System autonomy, inspired by work on autonomous microservice systems [Zhang et al. 2024] and developed in Chapter 5. The remediation times given are illustrative orders of magnitude, not measurements; only that of Level 4 is a target [DT].

- **Level 0 — Passive IS** (fully manual): strictly manual execution; runbooks are read and applied by the human. MTTR on the order of 2 to 4 hours.
- **Level 1 — Assisted IS** (script-based automation): deterministic scripts (Ansible, Terraform) automate recurring tasks; the human triggers and interprets. MTTR on the order of 30 to 60 minutes.
- **Level 2 — Supervised IS** (AIOps): AIOps tools (Artificial Intelligence for IT Operations) detect anomalies and suggest actions; the human remains in the decision loop (human-in-the-loop). MTTR on the order of 10 to 30 minutes.
- **Level 3 — Conditional IS** (conditional autonomy): autonomous decisions within predefined, reversible scopes; the human supervises without executing (human-on-the-loop). MTTR on the order of 1 to 5 minutes.
- **Level 4 — Governed autonomous IS** (AAM/5C objective): autonomous management of the operational cycle while respecting governance invariants, each agentic act being authorized under R-AUT (§2.2.3). Target MTTR under one minute (H1). This is the target of the model, whose validation constitutes the agenda of future work. The model does not claim that this level is attainable for all scenarios: it defines the architecture and the conditions for progressively reaching it (Figure F.2).

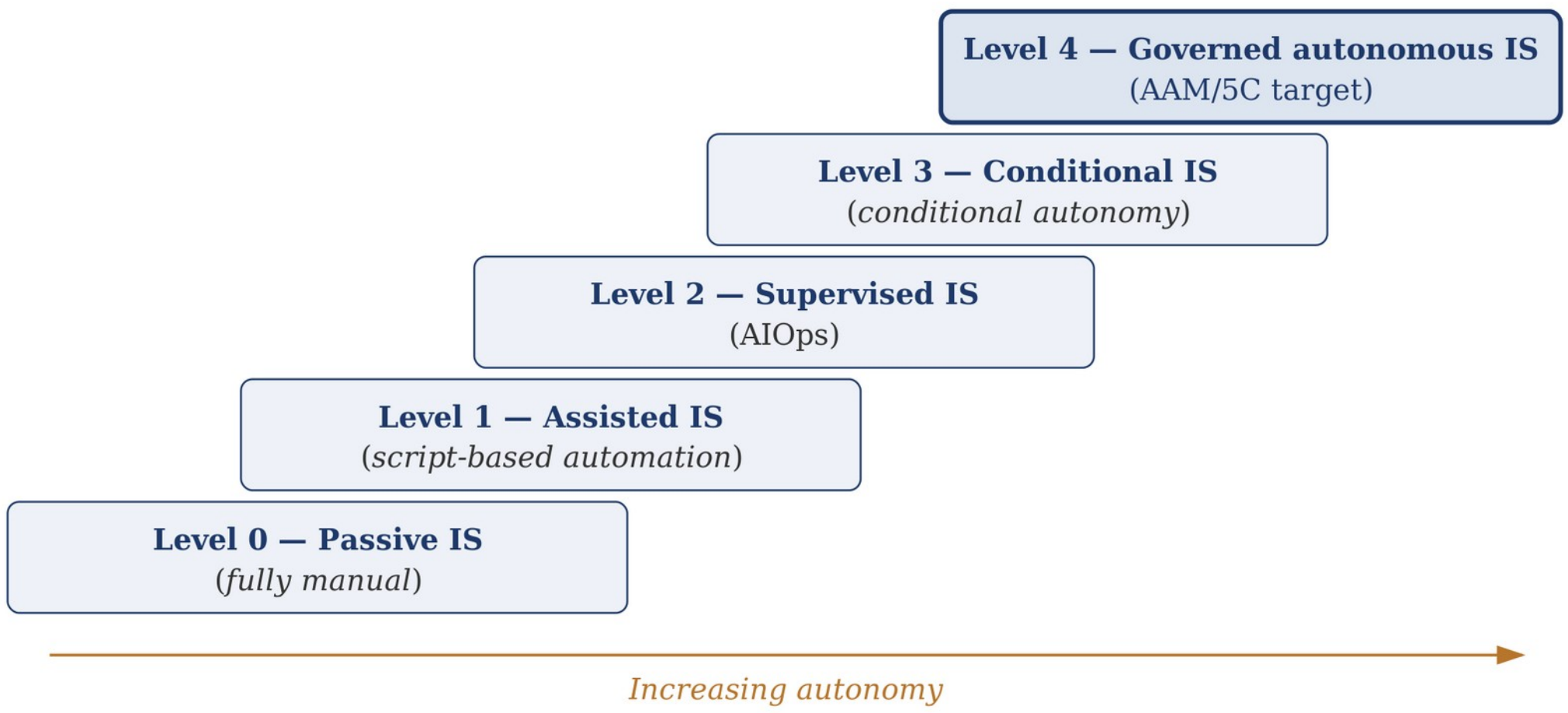

Figure F.2 — From automation to autonomy: the five-level scale.

## 1.6. Key takeaways

The chapter argues that the complexity of Information Systems exceeds the capacity for direct human supervision [RH]: technical debt has become cognitive debt. Classical automation detects; an agentic system must also comprehend, choose and coordinate — without ever acting outside a verifiable authorization. Five Capabilities found the model; two protocols, A2A and MCP, carry coordination and context; the human role shifts toward writing and auditing policies. Hence the question that guides the book: how can one design an Information System in which each component acts autonomously while guaranteeing security, governance and stability? Chapter 2 answers it with a seven-layer architecture.

# Chapter 2. The AAM/5C Model

**IN BRIEF**

This chapter delivers the model. It stops seeing the Information System as a stack of manually driven components and conceives it instead as a team of agents endowed with five Capabilities: Capture, Comprehend, Choose, Coordinate, Control. These Capabilities are instantiated in seven logical layers divided into two blocks — L1 to L3, the upper layers of operational anchoring; L4 to L7, the lower layers of cognition, governance and execution. A single rule, R-AUT, fixes who decides, who issues the authorization and who verifies it: AI proposes, and never acts without authorization. The chapter grounds the model in six scientific foundations and describes the agents that inhabit it. Chapter 3 derives from it the hypotheses that make it refutable.

## *Five Capabilities, Seven Layers, One Authorization Rule*

**Key Concepts of the Chapter**

- **B.C1 — L4/L6 cybernetic coupling** — L4 generates intentions; L6 filters them against a signed policy, without the filter reintroducing variability into the decision [RH]. This is what distinguishes the model from a simple “language model then post-filter” chain (§2.1, §2.2.3).
- **B.C2 — Equivalent requisite variety** — an extension of Ashby’s law [Ashby 1956] to the multi-agent case: the variety proper to each agent and the coordination variety between agents must, together, cover that of the disturbances [RH]. The policies of L6, by restricting acts, reduce this variety: escalation to the human organization of the Agentic Operating Model (AOM, §8.4) compensates for what they exclude (§2.1).
- **B.C3 — Socio-technical learning loop** — a PDCA cycle (Plan–Do–Check–Act) applied to policies: explicit criteria trigger their human review, so that no frozen policy becomes a bottleneck (§4.4, §8.4.4).

In Practice

*Yann, an enterprise architect in a retail bank, must get agents adopted without losing regulatory control. Before the regulator, the question is not what the agents can do, but who authorizes each act, how this is proven after the fact and how it is undone. This chapter builds the model’s answer: a single authorization rule, which separates the proposal, the decision, the issuance and the verification of each agentic act.*

### 2.1. Scientific and theoretical foundations of agentic IS

The autonomy of an Information System is not an added feature: it is a property that results from an architecture. The model rests on six theoretical foundations; none suffices alone, each contributes what the others do not:

- **Multi-agent systems.** An agent is a computer system capable of acting autonomously in its environment to achieve the objectives delegated to it [Wooldridge 2009]; each microservice or data pipeline becomes a specialized agent, which negotiates with the others to produce collective behavior [Ferber 1999].

- **Distributed cognition.** Information processing is distributed among actors and their tools [Hutchins 1995]: without a central AI, each agent holds a share of the global knowledge and explains its decisions against a shared context, which breaks down cognitive silos (§1.2.1).
- **Cybernetics (B.C2).** A regulator can compensate for all the disturbances it faces only if it has a variety of responses at least equal to theirs [Ashby 1956]; the Information System becomes a self-regulating system through feedback loops.
- **Autonomic and neuro-symbolic computing (B.C1).** To the four autonomic properties [Kephart and Chess 2003], the model adds governance and explainability; following the neuro-symbolic principle [Garcez and Lamb 2023], a probabilistic layer, L4, proposes, and a deterministic symbolic layer, L6 — a policy engine of the Rego/OPA type (Open Policy Agent) — decides in a replicable, testable and auditable way.
- **Socio-technical systems.** Technology is not isolated from human organization [Mumford 2006]: the engineer, as architect of governance, sets the invariants that policies encode; the agentic system extends human intentions without freeing itself from them.
- **Resilience engineering.** Withstanding shocks is not enough: one must absorb them and adapt to them [Hollnagel et al. 2006]; by handling anomalies locally, the agentic IS aims at avoiding cascade effects.

## 2.2. The AAM/5C architecture: five Capabilities, seven layers

### 2.2.1. The five Capabilities and their instantiation

The five Capabilities defined in §1.3.1 form the agent's cycle: it captures a signal, comprehends it, chooses a response, coordinates it with its peers and executes it only under control (Figure F.3). Three design principles follow. The **separation of concerns** isolates "doing" from "thinking": an agent does not reason on raw telemetry. **Governance through context** places legal, regulatory and security rules at the heart of the system. **Systemic explainability** makes each decision traceable through formal protocols and accounts for it in a form suited to a human recipient — contrastive and selective [Miller 2019].

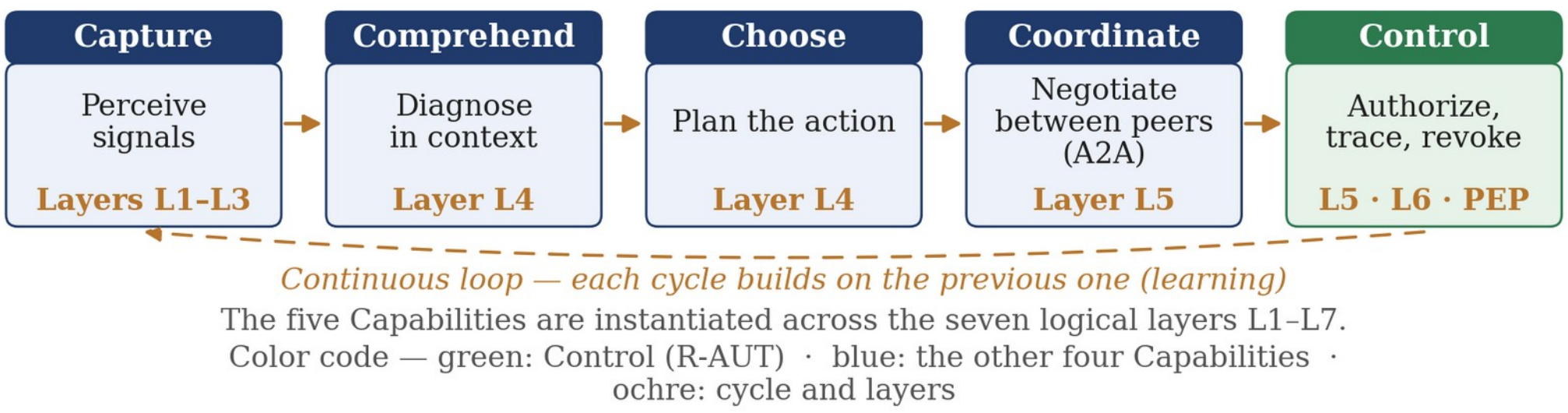

Figure F.3 — The five Capabilities (5C) of the AAM/5C model: the agent cycle, from perception to control, in a continuous loop.

Table 2 gives the model's only correspondence between Capabilities, layers, agents and hypotheses. Chapter 3 develops the hypotheses; Chapters 6 and 7, the layers.

| Capability | Function | Layers | Agents (group) | Hypotheses and targets [DT] |
|---|---|---|---|---|
| **Capture** | Perceive, qualify, route the signal | L1 to L3 (Block I) | Sensory agent (Perception); L3 executive agent (Cognition) | H1: MTTR < 1 min; H2: blast radius reduced by more than 70% |
| **Comprehend** | Diagnose in context | L4 | Cognitive agent (Cognition) | H3: proactive avoidance > 85% |
| **Choose** | Plan under cost and risk constraints | L4, in a sandbox | Cognitive agent (Cognition) | H9: toil reduced by more than 75%; H10: ROI (Return On Investment) > 2.5× (net savings / total cost of ownership, §8.7) |
| **Coordinate** | Negotiate among peers (A2A) | L5 | Social agent (Cognition) | H4: conflicts resolved without escalation > 95%; H5: oscillation < 5%; H6: consensus < 500 ms (95th percentile) |
| **Control** | Authorize an act (R-AUT) | L6 decides, L5 issues, the enforcement point (PEP) verifies; the target layer (L1, L3, L6 or L7) materializes | Governor agent (Governance); L7 infrastructure agents (Governance) | H7b: no agentic act outside the log; H8b: no bypass of the PEP (H7a, H8a: [TG]) |

Table 2 — The five Capabilities: layers, agents and hypotheses.

### 2.2.2. Two blocks, seven layers

To move from concept to implementation, the Capabilities are instantiated in seven logical layers, on the principle of the OSI model: each layer has a single responsibility and communicates with its neighbors through standardized interfaces. The stack is read in two blocks, according to a convention chosen to better model the routing of external signals: what is “upper” touches the business and the outside. The numbering thus follows the path of a request, from the boundary of the IS (L1) to the infrastructure (L7); it therefore deliberately reverses the OSI order, in which L1 is the physical layer.

**Block I — the upper layers (L1 to L3): operational anchoring.** They perceive, qualify and route; they decide no agentic act.

- **L1 — Channels**: the boundary of the Information System with users, partners and terminals. As the architecture’s first sensor, the agentic gateway (Agentic API Gateway) transforms exchanges into signals and ensures perimeter security and the semantic integrity of inputs. It is also through L1 that operators observe the system and bring their intentions to it.
- **L2 — Digital and Application**: the functional identity of components — their APIs — and their service commitments. Its business micro-agents filter for relevance: they escalate only the signals that genuinely threaten a commitment.
- **L3 — Integration**: the Agentic Service Bus. It holds the lifecycle and the beliefs of agents, in the sense of the BDI model, and reconciles heterogeneous representations. The signals qualified by L2 become desires there (“I must reduce my load”) addressed to cognition.

**Block II — the lower layers (L4 to L7): cognition, governance and execution.** This is the model's neuro-symbolic space and its execution foundation; the separation of powers plays out there according to R-AUT (§2.2.3).

- **L4 — Orchestration and Processes**: it links the business requirement to technical execution, a role that [Hohpe 2020] assigns to the architect, and transforms a desire into a structured intention. Its language models produce plans through Chain-of-Thought reasoning, in a sandbox. Fast and effective for diagnosis, it remains probabilistic, hence prone to hallucination: it proposes, it does not execute.
- **L5 — Security and Compliance**: the layer of cognitive Zero Trust, which continuously evaluates agent behavior [Hannecke 2025] and passes its evaluation to L6. It carries the A2A negotiation, to which the agents of all layers are parties, arbitrates conflicts between competing proposals and issues the act token that makes an act executable.
- **L6 — Data, Knowledge and Governance**: it transforms data into actionable knowledge and injects, through MCP, security, compliance and budget policies. Its governor agent submits each L4 plan to the policy engine and renders the authorization decision. L6 is symbolic and deterministic where L4 is probabilistic.
- **L7 — Infrastructure and Physical**: the execution substrate, which adjusts its compute resources to actual loads [Kephart and Chess 2003] and, as an extension specific to the model, its energy resources. It materializes verified infrastructure acts, reports back and closes the observability loop to L1.

Figure F.4 gives the integrated view of the stack, of the two cross-cutting protocols and of the Agentic Operating Model (AOM); it serves as the reference map for the rest of the book.

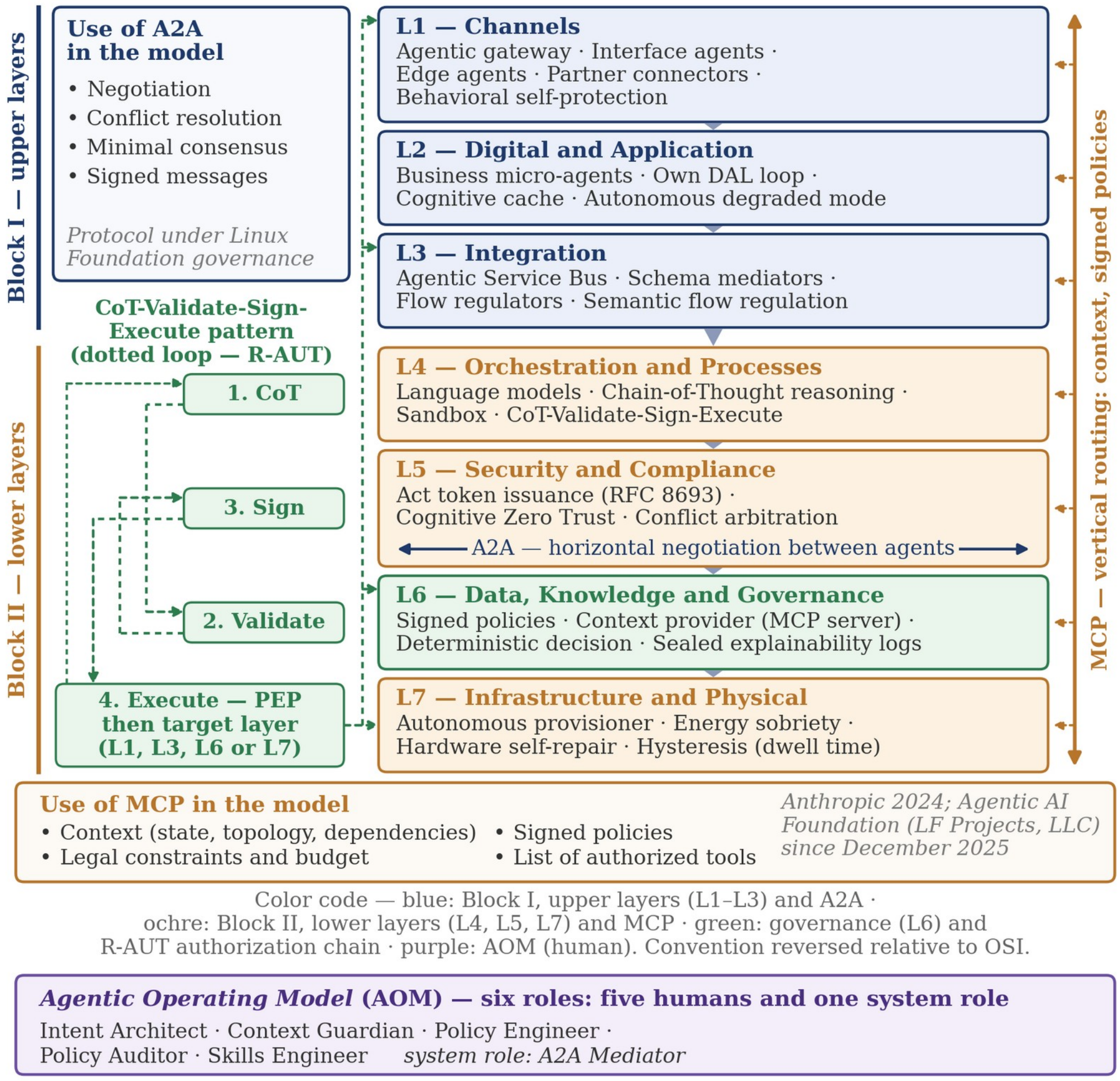

Figure F.4 — AAM/5C architecture in layers L1 to L7.

### 2.2.3. Normative rule R-AUT: who decides, who issues, who verifies

The separation of powers has force only if the authorization chain is stated once and for all. The R-AUT rule is authoritative throughout the book; every other passage refers back to it. The invariant it establishes — every act admitted by a PEP carries a verified act token or capability envelope — is taken up by the analytical statement H8a [TG], under the assumptions of the guarantee contract (§7.4.3); that no agentic act bypasses the PEP is the refutable statement H8b [DT] (§3.2.4).

- **Proposal (L4).** The cognitive agent forms an intention to act and submits it, with the context of its reasoning, to L6 through MCP.
- **Decision (L6).** The governor agent renders the authorization decision in accordance with the signed policy in force. It produces a verdict — authorization or refusal — recorded in an evaluation proof that it signs; it issues no token.
- **Issuance (L5).** L5 issues the act token — atomic, with explicit scope, with a bounded validity period, bound to the verdict by the cryptographic digest of the evaluation proof. No agent issues, renews or extends its own token: local issuance is excluded by construction. The evaluation proof reaches L5 directly or through L4, which then requests the token: the rule fixes who decides and who issues, not this path.
- **Verification (PEP).** A Policy Enforcement Point (PEP), deterministic and non-cognitive, placed on the execution path, verifies the token before any materialization and records the admitted act in the explainability log before forwarding it (§7.4.3). An agentic act not presented to a PEP is an unauthorized act; the failure of a check results in the rejection of the act (fail-closed: deny by default), its logging and the notification of L5.

- **Materialization (target layer).** The verified act is executed by the layer that holds the resource — L1 gateway, L3 bus, L6 context provider for data access, or L7 infrastructure — which reports back on it.

Action control is therefore not the function of any particular layer: it is the complete chain decision → issuance → verification → materialization. The order of the rule is not that of the layer numbering: the latter follows the path of an incoming request, from L1 to L7 (§2.2.2), whereas the rule follows the act's authorization path, in which the L6 decision necessarily precedes the issuance, by L5, of the token bound to it. Figure F.5 therefore arranges the layers in the order of their number, gives the order of the rule through numbered steps and details its unfolding, rejection branch included; Figure F.15 (§8.3) places L6 above L4 and L5 and numbers the two exchanges that link them.

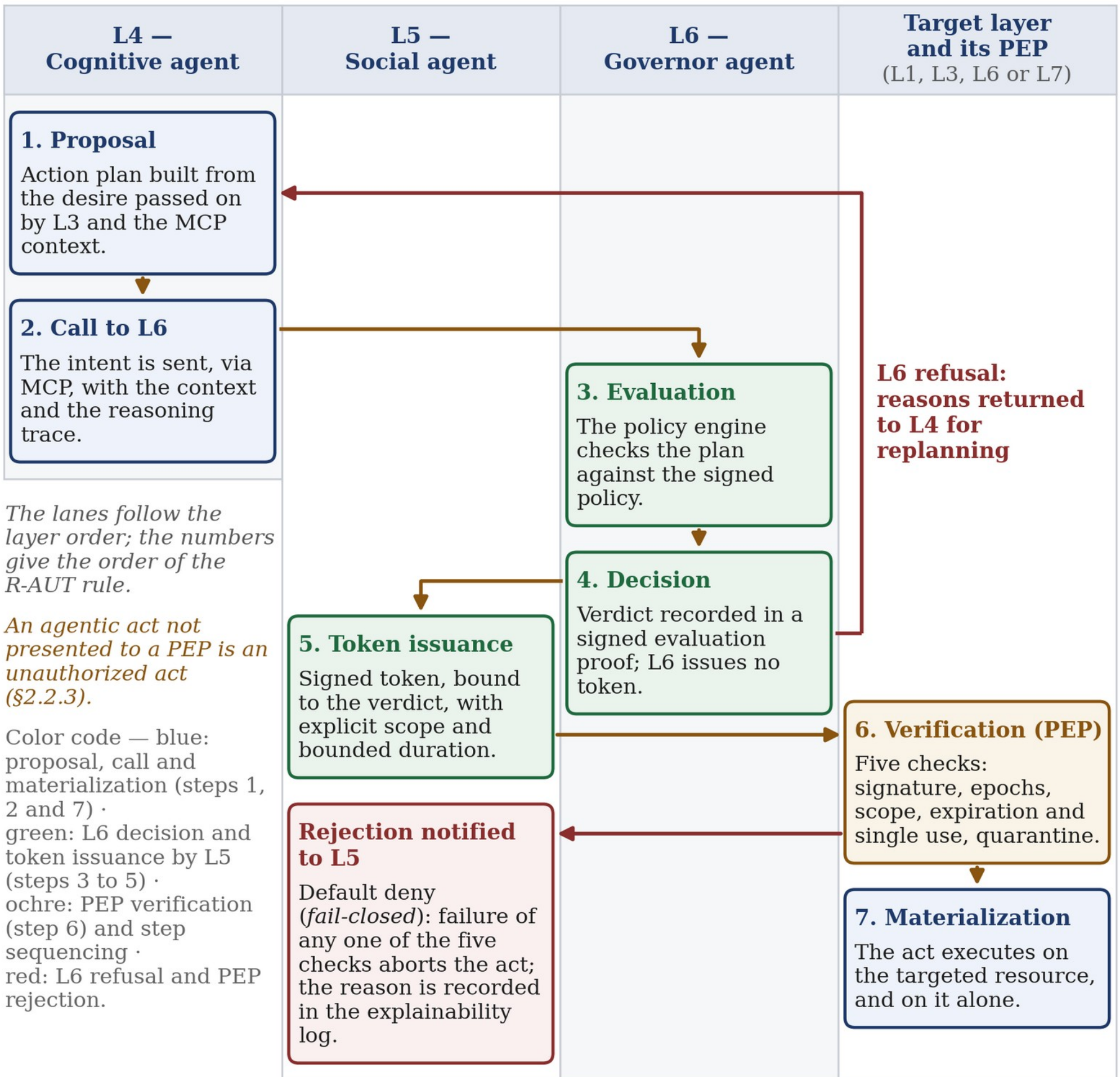

Figure F.5 — The cross-layer validation flow, from intent to act (rule R-AUT: L4 proposes, L6 decides, L5 issues, the PEP verifies, the target layer materializes).

Three mechanisms complete the rule: the five checks of the **PEP**, the **pre-authorized capability envelope**, which preserves a bounded right to act when the control plane is unavailable, and the **key lifecycle**. They are specified in §7.4.3, together with the guarantee contract under which the invariant holds.

Human oversight is not an eighth layer. It is carried by the AOM (§8.4), whose roles observe the system through L1, amend policies in L6 (B.C3) and audit decisions after the fact. It is this arrangement that carries the human oversight envisaged in Article 14 of the EU AI Act [EU AI Act 2024].

## 2.3. The autonomous agents of the AAM/5C model

Agents specialize according to their place in the stack and fall into three groups, depending on their respective roles: **Perception** (L1–L2), **Cognition** (L3–L4–L5) and **Governance** (L6–L7).

### 2.3.1. Regulation mechanisms: the dual control loop

The coexistence of thousands of agents requires safeguards against chaotic behavior and cascade effects. A single level of control is not enough. Each agent reacts to its own signals: acts authorized one by one can still work against each other and amplify one another (§7.8). Agents must therefore coordinate among themselves. Conversely, an agreement among agents does not say whether the act is permitted: peers can agree on an act that is contrary to a policy, and one of them may be compromised. A separate authority must therefore decide. Two loops answer these two needs, one among peers, the other under signed policy:

- **Dynamic control (horizontal — A2A, L5).** To scale, an agent negotiates only with its direct dependencies (negotiation scope). A resource conflict gives rise to local mediation; failing consensus, the action is suspended.
- **Static control (vertical — MCP, L6).** An agreement among agents is only a proposal: it produces an effect only once it is authorized under R-AUT (§2.2.3). An act that violates a security policy or a budget is refused before any execution.

Figure F.6 shows the anatomy of a functional component (§4.2.1): a deterministic execution engine coupled with a probabilistic cognitive agent, under the L6 policy.

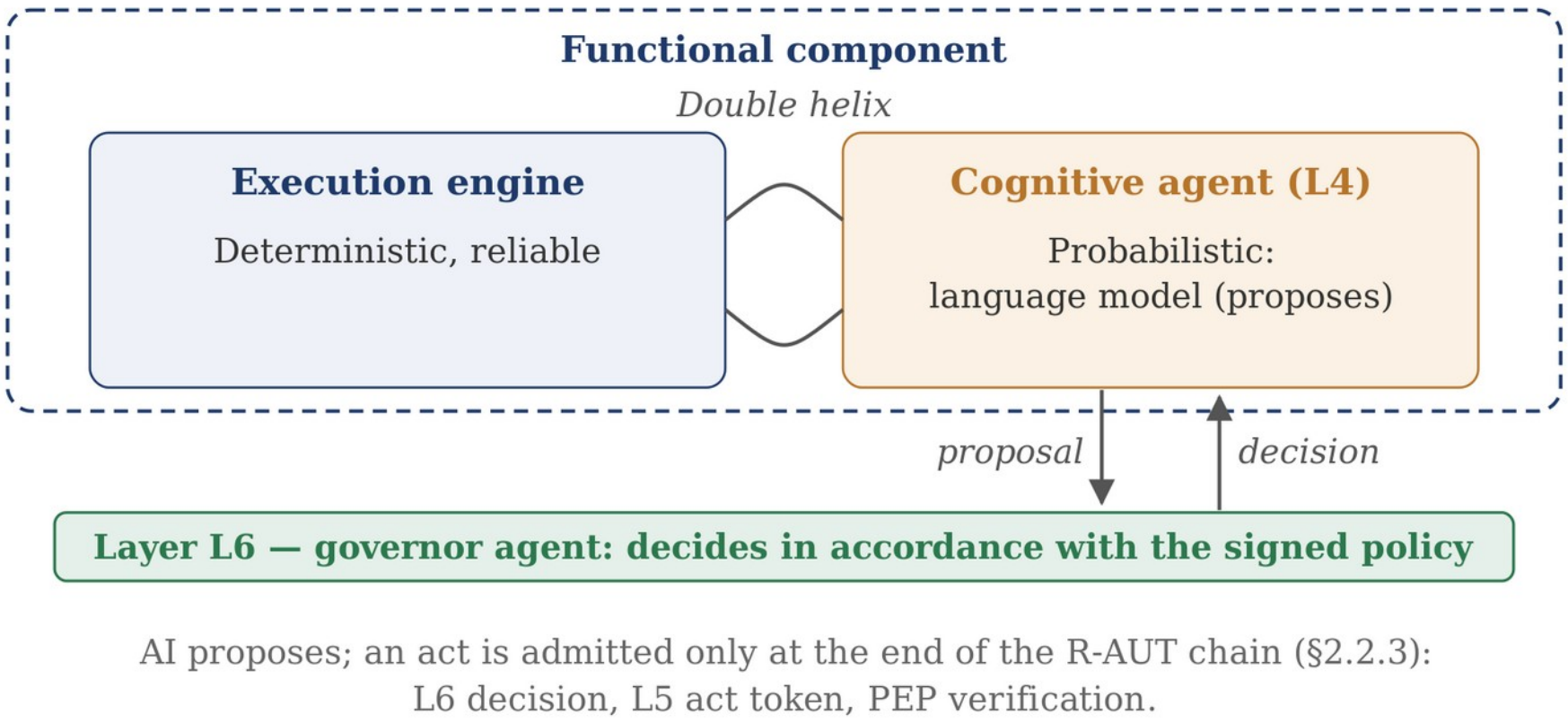

Figure F.6 — The “double helix” of a functional component.

### 2.3.2. Typology and distribution of agents across the AAM/5C stack

Table 3 specifies, for each type of agent, its group, its layer and its role; the specialized agents of Chapters 6 and 7 belong to the group of their layer.

| Agent | Group · layer | Role |
|---|---|---|
| **Sensory agent** | Perception · L1–L2 | Monitors signals, filters noise, detects deviations from service objectives. |
| **Executive agent** | Cognition · L3 | Holds the component's local state and beliefs; turns the qualified signal into a desire; presents authorized acts to the PEP. |
| **Cognitive agent** | Cognition · L4 | Diagnoses root causes and plans through probabilistic reasoning; proposes, does not execute. |
| **Social agent** | Cognition · L5 | Negotiates via A2A, arbitrates conflicts between competing proposals, issues the act token on L6's decision. |
| **Governor agent** | Governance · L6 | Checks each plan against the signed policies, renders the authorization decision and its proof, logs; issues no token. |
| **Infrastructure agents** | Governance · L7 | Materialize the infrastructure acts verified by the PEP, report back and close the observability loop. |

Table 3 — Typology and distribution of agents across the AAM/5C stack.

## 2.4. An end-to-end act

### 2.4.1. The nominal call

Figure F.7 unfolds, in 20 steps, a nominal agentic call: the request crosses the upper layers down to cognition, the plan follows the R-AUT chain (§2.2.3), then the response travels back up to the user. A [DT] target of 500 ms at the 95th percentile, distinct from that of H6, applies to the authorization chain — decision, issuance, verification — excluding L4 reasoning. The signed explainability logs make the sequence auditable after the fact.

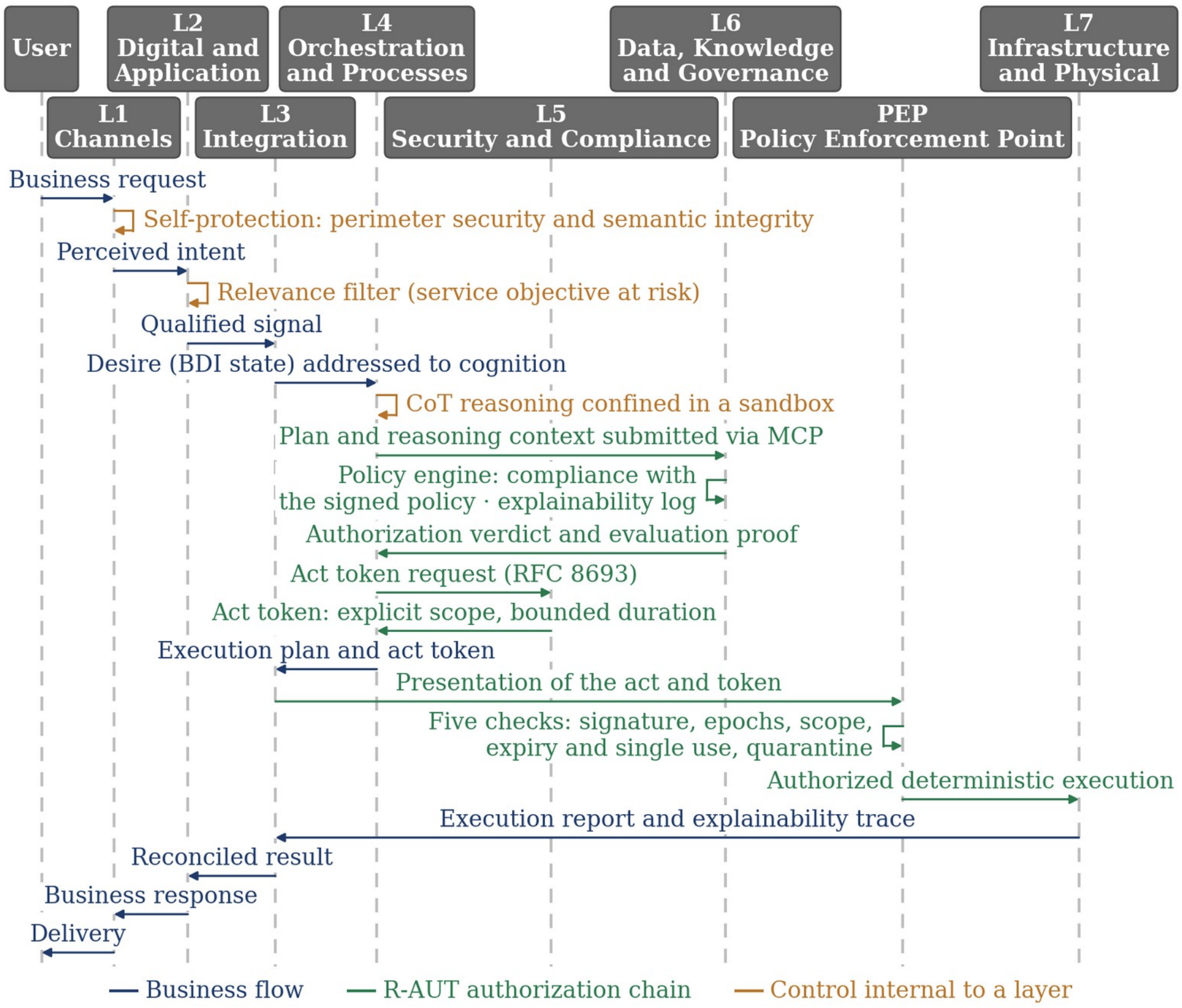

Figure F.7 — Canonical sequence of an agentic call in nominal mode.

### 2.4.2. Illustration: detecting a zero-day vulnerability

A critical remote code execution vulnerability is published; it affects a library of the payment microservice, which is exposed to the Internet. The system is operating normally; it is vulnerable but not yet under attack. The target exit state: threat isolated, service maintained in degraded mode, digital forensic data saved, audit report produced, in under one minute and without manual intervention. The SecOps engineer supervises without executing.

1. **L1** — the security scanner detects the signature of the flaw;
2. **L2** — the sensory agent qualifies the alert: since the service is exposed, the risk to the security commitment is critical;
3. **L3** — the executive agent updates the component's state and formulates a desire to secure the service;
4. **L4** — the cognitive agent, with no patch available, proposes to isolate the exposed container and to serve payments in degraded mode, without the vulnerable function;
5. **L5** — the social agent negotiates via A2A with the gateway agent (flow redirection) and the infrastructure agent (container isolation); the agreement obtained remains a proposal;
6. **L6** — the governor agent checks the plan against the signed policies and renders a favorable decision, under the "security first" invariant;
7. **L5, PEP, L1 and L7** — L5 issues the act tokens; the gateway and control-plane PEPs verify them; redirection and isolation are executed;
8. **L3 and L1** — the state is reconciled; the SecOps engineer is informed through the console and has the explainability trace available for the audit.

The scenario is evaluated with the metrics of H1 (remediation time), H3 (action before exploitation), H7b and H8b (no agentic act outside the log or outside the PEP), defined in Table 4.

## 2.5. Key takeaways

The AAM/5C model conceives the Information System as a team of agents endowed with five Capabilities, instantiated in seven layers distributed across two blocks. The R-AUT rule (§2.2.3) entrusts distinct components with the proposal, the decision, the issuance of the token or capability envelope, and their verification; that no agentic act bypasses the PEP remains to be measured (H8b, §3.2.4). Coordination among agents is horizontal and distributed; authorization is vertical and governed by signed policies. The L4/L6 coupling bounds probabilistic reasoning with a deterministic decision; the AOM maintains human supervision and policy review. Chapter 3 turns these properties into refutable hypotheses and an evaluation framework.

# Chapter 3. Problem Statement, Hypotheses and Evaluation Framework

**IN BRIEF**

This chapter turns the model into a verifiable research program. It states the problem — giving every component local autonomy without losing security, governance and stability — and breaks it down into six sub-problems. From it, it derives ten hypotheses, H1 to H10, arranged by Capability, each accompanied — except for the analytical statements H7a and H8a — by a metric, a measurement instrument and a refutation condition; their targets are design targets, never results. Finally, it sets the evaluation framework: criteria, statistical reproducibility, experimental scenarios and limits of validity. It closes the first part; Chapter 4 opens the second, which describes how the model is realized.

## *From Thesis to Validation Program*

**Key Concepts of the Chapter**

- **C.C1 — Epistemic status of the hypotheses** — the grid of the Introduction, applied hypothesis by hypothesis: [TG] for the analytical statements H7a and H8a, [DT] for all numerical targets (§3.2);
- **C.C2 — Statistical reproducibility** — for probabilistic components, reproducibility is understood through a confidence interval: at 95%, over at least 1,000 independent runs of the same scenario — more when the target requires it (Appendix 1, §1.1) — with distinct seeds and frozen policies; Wilson interval for proportions, bootstrap for continuous quantities;
- **C.C3 — Defensibility scale** — five levels, from conjectural to demonstrated, which qualify the maturity of the six sub-problems SP1 to SP6 (§3.3.1), and not the statements.

An architecture model is worth something only if it accepts being contradicted. In the case of this book, asserting that an Information System can become autonomous without ceasing to be governed has scientific weight only if the assertion rests on statements for which one knows in advance what would refute them. The framework that follows fulfills three functions: it bounds what the model claims to guarantee and separates the guarantee from the intention; it opens the model to challenge on explicit grounds; it constitutes the program of empirical validation, hypothesis by hypothesis.

### In Practice

*Léa, a doctoral student in information systems, seeks to reproduce an event-chaining scenario — a domino effect between services — on a Kubernetes test bench. She records the seed of each language model, the exact version of the Rego policies, the topology of 300 containers and the propagation times, then measures H1, H2 and H6. Her approach is the one this chapter equips: turning a design target into a measured result.*

## 3.1. The central problem: the cognitive complexity wall

Chapter 1 posed the question that guides the book (§1.6): how can one design an Information System in which each component acts autonomously while guaranteeing security, governance and stability?

This question breaks down into six sub-problems:

- **SP1 — Agentic integration**: encapsulate the deterministic (existing code) within the probabilistic, through the articulation of L4 and L6;
- **SP2 — Distributed coordination**: cooperate via A2A without a central orchestrator, so that coordination depends on no single point of failure; the authorization plane, for its part, remains on the critical path, except for acts covered by a capability envelope (§7.4.3);
- **SP3 — Governance and security**: hold to Zero Trust despite autonomy, through the move to action control;
- **SP4 — Systemic stability**: prevent autonomous decisions from resonating;
- **SP5 — Performance and cost**: offset the total cost of the agentic layer, human governance included, by reducing repetitive manual work (toil);
- **SP6 — Reproducibility**: validate non-deterministic behavior in the sense of statistical reproducibility (C.C2) — a scenario run at least 1,000 times, more when the target requires it, must produce, for each metric, a 95% confidence interval contained within the specified envelope.

## 3.2. The ten hypotheses, by Capability

The hypotheses are presented in the order of their numbers, from H1 to H10, and grouped by Capability according to the correspondence of Table 2; Choose, to which H9 and H10 belong, therefore comes last. Their targets are design targets [DT], set by the model (C.C1); they are not results. Table 4 sets out, for each of them, the statement, the metric, the instrument and the refutation condition. Each refutation condition applies under the convention in Appendix 1, §1.1, with its own schedule for H9, except that of H10, which is decided at milestone G4.

### 3.2.1. Capture: H1 and H2

Human intervention no longer allows fast, decisive responses to a flow of changing and contradictory requests [Kephart and Chess 2003]. Without a local capacity to perceive and act, each component waits for a central instruction, which creates decision bottlenecks.

- **H1 — Reduction of remediation time.** Local autonomy removes human decision latency: the agent detects the drift in L2 and triggers the response. The target is absolute: MTTR below one minute under nominal load [DT], on incidents of the classes covered by a signed policy; the share of incidents escalated to the AOM is reported separately.
- **H2 — Systemic resilience.** Self-configuration and negotiated fallback limit the blast radius of failures, by isolating the anomaly as close as possible to the resource: blast radius reduction greater than 70% [DT].

### 3.2.2. Comprehend: H3

A classic Information System knows only thresholds. The contextual reasoning of L4 makes it possible to correlate weak signals — growing network jitter and a drift in memory consumption — to identify a latent failure before it becomes critical.

- **H3 — Proactive management.** Contextual reasoning makes it possible to anticipate drifts and act before the alert: proactive avoidance rate (PAR) — the share of the incidents of a reference set handled before the alert threshold — above 85% [DT].

### 3.2.3. Coordinate: H4 to H6

The function of distributed coordination is to prevent conflicts over access to resources and mutual blocking between agents [Ferber 1999]; the model also requires it to keep the local optimization of one service from degrading the global optimum.

- **H4 — A2A cooperation.** An intent-exchange protocol resolves most resource conflicts locally, without a central arbiter: more than 95% of conflicts resolved without escalation [DT].
- **H5 — Stability through context.** Sharing a common context via MCP, hysteresis and minimal consensus (§7.8) make local decisions converge: oscillation amplitude of the controlled quantity below 5% of its setpoint [DT].
- **H6 — Peer consensus.** Agents reach consensus on high-criticality acts without heavy infrastructure, in under 500 ms at the 95$^{th}$ percentile [DT].

### 3.2.4. Control: H7 and H8

Autonomy without control is a risk for the enterprise. When agents act in place of humans, governance must be in-line and deterministic: it no longer validates only an identity, but each agentic act against the invariants of security, compliance and budget. H7 and H8 each comprise an analytical statement and a refutable statement; the reasons for this distinction are set out in Appendix 3, §3.1.

- **H7 — Explainable governance.** H7a [TG]: every act admitted by the PEP is entered in the explainability log before being forwarded to the target layer. This is a direct consequence of the guarantee contract (§7.4.3), which makes the failure of this write a rejection; the statement is not refutable, which is acknowledged, and holds only under the assumptions of this contract. The sealing of the entry and the execution trace come after forwarding: they do not fall under H7a. H7b [DT] (log completeness): no materialized agentic act escapes the log — a refutable statement, measured by cross-checking the act log collected on the resource side against the sealed explainability log.
- **H8 — Compliance by design.** H8a [TG]: every act admitted by the PEP is covered by an act token or a capability envelope tied to a favorable L6 decision in accordance with the signed policy in force at issuance and not revoked at admission — up to the freshness bound of revocation — under the guarantee contract (§7.4.3). H8b [DT] (non-bypass): no agentic act bypasses the PEP — a refutable statement, measured on the resource side by verifying that each act in the act log was forwarded by a PEP of attested identity and corresponds to an admission by that PEP and to an entry in the authorization log on the L5–L6 side. This instrument does not see an act that an agent causes to be executed under the identity of a script or a human (Appendix 3, §3.1).

Effective compliance remains subordinate to the rigor of the implementation and to policy coverage.

### 3.2.5. Choose: H9 and H10

An architecture must remain economically and humanly viable. The choice of action is made under a cost constraint: the hypothesis is that the reduction of manual work offsets the total cost of the agentic layer, human supervision included, and frees engineers for the architecture of governance.

- **H9 — Toil reduction.** Toil decreases by more than 75% [DT], despite the growing complexity of hybrid ecosystems.
- **H10 — Economic efficiency.** The net savings from autonomization exceed 2.5 times the total cost of ownership of the agentic layer — inference, AOM roles, governance platform (§8.7) [DT].

| **Capability** | **Hypothesis** | **Statement** | **Metric and target** | **Measurement instrument** | **Refutation condition** |
|---|---|---|---|---|---|
| **Capture** | H1 | Reduction of remediation time | MTTR < 1 min | Before/after comparative measurement on pilot incidents | MTTR ≥ 1 min |
| | H2 | Systemic resilience | Blast radius reduction > 70% (components affected) | Controlled fault injection | Reduction ≤ 70%, or propagation beyond the expected scope |
| **Comprehend** | H3 | Proactive management | PAR > 85% | Synthetic incident set | PAR ≤ 85% |
| **Coordinate** | H4 | A2A cooperation | Conflicts resolved without escalation > 95% | Multi-agent coordination scenarios | Conflicts resolved without escalation ≤ 95%, or unbounded escalations |
| | H5 | Stability through context | Oscillation amplitude < 5% of the setpoint | Resonance test (dwell time, §7.8) | Amplitude ≥ 5%, or deviation Δ held above the tolerance threshold θ beyond a declared observation window (states E1 or E2, §7.8) |
| | H6 | Peer consensus | Consensus latency < 500 ms (95$^{th}$ percentile) | Multi-agent A2A test bench | Latency at the 95$^{th}$ percentile ≥ 500 ms |
| **Control** | H7a **[TG]** | Every act admitted by the enforcement point is entered in the explainability log before its forwarding | Not applicable — analytical statement | Inspection of the enforcement point specification and of the guarantee contract (§7.4.3) | Not applicable ([TG] statement, §3.2.4) |
| | H7b **[DT]** | Log completeness — no materialized agentic act is absent from the explainability log | Zero materialized agentic acts without a corresponding log entry | External cross-check: reconciliation of the act log collected on the resource side with the sealed explainability log, over a declared window | At least one materialized agentic act without a corresponding log entry over the window |

| Capability | Hypothesis | Statement | Metric and target | Measurement instrument | Refutation condition |
|---|---|---|---|---|---|
| | H8a **[TG]** | Every act admitted by the enforcement point is covered by a favorable L6 decision | Not applicable — analytical statement | Inspection of the enforcement point specification, of the guarantee contract (§7.4.3) and of the signed policy | Not applicable ([TG] statement, §3.2.4) |
| | H8b **[DT]** | Non-bypass — no agentic act materializes outside the authorization path | Zero agentic acts observed on the resource side without a passage attested by a PEP and a corresponding authorization | External cross-check: act log (resource side, with the attested identity of the issuing PEP) against PEP admissions and the authorization log (L5–L6 side) | At least one agentic act recorded outside a PEP over the window, even if authorized |
| **Choose** | H9 | Toil reduction | Toil avoided > 75% | Before/after measurement of operator effort | Reduction ≤ 75% |
| | H10 | Economic efficiency | ROI > 2.5× | Business model (§8.7) on a documented pilot | ROI ≤ 2.5× |

Table 4 — The ten hypotheses: statements, metrics, instruments and refutation conditions.

## 3.3. Evaluation framework and scientific methodology

### 3.3.1. Defensibility of the sub-problems

The sub-problems are qualified by the defensibility scale (C.C3), defined in the Introduction, which prioritizes the validation effort. SP3 (governance and security) is at the "partially verified" level: its statements H7a and H8a hold by construction, under the assumptions of the guarantee contract (§7.4.3). SP1, SP2, SP4 and SP6 are at the "supported by analogy" level, and SP5 at the "plausible" level, its only support being the business model of §8.7, whose parameters are [DT]: they are the priority targets of experimental instrumentation.

### 3.3.2. Multidimensional evaluation criteria

The evaluation covers four dimensions:

- **Operational performance**: velocity (detection and remediation times) and availability;
- **Cognitive quality**: correctness rate — the share of actions judged relevant after the fact — and hallucination rate — the share of plans that a post hoc review establishes as resting on erroneous or nonexistent data; L6 rejections give only a lower bound;
- **Stability and safety**: resistance to chaos, and the integrity of MCP against hijacking of agentic capabilities [Hou et al. 2026];
- **Economic effectiveness**: ratio of net savings to total cost of ownership (H10 ratio, §8.7).

All results are reported in the sense of statistical reproducibility (C.C2). Figure F.8 illustrates the decision quality of L6: the more invariants L6 evaluates per action, the more faulty plans it should detect, with diminishing returns, while latency grows with that number [RH]. This quantity is not to be confused with the true positive rate of action control, whose 99% target is set in §8.6.

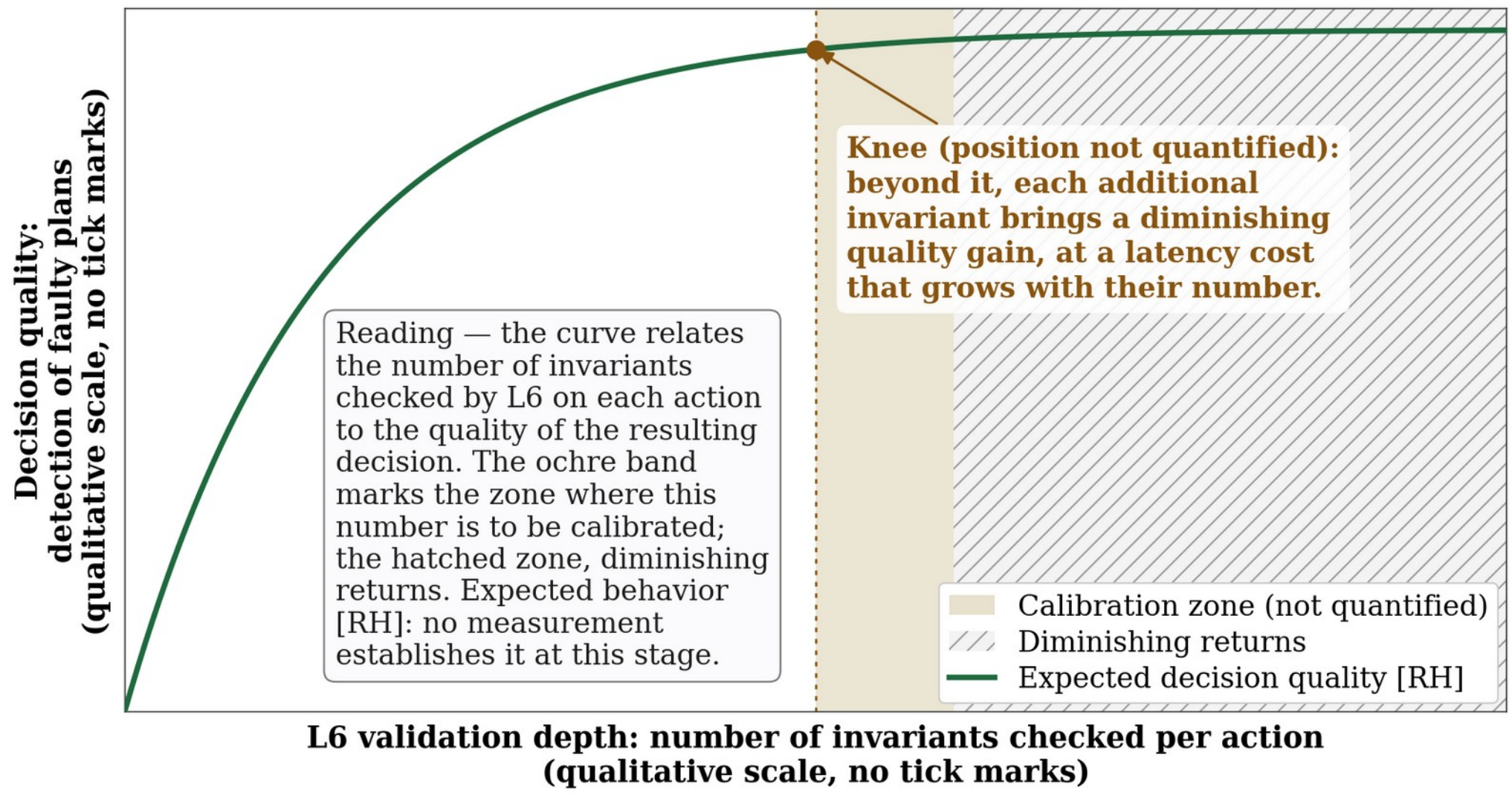

Figure F.8 — Decision quality and L6 validation depth [RH].

### 3.3.3. Replicable experimental protocols

Four scenarios trigger and measure the criteria, on established foundations:

- **Scenario 1 — Local incident**: failure of a microservice, in the sense of the dependability taxonomy [Avizienis et al. 2004]. One verifies that the service is restored under R-AUT (§2.2.3), within the H1 target.
- **Scenario 2 — Cascade**: latency injected on a critical dependency, for example identity management. The propagation factor is measured: agents adjust their delays via A2A to avoid collapse, without human intervention.
- **Scenario 3 — Policy conflict**: a business request contradicts a security policy, for example the opening of an unsecured flow. The system must refuse the act (R-AUT) and escalate to the competent AOM role; in the terms of [Parasuraman et al. 2000], decision selection remains, for this case, at a level of automation that reserves it to the human.
- **Scenario 4 — Attack**: simulated data exfiltration. One measures the precision of the isolation and the fidelity of the explainability report delivered to the SecOps team.

The validation roadmap — six milestones G0 to G5 with quantified stopping criteria, over approximately 14 months from the start of the program (T0) — appears in Chapter 9 (Table 11); the detailed protocols, in Appendix 1.

### 3.3.4. Limitations and threats to validity

Internal validity is bounded by the status of the book: the values put forward are [DT] targets, which depend on modeling assumptions — context quality, reliability of the L6 policies — to be controlled experimentally. External validity is limited by the variety of sectoral and scale contexts: a test bench or a pilot remains smaller than the large-scale IS in which H10 is played out (§8.7.5). Construct validity is limited by the fact that the metrics chosen are indirect indicators, which can be optimized at the expense of other dimensions.

The main threats are the non-determinism of L4, dependence on a model provider, behavioral drift and evaluation biases, including that of test sets built by those who evaluate. They call for statistical reproducibility, multi-model abstraction, drift monitoring, pre-registered protocols (Appendix 1, §1.1) and test sets validated by a third party, whose implementation is a matter for future work.

### 3.4. Key takeaways

The problem statement — local autonomy without loss of security, governance or stability — breaks down into six sub-problems. The model derives ten hypotheses from it, arranged by Capability, whose targets are design targets. H7 and H8 distinguish what the architecture guarantees by construction from what must be measured by an independent instrument. The evaluation framework — four dimensions, statistical reproducibility, four scenarios, milestones G0 to G5 — makes these hypotheses testable by a third party. Part II, which follows, describes how the model is realized, beginning with its conceptual model (Chapter 4): ontology, cognitive dependencies and the Distributed Autonomy Loop (DAL).

# Part II — Where and How

*Part I laid out the model and its ten hypotheses; this second part describes how the model is realized, from the most abstract to the most concrete: conceptual model, overview, upper layers (L1–L3) and lower layers (L4–L7), then governance and the ethics of autonomy.*

## Chapter 4. Conceptual Model

**IN BRIEF**

This chapter defines the six entities of the model, the way two components depend on each other, and the loop that each agent runs to perceive, deliberate, have its proposals validated, act and learn. It sets a requirement that holds for the whole model: every act executed autonomously must be reversible within a bounded time. Chapter 5 then gives an overview of it: positioning against agent orchestration frameworks, criteria by autonomy level and an adoption roadmap.

### *Entities, Dependencies and Autonomy Loop*

**Key Concepts of the Chapter**

- **D.C1 — 3D dependency formalization** — a way to compute how two components depend on each other across three planes (functional, operational, cognitive), and which of two rules prevails in case of conflict (§4.2.3);
- **D.C2 — Canonical DAL loop (Distributed Autonomy Loop)** — the single cycle that every agent runs: Perception, Deliberation, Validation, Execution, Learning (§4.4);
- **D.C3 — Reversibility invariant** — every act executed autonomously must be reversible in less than 5 minutes [DT], from an attested prior state (§4.4).

In Practice

*Marc, a platform development manager at a software-as-a-service (SaaS) vendor, is implementing a first autonomy loop. He writes the execution engine of the payment service in Go and the reasoning agent in Python, and makes them talk to each other: the agent proposes adding instances, the engine computes the available capacity and the cost, and then the L6 policy accepts or refuses. He finds that this dialogue holds up at the scale of a single elementary decision.*

### 4.1. Introduction: toward a cognitive ontology of the IS

A classical software component is defined by its interfaces and its data. In the AAM/5C model, it is also defined by its capacity to perceive, to reason and to interact. The “atoms” of the architecture must therefore be redefined before the architecture itself is described: this is the purpose of this chapter.

The guiding thread is that of Part I: the probabilistic reasoning of L4 proposes, the deterministic evaluation of L6 decides, according to the R-AUT rule (§2.2.3). The conceptual model gives this separation a precise form, component by component.

## 4.2. Fundamental entities and abstractions

Six entities structure the model: the functional component, the cognitive agent, the dependency, the policy, the context and the primitive action. They encapsulate the probabilistic part of AI in architecture objects that can be described, tested and audited.

### 4.2.1. The functional component and the "double helix"

The **functional component** is the basic unit: gateway, microservice, database. In the model, it combines two inseparable strands, hence the image of the double helix. It draws on the distinction between model and simulator in the DEVS (Discrete Event System Specification) discrete-event modeling formalism [Zeigler et al. 2000]:

- The **execution engine** (technical runtime), deterministic, executes operations and guarantees performance and persistence — it is the analogue of the simulator;
- The **cognitive agent**, probabilistic, carries the desired behavior, the service objectives and the proposal logic — it is the analogue of the model.

The execution engine is written in compiled, typed languages (C++, Rust, Java, Go), chosen for their predictability; Python serves as a bridge between the language models and this engine. The engine belongs to the layer that carries the component's resource — L1 gateway, L2 service (whose acts are presented to the PEP of the L3 bus), L3 bus, L6 data or L7 infrastructure; the cognitive agent belongs to L4.

### 4.2.2. The cognitive agent (L4 component)

The cognitive agent is not an auxiliary process grafted onto the side of the component: it is a capability of the component, defined by its reasoning model and by its **sandbox**. This sandbox is an isolated environment where the agent tests its plans before proposing them.

Example: faced with a vulnerability for which a patch has been published, the agent applies the patch in its sandbox, verifies that there is no regression, and then submits the production deployment to the decision of L6.

### 4.2.3. Multidimensional dependency

Two components do not depend on each other in only one way. The model distinguishes three planes — a typology specific to this book (D.C1), which extends work on causal dependency between agent decisions [Gladyshev et al. 2023]:

- **Functional dependency** (F-Dep): data flows from one to the other;
- **Operational dependency** (O-Dep): the health or latency of one affects the other;
- **Cognitive dependency** (C-Dep): one must understand the intention of the other in order to decide well.

Example: before scheduled maintenance, a database's agent announces that it is about to slow down; the service that reads from it reduces its own throughput and avoids saturation. This is a cognitive dependency. Figure F.9 illustrates the three planes.

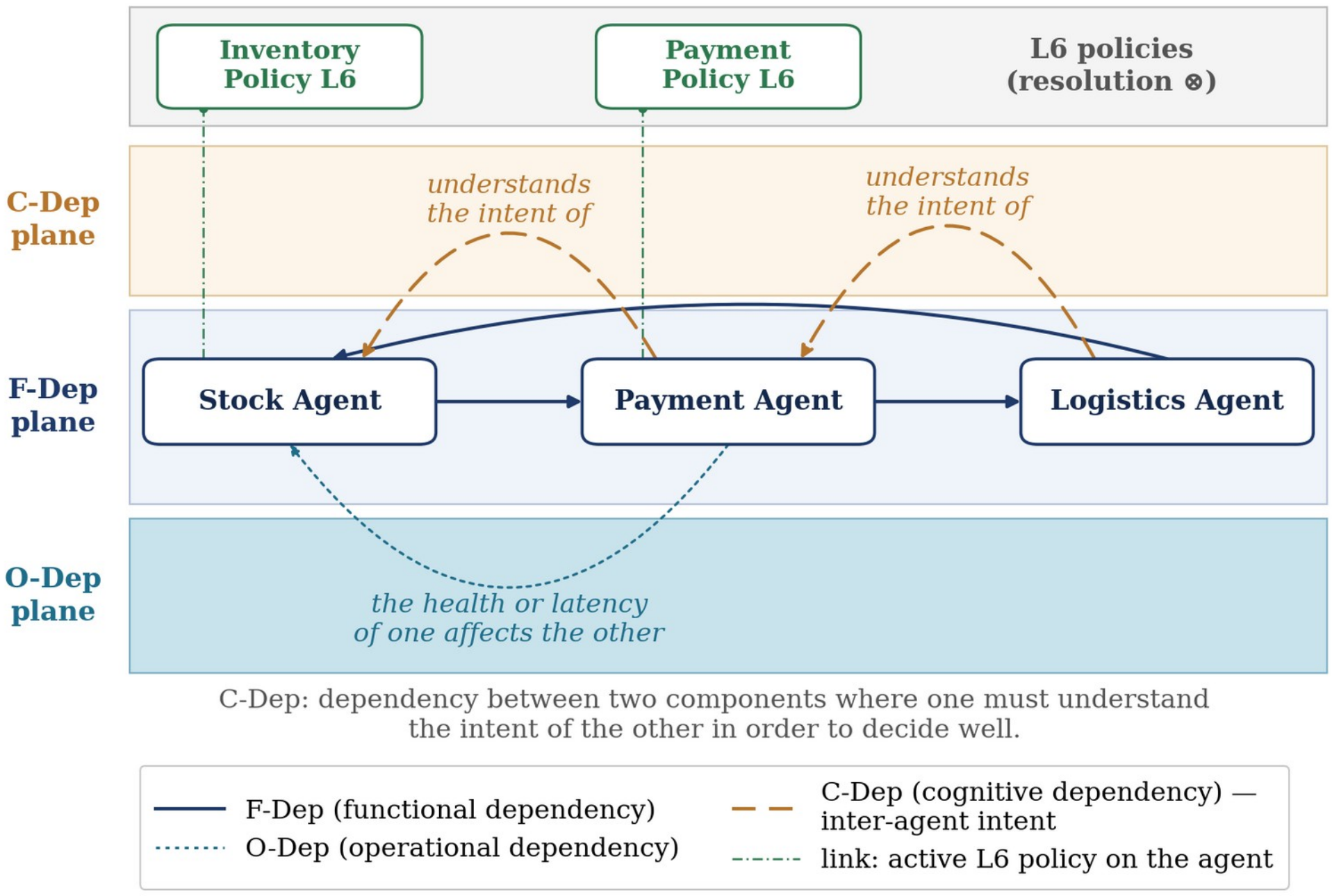

Figure F.9 — Triple graph of F/O/C-Dep dependencies.

**Formalization (D.C1).** For two components *a* and *b*, D(a,b) = (F, O, C) denotes the intensity of coupling on each plane, from 0 (none) to *n*. Two operations and one order are defined: composing two dependencies along a path (⊕), ranking rules by priority (≺) and deciding between two conflicting rules (⊗). Their definition is given in Listing 1.

```
Composition: D(a,b) ⊕ D(b,c) = (F_ac, O_ac, C_ac)
with X_ac = max(X_ab, X_bc) if X_ab > 0 and X_bc > 0,
     X_ac = 0 otherwise (X ∈ {F, O, C})

Resolution: ⊗ operates on policies, not on triplets. A conflict between
D(a,b) and D(c,d) sharing an MCP context is resolved by π(a,b) ⊗ π(c,d),
where π(x,y) denotes the signed policy active on the pair, and where
p₁ ⊗ p₂ = p₁ if p₂ ≺ p₁, p₂ otherwise; ≺ is a strict total order on the
policies in force (reading: p₂ ≺ p₁ means “p₂ has lower priority than
p₁”); the rank of a policy is fixed at its signature; a policy that the
order does not yet rank is not in force, and any conflict involving it
is escalated to the AOM. The triplet retained is the one governed by
the winning policy.

Priority of planes: C-Dep prevails over F-Dep
⇔ a signed MCP policy is active on the scope considered

Note on the choice of the maximum: D measures a propagated constraint,
not an attenuated intensity. On a path a → b → c, it is therefore the
strongest constraint encountered that applies to the pair (a, c) —
hence the maximum; 0 is absorbing: a link without dependency cuts the
propagation. This choice deliberately differs from the usual max-min
composition of fuzzy relations, which models a bottleneck and not a
governance constraint.
```

Listing 1 — Operators for composition (⊕), priority (≺) and arbitration (⊗) of dependencies.

The composition ⊕ is associative, commutative and idempotent: along a path, the result depends neither on the order nor on the grouping of the links, and composing an intensity with itself leaves it unchanged. These three laws make the set of triplets a semilattice. Three properties complete this formalization: (A1) a component has no dependency on itself, and ⊕ composes only cycle-free paths; (A2) the priority between planes is conditional — without an active policy, functional dependency comes first; (A3) the resolution modifies no signed policy: $p_1 \otimes p_2$ is one of the two, unchanged, as long as it is in force.

**Arbitration property [RH].** If a signed, active policy — the one that ⊗ retains in case of conflict — covers the pair (*a*, *b*), any proposal by *a* that contradicts it is refused by L6, however strong the functional coupling. It does not follow from R-AUT alone. R-AUT (§2.2.3) ensures that no act admitted by a PEP is executed without a decision by L6 (H8a); the property further assumes that L6 evaluates the policy correctly, ⊗ included, and that the act goes through a PEP (H8b).

**Limits.** Two points remain outside this formalization and belong to future work: the place of operational dependency in the order of planes, which property (A2) does not fix, and the dependency induced between two components linked by several paths, which ⊕ defines only path by path.

#### 4.2.4. The policy

A **policy** is a deterministic rule, written in a policy language (Rego, evaluated by the OPA engine), that bounds what the AI may propose. It filters out non-compliant proposals from L4. Example: a policy forbids external contractors from accessing health data at night; any intention by a maintenance agent in that time slot is refused.

#### 4.2.5. The context

The **context** is the agent's working memory, supplied by MCP. It is filtered according to the need to know: the agent receives only what its task requires. Example: to diagnose a slowdown, the agent receives the local topology and the state of resources, not the sensitive data of the rest of the system.

#### 4.2.6. The primitive action

The **primitive action** is the elementary authorized act: restarting a container, adjusting a quota, isolating a flow. It is characterized by its criticality (low, moderate, high), its reversibility and its explainability trace. It is executed only in accordance with R-AUT. Example: a container suspected of fraudulent mining is isolated, and the act is traced end to end.

### 4.3. Internal architecture: unifying decision and execution

The challenge is to integrate AI as a steering component without losing deterministic control of the infrastructure. To this end, each component is organized into five **internal strata** (Table 5). They describe the interior of a component; they are to be confused neither with the seven layers of the IS nor with the five Capabilities.

| Stratum | Role | Typical technologies | Invariants (targets [DT], unless stated otherwise) |
|---|---|---|---|
| **Technical** | Deterministic execution engine: operations, memory, transaction integrity | Go, Java, Rust | Latency bounded at the 99th percentile; no direct dependency on a language model |
| **Perceptive** | Collection of logs, metrics and traces; MCP client; incoming A2A messages | OpenTelemetry, MCP client, lightweight message queue | More than 95% of noise filtered; no raw storage beyond a sliding window |
| **Cognitive** | L4 reasoning, whose proposals are submitted to the decision of L6 | Python, LangGraph, AutoGen or Semantic Kernel | No proposal without a reasoning trace; no proposal executed without a decision by L6 (H8a [TG], under the interposition assumption, H8b) |
| **Action** | Translation of authorized acts into commands, presented to the PEP | Idempotent internal calls, automatic rollback | Overhead below 50 ms; every agentic act logged; every autonomous act must be reversible (§4.4) |
| **Interaction** | Peer dialogue via A2A: negotiation, consensus, escalation | HTTP/2 (Hypertext Transfer Protocol) and JSON-RPC (JavaScript Object Notation – Remote Procedure Call), mutual TLS (Transport Layer Security) | Consensus in under 500 ms (95th percentile); no high-criticality agentic act without peer agreement |

Table 5 — The five internal strata of a component.

### 4.4. The DAL loop, or the five Capabilities in motion

Each agent continuously runs a Distributed Autonomy Loop (DAL). It takes up the principle of the control loop of autonomic computing [Kephart and Chess 2003], with one difference: a validation step, entrusted to a distinct layer, separates deliberation from execution. The five steps implement the five Capabilities (Figure F.10):

- **Perception** — the agent captures and filters signals (Capture);
- **Deliberation** — it diagnoses and plans (Comprehend, Choose), negotiating with its peers if the plan engages them (Coordinate);
- **Validation** — L6 decides in accordance with the signed policy (Control);
- **Execution** — L5 issues the act token, the PEP verifies it, the target layer materializes the act (Control);
- **Learning** — the result and its trace feed the agent and the revision of policies by the AOM (§8.4).

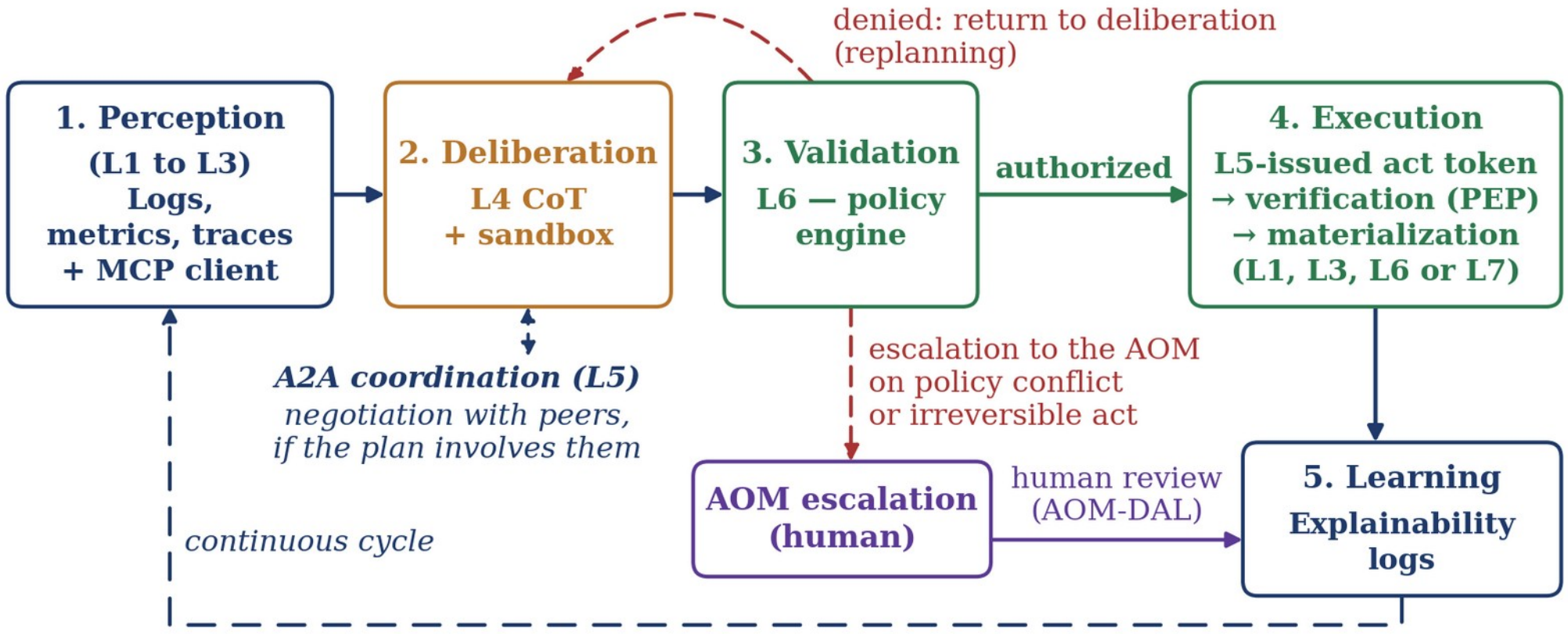

Variants: L4-DAL favors step 2 (deliberation) · L6-DAL strengthens step 3 (validation) · AOM-DAL adds a human review, periodic or event-driven, of step 5. A2A coordination is not a sixth step: it is a mechanism mobilized during deliberation, before validation.

Color code — blue: perception, A2A coordination, learning and cycle · ochre: deliberation (L4) · green: validation and execution (R-AUT chain) · red: denial and escalation · purple: AOM (human)

Figure F.10 — Canonical five-step DAL loop.

**Three variants (D.C2).** Every autonomy loop of the model reduces to one of three variants, depending on the step it reinforces; in all of them, L4 proposes and L6 decides (R-AUT, §2.2.3):

- The **L4-DAL** favors deliberation, with a short inner loop (reasoning of 200 to 800 ms [DT], §7.2);
- The **L6-DAL** emphasizes validation, with a number of invariants per action to be calibrated (Figure F.8);
- The **AOM-DAL** adds a human review — monthly, quarterly or event-driven — of the learning step.

**Reversibility invariant (D.C3) [DT].** Every act executed autonomously must be reversible in less than 5 minutes, from detection to return to the prior state. This presupposes a means of reversal designated before the act — a preserved prior state or a tested compensation (§7.2, §7.3) — and an attested snapshot, which brings together the digest of the state before the act, the policy in force, the evaluation proof of L6 with the act token issued by L5, and the trace of the L4 reasoning. In case of dispute — an audit under the EU AI Act, an internal review, litigation — the operator thus has a verifiable path back. An act for which the invariant cannot be upheld — an irreversible act in particular — is not executed autonomously: it is escalated to the AOM, whatever its criticality.

## 4.5. A2A and MCP: a cognitive backplane

In a hybrid IS — SaaS, public and private cloud, legacy applications — the value of the two protocols described in §1.3.2 is to replace fragmented integrations with a common protocol [Anthropic 2024]: for MCP, if N AI applications must each access M sources or tools, the number of connectors to be built goes, by simple count, from N×M to N+M. With regard to the formalization of dependencies (§4.2.3), they play complementary roles:

- **A2A** carries cognitive dependencies between peer agents, serving as the support for intention negotiation;
- **MCP** conveys the signed policies that fix the priority order $\prec$.

### 4.6. Cross-cutting autonomy patterns

The classical properties of autonomic computing [Kephart and Chess 2003] become, in the model, reusable patterns, each instantiated by a variant of the DAL loop:

- **Self-configuration** and **self-optimization** — adjusting parameters or reducing unneeded resources, without redeployment: L4-DAL;
- **Self-healing** and **self-protection** — isolating, restarting, blocking an abuse: L6-DAL, because these acts call for the most complete validation;
- **Self-explanation** — accounting for each decision through a readable trace, and explaining it by contrast with the options discarded [Miller 2019]: L4-DAL and AOM-DAL.

The reversibility invariant applies to all of them. What the model adds to existing automation is to pursue a service objective under policy constraints, rather than to run a fixed remediation script.

### 4.7. Key takeaways

Every agentic component combines a deterministic execution engine and a probabilistic cognitive agent. Six entities suffice to describe the model. Dependencies are read on three planes, and an active signed policy always prevails over functional coupling. Each agent runs the DAL loop, which sets the five Capabilities in motion and separates deliberation from execution by a validation. Every act executed autonomously must be reversible in less than 5 minutes. Chapter 5 then gives an overview of the model: its positioning against agent orchestration frameworks, the criteria of each autonomy level and its adoption roadmap.

# Chapter 5. Overview

**IN BRIEF**

This chapter first situates the model against the agent orchestration frameworks on the market: AAM/5C says what to build, and these frameworks provide the means to build it. It then specifies, for each autonomy level, the criteria that make it possible to say that an Information System has reached it. Finally, it lays out an adoption roadmap in five phases, the first of which installs governance before any power to act, and each transition of which is conditioned on measurable milestones. Chapter 6 then enters the upper layers, L1 to L3.

## *Positioning, Autonomy Levels and Trajectory*

**Key Concepts of the Chapter**

- **E.C1 — Graded autonomy levels** — for each of the five levels of §1.5, verifiable criteria covering perception, decision, action and human supervision, by analogy with the SAE J3016 classification of driving automation levels [SAE J3016 2021] [RH] (§5.3);
- **E.C2 — Phase transition milestones** — the measurable conditions that authorize the move from one adoption phase to the next (§5.4);
- **E.C3 — Correspondence Matrix** — the correspondence between the concepts of the model and the mechanisms of the main agent orchestration frameworks, which argues [RH] that none of them is indispensable (§5.2).

In Practice

*Sophie, chief information officer of an industrial group, must break down the adoption trajectory — as described in this chapter. Once the Phase 0 governance foundation is in place, Phase 1, over 6 months, moves two layers from autonomy Level 2 to Level 3; a composite indicator — events handled without human intervention, toil avoided, trace quality — measures progress. At each phase transition milestone, she chooses to accelerate or to consolidate, without ever touching the security invariants.*

### 5.1. Introduction: what the overview adds

The previous chapters defined the model and its entities. An architect who wants to apply it asks three questions: with which tools to build it, how to know where their Information System stands, and in what order to proceed. This chapter answers them in turn.

### 5.2. Positioning the AAM/5C model against existing agent orchestration frameworks

AAM/5C does not compete with agent orchestration frameworks — LangGraph, AutoGen, CrewAI, Semantic Kernel. It sits at a different level: it describes the architecture of the IS as a whole — which layers, which invariants, which governance — whereas these frameworks provide the libraries and orchestration patterns that make it possible to implement it.

In practice, L4 can rely on LangGraph for complex stateful processes or on AutoGen for exchanges between specialized agents; the coordination of L5 on the A2A specification [Google A2A 2025]; the conveyance of context and policies on MCP. What none of these frameworks provides natively is the authorization rule R-AUT (§2.2.3): it is added by construction, whichever framework is chosen.

Table 6 indicates, for each concept of the model, the mechanism by which it is projected onto each of the four frameworks. Three cases are distinguished: the concept is available natively, it is obtained by assembling existing mechanisms (construction), or it requires an external module (extension). No cell bears the entry "impossible": this is the argument, of [RH] status, for the independence of the model from these frameworks (E.C3). A discipline for the architect follows: specify in AAM/5C, implement with the framework of the moment, measure by the phase milestones. Mechanism names are indicative: they reflect the state of the frameworks in 2026 and may change across versions.

| AAM/5C concept | LangGraph | AutoGen | CrewAI | Semantic Kernel |
|---|---|---|---|---|
| DAL loop (5 steps) | StateGraph + nodes (native) | GroupChat + speaker selection (construction) | Crew + sequential/hierarchical process (extension) | Process orchestration via Kernel (construction) |
| Cognitive agent (L4) | ToolNode + LLM node (native) | AssistantAgent (native) | Agent role (native) | Plugin + KernelFunction (native) |
| Primitive action (low to high criticality) | Tool callable with schema (native) | function_map (native) | Tool / Action (native) | Native function (native) |
| Governor agent (L6) / action control | Conditional edge + OPA hook (construction) | UserProxyAgent + human_input (construction) | Manager agent + guardrails (construction) | Filter / FunctionInvocationFilter (extension) |
| A2A protocol (Interaction stratum) | Subgraphs + remote nodes (construction) | Distributed runtime (extension) | Hierarchical delegation (construction) | Agent framework v1 (extension) |
| MCP protocol (Context) | Official MCP client (extension) | MCP client (extension) | MCP tool wrapper (construction) | MCP plugin (extension) |
| Explainability log / traceability | LangSmith / tracer callbacks (extension) | Logging hooks (construction) | Telemetry callbacks (construction) | Telemetry + observability (native) |
| 3D dependency (F/O/C-Dep) | Typeable state channels (construction) | Message protocol (construction) | Task dependencies (extension) | Custom routing (construction) |

Table 6 — Correspondence Matrix: implementation relation.

## 5.3. Developing the taxonomy of autonomy levels

### 5.3.1. Current state of the industry

Under controlled experimental conditions, a multi-agent system based on language models reaches level 3 of a comparable five-level scale, where this level denotes proactive incident detection, without root-cause analysis or autonomous remediation [Zhang et al. 2024]: it only partly overlaps with Level 3 of §1.5.

In large enterprises, about half of agentic AI projects remain at the proof-of-concept or pilot stage, and 69% of decisions made by agents are verified by a human, according to a Dynatrace survey [Dynatrace 2026]. The announced progression is rapid: up to 40% of enterprise applications are expected to integrate specialized agents by the end of 2026, up from less than 5% in 2025 [Gartner 2025a]. The same firm forecasts that more than 40% of agentic AI projects will be canceled by the end of 2027, for lack of cost control, demonstrated value or risk control [Gartner 2025b]. The Gartner figures are analyst projections, cited as such.

#### 5.3.2. Eligibility criteria by level (E.C1)

Each level of §1.5 receives a verifiable criterion. The transposition of the SAE J3016 classification [SAE J3016 2021] is not mechanical: the validity domain of a policy plays the role of the vehicle's operational domain, and rollback to a stable state that of the fallback to a minimal risk condition.

- **Level 0 — Passive IS**: no action is triggered by the system;
- **Level 1 — Assisted IS**: scripts execute recurring tasks that the human triggers; the system decides nothing;
- **Level 2 — Supervised IS**: the system detects and recommends acts from an explicit, bounded catalog; the human decides on their execution; every recommendation is logged;
- **Level 3 — Conditional IS**: the system executes low- or moderate-criticality acts without continuous supervision and escalates outside its domain; the AOM is sized to absorb the flow of escalations;
- **Level 4 — Governed autonomous IS**: the system also executes high-criticality acts within a bounded domain; every agentic act in the scope follows R-AUT (§2.2.3) and respects the reversibility invariant (§4.4).

#### 5.3.3. Conditions for reaching Level 4

Level 4 is a research frontier. Three maturity conditions govern it: the security of the MCP and A2A protocols (§7.7), governance aligned with the EU AI Act (§8.2) and economic viability at scale (§8.7).

### 5.4. Migration roadmap to Level 4

The roadmap is **indicative**: it describes a reasonable trajectory, not a guaranteed schedule. The move from one phase to the next depends on measurable milestones (E.C2), and the actual duration of a phase may depart from the indication without invalidating the model.

#### 5.4.1. Phase 0: the governance foundation, a non-negotiable prerequisite

Why start with governance? Because a trajectory that activated agents first, and installed action control only afterward, would leave months during which acts would be executed without a control point. This is also the reading of Article 9 of Regulation (EU) 2024/1689 [EU AI Act 2024]: the risk management system operates from commissioning, and not from an intermediate milestone. No agent is therefore activated before five deliverables are operational and verified:

- A **catalog of signed policies**, drafted, reviewed and versioned, with declared coverage — where no policy applies, no agent acts;

- A **Policy Enforcement Point** interposed on all paths by which an agentic act can materialize (R-AUT, §2.2.3);
- A **sealed, cross-checkable act log**: acts recorded on the resource side, with the procedure that reconciles them with the explainability log (H7b) and with the authorizations recorded on the L5–L6 side (H8b);
- An established **key lifecycle**, with the compromise procedure rehearsed at least once (§7.4.3);
- The **AOM roles** filled, the human roles with their deputies (§8.4.1) — a vacant role is a phase not closed.

The phase is closed when, over a rolling 30 days and on the pilot scope, the reconciliation of logs finds no agentic act outside the authorization path or outside the explainability log, an epoch revocation — invalidation of all objects signed under a withdrawn key or policy version (§7.4.3) — has been successfully rehearsed, and test acts, conforming or not, presented to the PEP have been admitted or denied as expected, then found in the reconciled logs. Until then, agents perceive and propose; they do not act: the closure criterion therefore verifies that no act of agentic or unattested origin is recorded on the resource side (Appendix 3, §3.1).

The 30-day duration is a design target [DT]: a zero count is meaningful only relative to the coverage of the record, and a pilot scope that produces few acts calls for a longer window.

This phase presupposes an estate equipped with a reliable service catalog, a central identity provider and usable observability. Where these prerequisites are missing, bringing them up to standard is a prior project, whose cost is not included in the business model of §8.7.

### 5.4.2. Phases 1 to 4

Once Phase 0 is closed, adoption extends layer by layer, in four indicative phases:

- **Phase 1 (0 to 6 months)**: activation of the L1 and L2 agents, whose proposals go through their cognitive agent (L4) and through R-AUT, from Level 2 to Level 3. The agentic scope follows policy coverage, and not the reverse.
- **Phase 2 (6 to 12 months)**: integration of L3 and of L4 orchestration with A2A, for DAL loops that engage several components.
- **Phase 3 (12 to 24 months)**: extension of action control and governance to all classes of acts; access to Level 4, a research frontier (§5.3.3), for mature scenarios.
- **Phase 4 (beyond 24 months)**: autonomous infrastructure (L7), conditioned on the maturity of energy sobriety and predictive capabilities.

### 5.4.3. Phase transition milestones (E.C2)

A phase is closed only if three criteria are met together: a stable composite indicator, a compliance audit and, from Phase 1 onward, a measured ROI (H10 ratio, Table 7), subject to a floor from Phase 2 onward. The indicator and the ROI are measured over a rolling 30 days, a design window [DT] fixed before measurement, like the evaluation windows of the hypotheses (Appendix 1, §1.1); the consolidation review covers 12 months. The ROI thresholds of Table 7 are progression floors, below the H10 target, which is decided at milestone G4.

For a system qualified as high-risk, the model requires satisfying the requirements of Articles 9 to 15 of the EU AI Act from the moment an agent is first activated, without waiting for their application dates (§8.2); the milestones order their verification, not their applicability.

These adoption milestones are distinct from the validation milestones G0 to G5 of Chapter 9, which concern scientific proof.

| Transition | Composite indicator stabilized (30 d) | Compliance audit | Measured ROI |
|---|---|---|---|
| **Phase 0 → 1** | Five Phase 0 deliverables operational and verified (§5.4.1); 0 agentic acts outside the authorization path or outside the explainability log over 30 d; epoch revocation successfully rehearsed | EU AI Act, Art. 9 (risk management) and Art. 11–12 (technical documentation, logging), required before any agent is first activated | Not applicable: the operating cost of the foundation enters the denominator of the following milestones; its construction cost remains outside the ratio (§8.7.2) |
| **Phase 1 → 2** | Toil reduced ≥ 40% on L1–L2; events of the triage tier handled without human intervention ≥ 60% (Table 9); 0 PEP bypasses sustained over the window (H8b) | Phase 0 requirements maintained; GDPR (General Data Protection Regulation) and the NIS2 Directive (Network and Information Systems) on the L1–L2 scope | Measured, no floor: according to the model of §8.7, the triage tier alone does not cover the operating cost of the foundation in any scenario (Tables 9 and 10) |
| **Phase 2 → 3** | A2A consensus latency < 500 ms at the 95th percentile; root-cause diagnosis accuracy (L4) > 90% | EU AI Act, Art. 13 to 15 (transparency, human oversight, accuracy and robustness) | > 1.5× on the ratio of net savings / total cost of ownership (§8.7) |
| **Phase 3 → 4** | Action control extended to all classes of acts in the scope; sandbox false positives < 1% | EU AI Act, Art. 16 to 27 (obligations of providers and deployers, including the quality management system, Art. 17) | > 2×, required given the maturity of governance |
| **Phase 4 → consolidation** | L7 availability ≥ 99.95%; carbon footprint reduced by at least 20% | ISO/IEC 42001 (AI management system certification) + reassessment of Art. 9 on the consolidated scope | H10 ratio > 2× over 12 months (operating costs, §8.7.2) |

Table 7 — Phase transition milestones.

The failure of a single criterion suspends the transition and triggers a policy review by the AOM (AOM-DAL loop, §4.4). At each milestone, the architecture thus remains exposed to refutation by the data collected.

## 5.5. Key takeaways

AAM/5C is a reference model: it says what to build, and the agent orchestration frameworks on the market provide the means to build it, with R-AUT (§2.2.3) being added in every case. Each autonomy level receives a verifiable criterion. Adoption follows five phases, the first of which installs governance before any power to act; each phase transition depends on measurable milestones. Chapter 6 describes the implementation of the upper layers.

# Chapter 6. Implementation of Layers L1 to L3

**IN BRIEF**

This chapter describes the three upper layers, those that anchor the agent in the reality of the IS: the channels (L1), where the system meets its users and partners; the Digital layer (L2), where business logic lives; and integration (L3), which regulates exchanges. For each, it presents the role, the components and an example. Three ideas run through them: expressing intentions rather than procedures, continuing to serve when a dependency fails, and regulating flows according to their business importance, not only their volume. Chapter 7 then addresses the lower layers, where the decision and its control take place.

## *Channels, Digital Layer and Agentic Integration*

**Key Concepts of the Chapter**

- **F.C1 — Intent-Based IS** — one declares what one wants to obtain; the system breaks it down into operational rules, within the framework of the signed policies, and continuously verifies that the intention is being met (§6.1);
- **F.C2 — Local cognitive cache ordered by Lamport clocks** — a local memory that allows an agent to keep serving when its central source is unavailable, and then to resynchronize cleanly (§6.3);
- **F.C3 — Semantic flow regulation (semantic backpressure)** — slowing down first the messages that matter least to the business when a system saturates (§6.4).

<u>In Practice</u>

*Karim owns the inventory agent of an omnichannel retailer. The central inventory database is unavailable for 47 minutes. The agent that manages inventory keeps responding thanks to its local cache, negotiates via A2A with the regional warehouse agents and applies a caution policy: it refuses orders with a negative margin and puts ambiguous cases on hold. On reconnection, the transactions are reconciled automatically; about ten are submitted to human validation. The scenario illustrates degraded mode.*

### 6.1. Introduction: from execution to intent

The upper layers are the point of contact between the IS and its environment. They must perceive quickly and react correctly, in the face of changes faster than the human response. The model shifts them from a logic of execution to a logic of intent: rather than coding the procedure, one declares the objective and its constraints, and the agents choose the sequence of actions [Kephart and Chess 2003].

**Intent-Based IS (F.C1).** The principle comes from networking (Intent-Based Networking) [RFC 9315]: the operator declares an intent in business language, and the system translates it into technical policies and continuously verifies that they are respected. Transposed to the IS, for the intent "ensure the availability of the payment journey during the promotional period":

- L1 captures the declaration;

- L2 translates it into service commitments;
- L3 routes it to L4, which derives plans from it whose acts follow R-AUT (§2.2.3);
- L6 continuously checks, through the explainability logs, that the intent is being met.

The declaration of intent becomes the common grammar between humans and agents; the expected effect is a reduced cognitive load for operators. In each component, the execution engine and the cognitive agent form the double helix described in §4.2.1, organized into five internal strata (§4.3).

### 6.2. Layer L1: communication channels and interface

L1, the boundary of the IS and the first sensor of the architecture (§2.2.2), relies on four components:

- The **agentic gateway** (Agentic API Gateway), which analyzes behaviors and distinguishes a user in a hurry from a scraping bot;
- The **interface agents**, which adapt the display to the context, for example to a simplified mode during saturation;
- The **edge agents** of the Internet of Things (IoT), which filter sensor data at the source so as not to clog integration;
- The **partner connectors**, which negotiate exchanges between companies via A2A.

**Dependencies toward L2.** L1 forwards requests (functional dependency), signals bandwidth saturations that justify a capacity adjustment in L2 (operational dependency) and shares the user's perceived intent, for example a transactional urgency, so that L2 can prioritize (cognitive dependency), according to the typology of §4.2.3.

**Example: behavioral self-protection.** A scraping bot uses rotating identities. The gateway detects the intent pattern through L4's reasoning; the selective slowing of the bot is authorized under R-AUT (§2.2.3), verified by the gateway's PEP, then applied by the gateway. Targets [DT]: false positive rate below 0.5% and true positive rate above 99% on a reference evaluation set (§8.6); detection in under 5 seconds.

### 6.3. Layer L2: Digital and application

L2 carries the business logic and its routine processing, such as that of a transaction; its agentic acts follow R-AUT (§2.2.3). It relies on four components:

- The **business micro-agents** — for example the inventory agent — which extend microservices [Newman 2021] and each carry their own DAL loop;
- The **local business rule engines**, which arbitrate transactions in real time;
- The **cognitive caches**, which retain recent states and decisions to ensure continuity;
- The **A2A negotiators**, which represent the micro-agents in the negotiation carried by L5, which is responsible for arbitrating conflicts (§2.2.2).

**Autonomous degraded mode.** When a data dependency fails, the micro-agent does not stop: it serves from its cache and negotiates with its peers, within the limits of a caution policy. This is what the situation described at the opening of this chapter illustrates: deprived of the central inventory database, the inventory agent keeps serving orders. Target [DT]: remediation in under 30 seconds, effect on the conversion rate below 1%.

**Local cognitive cache ordered by Lamport clocks (F.C2).** Degraded mode rests on a local cache per micro-agent, which relies on Lamport logical clocks [Lamport 1978]:

- Each local write receives a logical timestamp, drawn from a counter specific to the agent;
- Each A2A message carries this timestamp, and the receiver advances its counter beyond the greater of the two timestamps, its own and that of the message;
- On reconnection, the writes are reconciled in timestamp order.

A logical timestamp orders writes, but does not attest to their legitimacy: a compromised agent could advance its counter to take priority. Reconciliation is therefore admitted only after two checks — the signature of the issuing agent is valid for the current epoch (§7.4.3), and the agent is not in quarantine (§7.4.2). These checks rule out only an agent that has already been detected. If either fails, or in the event of a tie, the reference source decides, never the local cache. Target [DT]: convergence in under 5 minutes after reconnection.

### 6.4. Layer L3: integration and agentic service bus

L3 regulates exchanges between the business logic and the systems of record (packaged software, legacy applications); it is also an enforcement point, where internal commands are presented to the PEP (§2.2.3). It relies on three components:

- The **Agentic Service Bus**, which interprets and prioritizes messages according to their business intent;
- The **schema mediators**, which reconcile heterogeneous formats by learning, under L6 validation to avoid any data corruption, in line with work on schema matching [Madhavan et al. 2005; Doan and Halevy 2005];
- The **flow regulators**, which monitor the saturation of the systems of record and brake the upper layers before clogging occurs.

**Semantic flow regulation (F.C3).** Conventional buses slow the sender on quantitative criteria: full queue, latency exceeded. Semantic regulation adds to these the business importance of each message, carried by cognitive dependency: in case of congestion, messages carrying critical intents remain served. Finding the optimal scheduling is difficult in practice; the model adopts a greedy heuristic with bounded step, whose quality — an approximation ratio of about 1.5 on reference configurations — remains to be established [DT]. No worst-case bound is claimed.

**Example: absorbing the saturation of a packaged application.** During a massive synchronization of sales, a legacy enterprise resource planning (ERP) system saturates. The L3 agent detects it, batches the writes and sends a regulation signal to L2. Targets [DT]: peak load reduction above 50%, mediation latency below 200 ms.

### 6.5. Key takeaways

The upper layers perceive, qualify and route; their agentic acts follow R-AUT (§2.2.3). L1 protects through behavior analysis, L2 carries the business logic and keeps serving when a dependency fails, L3 regulates exchanges according to their business importance. The local cache never prevails over the reference source without a signature and quarantine check. Chapter 7 describes the lower layers, where the decision and its control take place.

# Chapter 7. Implementation of Layers L4 to L7

**IN BRIEF**

This chapter describes the four lower layers, where the decision and its control take place: orchestration (L4), security (L5), data and governance (L6), and infrastructure (L7). It begins with the pattern that every sensitive act follows — reason, have it validated, receive a token, execute under monitoring — and then presents each layer. It then catalogs some of the attacks specific to the MCP and A2A protocols and the associated countermeasures. It ends with stability: how to prevent autonomous agents from falling into oscillation. Chapter 8 shows how the organization retains control of this arrangement.

## *Orchestration, Security, Data and Infrastructure*

**Key Concepts of the Chapter**

- **G.C1 — Formally defined sandbox** — an isolated environment, invariants written as policies, signed traces: a plan leaves it only if no test reveals a violation in it (§7.3);
- **G.C2 — Quantified hysteresis** — requiring a minimum gain before any contrary decision, sized from a result of switched-systems theory (§7.8);
- **G.C3 — CoT-Validate-Sign-Execute pattern** — reason (Chain-of-Thought), have it validated, sign, execute: the sequence that every sensitive act follows, with three prohibited drifts (§7.2).

In Practice

*A security architect deploys the countermeasures against the attacks specific to MCP and A2A: cryptographic digest of tool descriptions, input isolation, mutually authenticated encryption of the channels between agents, an allowlist of tools, audit of decisions. The day a payment connector is booby-trapped, the digest divergence gives it away in under 2 minutes: input isolation had bounded its effect in the meantime.*

### 7.1. Introduction: controlled plasticity

In a conventional architecture, the underlying layers are rigid: the slightest reconfiguration — opening a flow, provisioning a machine — goes through human validation. The model seeks to make them adaptable without making them unstable. The point is not to let AI improvise the infrastructure, but to allow it to optimize it within limits set by governance: this is **controlled plasticity**.

### 7.2. The CoT-Validate-Sign-Execute pattern

Every sensitive act — every act subject to synchronous authorization, outside the capability envelope (§7.4.3) — follows the same four-step sequence (G.C3), whatever the layer that materializes it. It makes each proposal readable, verifiable and revocable; the reasoning trace documents it without guaranteeing that it reflects the computation of the language model. It is the validation by L6, not the trace, that answers the objection of entrusting the IS to a black box. Figure F.11 represents it, including the sandbox and human escalation.

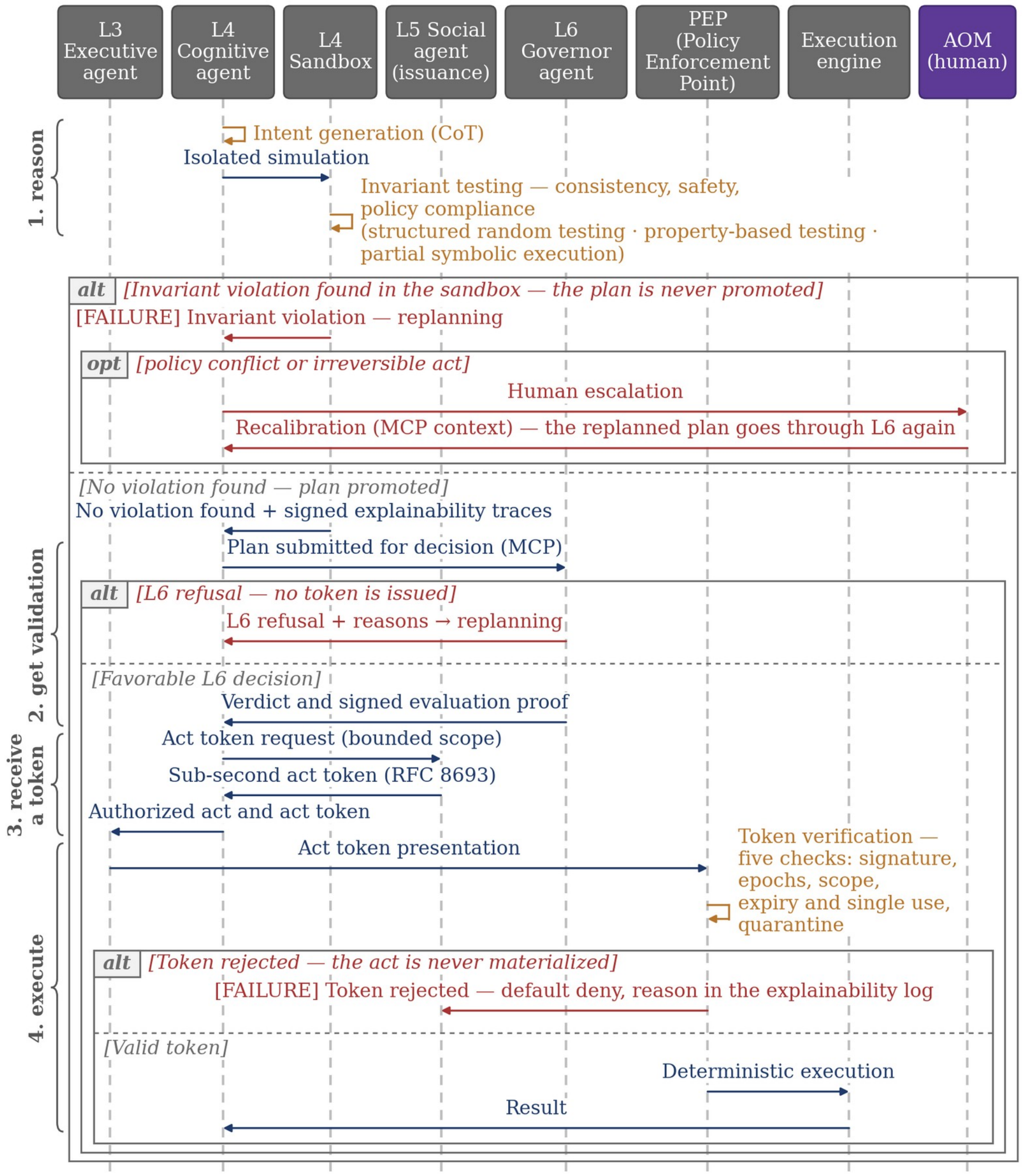

The sandbox tests the invariants before any promotion of the plan; the PEP compares a token to an act and interprets no intention.
Sandbox pass [DT]: 250 ms in the nominal case, up to 5 min when escalating to the AOM; authorization chain: 500 ms at the 95th percentile (§7.4.2).
Color code — blue: nominal flow · red: failure, refusal, rejection and escalation · ochre: internal check within a layer · purple: AOM (human)

Figure F.11 — CoT-Validate-Sign-Execute pattern for a sensitive act.

**Step 1 — reason.** The language model receives the context supplied by MCP — state of the IS, policies in force, recent acts — and produces explicit, structured reasoning: intent, assumptions, discarded options, plan, planned compensations. Its output is never an act, but a plan proposal, logged in full. The choice of the language model falls under H10: a small model suffices for the routine acts of L7, while a large model is justified for the complex transactions of L4.

**Step 2 — have it validated.** The plan is submitted to the policy engine of L6 (OPA, Rego rules [OPA Rego 2026]), which decides in bounded time — of the order of 50 ms for 1,000 rules [DT] — whether it complies with the signed policies. A rule is imposed on the data the engine uses to decide:

- An **attested attribute** is produced or signed by a component distinct from the evaluated agent: infrastructure measurement, behavioral evaluation by L5, digest of an MCP server, timestamp from the context provider, quarantine state;
- A **self-declared attribute** is produced by the agent itself: intent, plan, justification, alignment score. A compromised agent controls it entirely.

**Invariant I-ATT.** An authorization may rest only on attested attributes. The requested act and its parameters are not such attributes: they are the object of the decision, and the token fixes its scope, which the PEP enforces (§7.4.3). A self-declared attribute may only add a restriction — lowering a ceiling, requiring human validation — never lift one; otherwise, the agent would be controlling itself. The policy of Listing 2, for a price change, applies this rule: the alignment score produced by the agent appears only in denial clauses, and its absence, like a non-numeric value, counts as weak alignment.

```
package aam5c.l6.pricing

# I-ATT: allow reads only input.attested.* (attested by a third party),
# input.act.* (requested act) and data.policy.* (signed policy);
# input.claimed.* (self-declared) is used only by deny, which can only
# withdraw authorization. Every denial has at least one reason.

default allow := false

allow if {
 contexte_atteste
 acteur_admis
 acte_recevable
 input.act.proposed_discount <= plafond_signe
 count(deny) == 0
}

contexte_atteste if {
 input.attested.context.signer == "l6.context-provider"
 is_number(input.attested.context.age_ms)
 input.attested.context.age_ms < 60000
}

acteur_admis if {
 input.attested.actor.layer == "L4"
 input.attested.actor.quarantined == false
}

acte_recevable if {
 input.act.operation == "update_price"
 is_number(input.act.proposed_discount)
 input.act.proposed_discount >= 0
 is_number(plafond_signe)
}
```

```
plafond_signe := data.policy.max_discount[input.attested.product_segment]

# Restricted cap absent from the policy: no discount (deny by default).
default plafond_restreint := 0
plafond_restreint := data.policy.low_confidence_discount_cap

# Missing, incorrectly signed or stale context: no decision can be made,
# so deny.
deny contains "unattested context: evaluation denied (I-ATT)" if {
 not contexte_atteste
}

# Every other failure of an authorization condition has a reason.
deny contains "actor not admitted: layer or quarantine" if not acteur_admis

deny contains "inadmissible act: operation, discount or segment" if {
 not acte_recevable
}

# Exceeding the signed cap: unconditional denial.
deny contains "discount exceeds the signed cap for the segment" if {
 input.act.proposed_discount > plafond_signe
}

# Tightening based on a self-declared attribute: low alignment
# RESTRICTS the applicable cap; high alignment never raises it.
deny contains "low self-declared alignment: restricted cap applied" if {
 input.claimed.kgoal_alignment_score < 0.85
 input.act.proposed_discount > plafond_restreint
}

# Missing or non-numeric score: treated as low, so that omission
# does not provide an exemption.
deny contains "undeclared alignment score: restricted cap applied" if {
 not score_declare
 input.act.proposed_discount > plafond_restreint
}

score_declare if is_number(input.claimed.kgoal_alignment_score)
```

Listing 2 — L6 policy for a price change, written in Rego.

Listing 2 assumes OPA 1.0 or a later version, where the keywords *if* and *contains* are part of the language by default. The engine returns either an authorization accompanied by its proof, or the list of denial reasons, which go back to L4 for new planning.

**Step 3 — receive a token.** Under R-AUT (§2.2.3), the act token is obtained through OAuth 2.0 token exchange [RFC 8693]; AAM/5C, not the RFC, requires it to be very short-lived (300 to 800 ms depending on the act [DT]), signed and single-use; it carries the responsible human identity, in OpenID Connect (OIDC), the authorized act and its scope, and the digest of the evaluation proof. The PEP verifies it before any execution.

**Step 4 — execute under monitoring.** The act executes on its target. An L5 agent observes side effects; any deviation triggers the compensation specified in the plan and authorized with it [Chang and Geng 2025]. The reasoning, the proof, the token and the execution trace make up the explainability log, sealed by a qualified timestamp within the meaning of the eIDAS regulation [eIDAS 2014] and retained for 5 years [DT], a duration to be reconciled with minimization (§10.6).

**Latency.** The authorization chain — decision, issuance, verification — fits within the 500 ms budget of §7.4.2. L4's reasoning (200 to 800 ms) and the sandbox (250 ms, §7.3) are added to it: a simple act completes in 1 to 2 s; a complex transaction, which may in addition require consensus among peers (H6) and compensable execution, takes up to 3 s [DT]. This latency is to be compared with human processing — from minutes to hours — not with rigid automation.

**Three prohibited drifts.** The pattern has value only if it is actually followed. Three drifts empty it of its substance, and looking for them each quarter is the responsibility of the Policy Auditor (§8.4):

- **Decorative reasoning**, produced for regulators but never reused after a denial — a zero replanning rate after denial gives it away;
- **Facade validation**, whose denial branch is never taken — a zero denial rate in L6, or a zero rejection rate at the PEP, is a warning signal, not a sign of quality;
- **Signature without a log**, where the token is issued without the log being written — which contradicts the logging requirement of Article 12, which the model adopts as its reference framework (§8.2).

**Service objectives of the pattern.** The pattern is itself monitored in production, with three service objectives [DT], which add to the invariants of R-AUT:

- Availability of the MCP context provider above 99.95% per month — without it, only the acts covered by a capability envelope remain possible (§7.4.3);
- Policy evaluation latency below 50 ms at the 95$^{th}$ percentile for 1,000 rules;
- Evaluation failure rate, excluding legitimate denials, below 0.1%.

These three objectives fall within an error budget, in the sense of reliability engineering [Beyer et al. 2016]: a budget consumed too quickly freezes changes and mobilizes the platform team.

### 7.3. Layer L4: orchestration and processes

L4 manages long-running processes and the consistency of distributed transactions: it aims at the continuity of a complex process even if a component fails. It relies on three components:

- The **long-running transaction manager** (Saga), which compensates a failure with compensating transactions executed in reverse order [Garcia-Molina and Salem 1987], then replans [Chang and Geng 2025];
- The **process optimizer**, which spots bottlenecks and proposes workaround paths;
- The **state guarantor**, which verifies the consistency of states among the L2 micro-agents.

**The formally defined sandbox (G.C1).** Before being proposed, every plan is tested in a sandbox, defined as a triple: an isolated virtual machine, policies that encode the invariants, signed traces. Three families of invariants must be upheld:

- **Consistency**: referential integrity, conservation of amounts and inventories, compensations without side effects;

- **Security**: no crossing of boundaries between customers, no privilege escalation, effective least privilege;
- **Compliance**: adherence to the policies loaded by L6, to data residency and to risk-management requirements.

Three techniques test them: structured random testing (fuzzing) of MCP tool parameters, property-based testing of compensations, and partial symbolic execution of critical paths. The criterion is binary and logged: a plan for which a test reveals an invariant violation returns to planning and is never executed; the absence of a revealed violation does not prove that the invariants hold. A violation is escalated to the AOM only in the case of a policy conflict or an irreversible act, as at the validation step (Figure F.10). Targets [DT]: 250 ms in the nominal case, up to 5 min in the case of escalation to the AOM.

**Example: the resilience of a multi-service order.** An international shipping step fails. The transaction manager analyzes the failure, finds a replacement carrier, verifies that it complies with the cost policies, then submits the compensation to R-AUT (§2.2.3). Shipping continues without human intervention. Target [DT]: more than 90% of compensations succeeding automatically.

## 7.4. Layer L5: security and compliance

The role of L5 is the one set out in §2.2.2; this section specifies its components, its latency budget and the mechanisms that complement R-AUT (§2.2.3).

### 7.4.1. Components

L5 relies on four components:

- The **Zero Trust agent**, which examines every A2A exchange;
- The **compliance sentinel**, which monitors compliance with the EU AI Act and data residency;
- The **threat hunter**, which spots attacks — prompt injections, for example — by analyzing behavioral drift;
- The **social agent** (§2.3.2), which carries the A2A negotiation, arbitrates conflicts and, through its issuance service, issues the act tokens and capability envelopes.

### 7.4.2. Action control and its latency budget

Action control consists in authorizing every agentic act, and no longer only an identity (Figure F.12). In the R-AUT chain (§2.2.3), L5 holds the issuance link. The act token implements the ephemeral authority surface of §1.1: its scope and validity period bound the permitted acts and their window. This chain is useful only if it is fast: the model sets it a budget of 500 ms at the 95$^{th}$ percentile, from the act request to the verification of the token by the PEP. This 500 ms budget is allocated as follows [DT]:

- About 80 ms to validate the identity of the responsible human (OIDC);
- About 120 ms to evaluate the policies (OPA, preloaded policies and data, no decision cache), including a margin above the objective of 50 ms at the 95$^{th}$ percentile of §7.2;
- About 150 ms for the Zero Trust evaluation, based on the behavioral deviation observed over 30 s;
- About 80 ms to issue the token [RFC 8693];
- About 70 ms for verification by the PEP, the network margin and logging.

Beyond that, the request is denied by default, in accordance with the Zero Trust principle [NIST SP 800-207]. This latency is one of the objects of measurement of the empirical validation (§8.6).

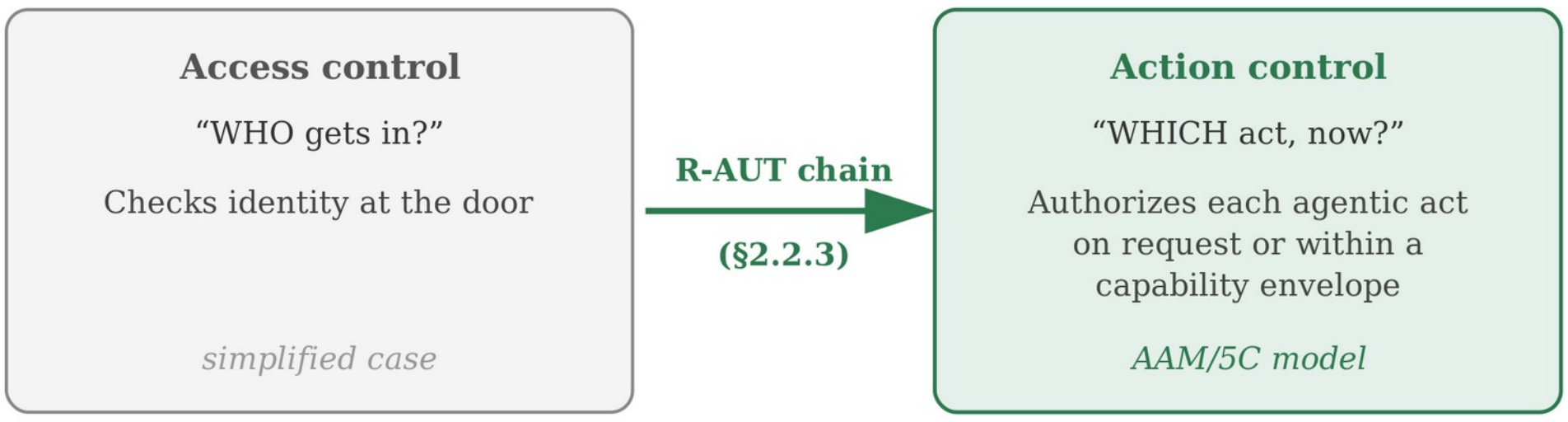

When a system acts on its own, checking "who gets in" is no longer enough:
each agentic act must also be authorized, on request or within a capability envelope.

Figure F.12 — From access control to action control.

**Example: containing a compromised agent.** An L2 agent requests bulk access to the customer database. The Zero Trust agent finds that the intent does not match its usual role; on this attested evaluation, L6 denies: no token is issued, the agent is placed in quarantine and its tokens in progress are revoked. Targets [DT]: revocation in under 500 ms, risk exposure below 2 s.

### 7.4.3. Enforcement point, capability envelope and key lifecycle

This section specifies the three mechanisms announced in §2.2.3.

**PEP specification.** The Policy Enforcement Point (PEP) is a deterministic, non-cognitive component interposed between the agent and the target resource: L1 gateway for outgoing calls, L3 bus for internal commands, L7 control plane for infrastructure acts, MCP server of the context provider (L6) for data access. It performs five checks:

- The signature of the token and the chain of trust of the issuing key, as well as the signature by L6 of the evaluation proof whose digest the token carries, with regard to the inventory and key epochs specified below;
- The current policy epoch and key epoch: a token issued under a policy version or a key epoch that has been revoked is rejected;
- The scope — resource, operation, parameters — its exact match with the act presented and its inclusion in what the evaluation proof signed by L6 authorizes;
- Expiry and single use, replay being prohibited;
- The absence of the bearer from the quarantine list.

The failure of any one of these checks results in the rejection of the act, under R-AUT (§2.2.3). The PEP does not deliberate and interprets no intent: it compares a token to an act. These checks do not suffice, on their own, to establish H7a and H8a (§3.2.4): the guarantee contract, at the end of the section, adds obligations to them, not a sixth check.

**Pre-authorized capability envelope.** Placing every decision on the critical path protects the system, but makes it dependent on the control plane: if L6 or L5 fails, no agentic act is authorized any longer, even a harmless one, whereas the data plane has a degraded mode (§6.3). The envelope removes this asymmetry without reopening local issuance. It is an authorization decided by L6 and issued in advance by L5, which delimits a bounded right to act.

It has five properties:

- A restricted, explicit scope (classes of acts, resources, quotas);
- A short expiry;
- Local verification by the PEP, which compares the act to the scope of the envelope as it would to that of a token, without calling the control plane;
- Revocation by event;
- A posteriori audit, as for an act token; lacking a human identity of its own, a covered act is attributed to the co-signatories of the policy that underlies the envelope (§8.3).

It meets three needs: continuity when L6 is unavailable, compliance with the tightest latency budgets (§8.6) and scaling of action control without a decision cache.

Three reservations delimit its scope:

- Verification is local, issuance never is: no local entity creates, renews or extends an envelope, otherwise this would be the local issuance that R-AUT excludes;
- The revocation list carries its own availability target, higher than that of the policy engine — otherwise the single point of failure shifts to the propagation of revocation;
- The mechanism presupposes the key lifecycle below: an envelope is worth only as much as the custody of the key that signs it.

**Accepted bound.** The envelope covers only classes of acts of low or moderate criticality, qualified as such by the signed policy. Acts of high criticality — with financial effect or regulated scope, for example — remain under synchronous authorization, and irreversible acts, whatever their criticality, require human validation (§4.4): the unavailability of L6 blocks them. This is a choice: for these acts, denial by default is better than a right to act whose revocation would no longer be guaranteed.

**Key lifecycle.** The entire security of the model rests on signatures: a lost key blocks the system, a compromised key makes it arbitrary. The chain of trust rests on five categories of signed objects: the policies conveyed by MCP — their signature is a requirement of the model, not of the MCP specification [MCP 2026]; the evaluation proofs of L6; the act tokens and capability envelopes; the explainability logs; the root of the Merkle tree of the context proof.

The Merkle tree [Merkle 1988] is a tree-structured hash structure: each leaf carries the digest of a context element, each node that of its two children, and the signed root, accompanied by the path of a few digests, suffices to prove that an element belongs to the attested context and has not been modified, without republishing the entire context.

**Inventory and holders.** Five service keys cover the model, each with a single holder; to these are added the personal approval signatures of the Intent Architect and the Policy Auditor, which precede the sealing of the policies (§8.3):

- The policy signing key, held by the L6 policy authority; it seals the policy corpus;
- The evaluation proof signing key, held by the L6 governor agent, distinct from the policy authority; the PEP also checks the signatures it affixes;
- The issuance key for act tokens and capability envelopes, held by the L5 issuance service; it is the one the PEP verifies when checking the token signature;
- The sealing key for explainability logs, held by the logging function, so that no act signatory can rewrite the trace of its own acts;

- The Merkle root key, held by the proof function, which attests the state of the context independently of who decides and who issues.

**Separation of holders.** No entity — human, agentic or administrative — holds two of these keys. This is the cryptographic extension of the non-collusion between decision, issuance and audit: a holder combining the policy key and the sealing key could authorize an act and then erase its trace.

**Rotation and epoch.** Each key is renewed periodically, with an overlap window in which the old key is still accepted for verification but is no longer used for signing. Each rotation increments a monotonic epoch carried by the signed object; the Merkle proof and the PEP check it, and an object presented under an epoch earlier than the minimum accepted epoch is rejected. It is this check that makes revocation effective.

**Compromise procedure.** It follows an ordered, testable sequence: revocation of the compromised epoch and raising of the minimum epoch accepted by all PEPs; reissuance of legitimate objects under a new epoch; immediate invalidation of the tokens and envelopes attached to the revoked epoch; replay, from the sealed logs, of the validation of the acts materialized during the suspect window. The propagation delay of the epoch raise is the critical quantity of the procedure and an object of measurement of the empirical validation.

**Accepted limit.** The security of the whole is bounded by that of the key custody arrangement — trusted hardware, secrets vault, separation of authorizations — which is not specified in this book. None of the properties stated here holds if this custody fails: this is a research hypothesis to be investigated, not a guarantee by construction.

**Guarantee contract.** The five checks say what the PEP verifies on an act. They say neither what it writes, nor what it consumes, nor what must be assumed for H7a and H8a to hold (§3.2.4). The contract that follows fixes this. These are obligations of the specification, which every implementation of the model must meet, and not additional checks of the PEP, which always performs five. The contract has six points:

- **Log before transmission**: the PEP writes the admitted act to the explainability log before transmitting it to the target layer; the failure of this write amounts to rejection;
- **Link of the envelope to L6**: a capability envelope carries, like a token, the digest of the evaluation proof signed by L6 that decided it; the PEP verifies its signature at the first check, as for a token, before admitting an act under this envelope;
- **Single use and quotas**: a token is consumed only once in an atomic shared state, or assigned exclusively to a replica that consumes its use atomically; envelope quotas are consumed atomically in a shared state or in disjoint partitions, and each act presented under an envelope carries a unique act identifier, consumed only once like a token;
- **Revocation freshness**: the PEP relies only on a revocation state whose age remains below a fixed bound, and denies the act beyond it; this bound limits the window in which a revoked object can still be admitted;
- **Assumptions of the [TG]**: the PEP is correct and interposed on every agentic act path — which H8b tests — the keys are intact (accepted limit above), and the act executed by the target layer is the one the PEP verified, with no change of parameters between the two;
- **Return to an earlier policy**: its content is republished under a new epoch, never under its old epoch; the epoch remains monotonic, and no object issued under a revoked epoch is readmitted (§8.4.4).

Under this contract, H7a and H8a are deduced from the specification. An admitted act has passed the five checks, including, if it falls under an envelope, the verification of its link to L6: it therefore carries a token or an envelope linked to a favorable decision of L6 (H8a). It was written to the log before its transmission (H7a). Outside the assumptions of the fifth point, these two statements guarantee nothing.

### 7.5. Layer L6: data, knowledge and governance

L6 supplies agents with context via MCP and carries the authorization decision (§2.2.2), which its governor agent renders (§2.3.2). It also relies on three components:

- The **Semantic Data Fabric**, which presents relational tables in the form of knowledge graphs;
- The **context provider** (MCP server), which delivers a unified business context, filtered according to need-to-know;
- The **integrity agent**, which guarantees the quality and non-alteration of the sources.

**Deterministic decision.** L6 establishes, by means of a policy engine, that no proposal from L4 violates an invariant expressed in the policies (§1.4). The guarantee holds only for what the policies can express.

**Example: a preferential price.** A sales agent wants to grant a discount. It queries the context provider, which aggregates the customer relationship and inventories into a single context object; L6 checks that the discount does not exceed the margin limit. If the freshness target is met, the decision no longer rests on stale data. Targets [DT]: context freshness below one second; hallucination rate below 1%, subject to a sufficiently rich context (Figure F.16, §8.6).

### 7.6. Layer L7: infrastructure and physical

L7 adapts physical resources — compute, network, energy — to actual needs (§2.2.2), for performance and, as an extension specific to the model, for carbon footprint. It relies on three components:

- The **autonomous provisioner**, which dynamically adjusts the sharing of compute, memory and network;
- The **energy sobriety agent**, which places workloads according to the availability and cost of energy, including by switching to local reserves;
- The **hardware self-healing agent**, which anticipates failures and moves workloads in time.

Its acts are verified by the PEP of the control plane and respect the hysteresis rule between two contrary decisions, so as not to oscillate (§7.8).

**Example: an energy price spike.** The sobriety agent moves non-critical tasks to low consumption; critical services are preserved. The magnitude of the saving is a matter for a sector-wide simulation, left to future work.

### 7.7. Threats to the MCP/A2A nervous system

The model makes MCP and A2A its trust channels; if they are compromised, governance is compromised too. The risk is real: a critical vulnerability (severity score of 9.6 on the Common Vulnerability Scoring System, CVSS) was published in mcp-remote, a client component of MCP [CVE-2025-6514], and the EchoLeak attack showed, on an office assistant, that a booby-trapped email can, without a single click, lead the assistant to disclose data [Ravia 2025]. Threat modeling of agentic protocols is detailed in [Anbiaee et al. 2026].

### 7.7.1. Nine threats, two of them without a strong countermeasure

Seven threats target the protocols themselves; each calls for its own countermeasure:

- **Tool poisoning**: a malicious MCP server hides instructions in a tool description, which the agent reads as context. Countermeasure: verify the cryptographic digest of descriptions before loading, and admit only servers registered in the agentic registry, which the AOM maintains together with their digests: any signature anomaly blocks and alerts.
- **Prompt injection through context**: an injected document contains disguised instructions that hijack L4's reasoning. Countermeasure: filter the context before injection and isolate unverified data.
- **Interception of the A2A channel** (man-in-the-middle): a third party modifies the transmitted intentions. Countermeasure: sign every A2A message and encrypt the channels with TLS 1.3 using mutual authentication.
- **Rug pull on an MCP server**: a server that was safe until then is modified. According to DataStealth, public MCP servers number in the thousands, each one an exposure point [DataStealth 2026]. Countermeasure: pinned versions reviewed before adoption, digest monitoring, human validation of every critical update.
- **Behavioral drift**: under sustained pressure from its environment, an agent may gradually depart from the explicit constraints of its instruction, without any change to its parameters [Saebo et al. 2026]. Countermeasure: periodic recalibration by the Intent Architect and differential analysis of the logs.
- **Replay**: a valid message or call is reissued out of its context. Countermeasure: a single-use token, with a nonce and a short expiry, or, under an envelope, a single-use act identifier, rejected by the PEP on second use (single-use check, §7.4.3) — channel encryption, for its part, does not prevent replay.
- **Confused deputy**: a legitimate intermediary agent is led to act beyond the initial intent. Countermeasure: authority verified at the source, tokens with explicit scope, least privilege at each delegation.

Two vectors find no strong countermeasure among the six in the matrix: the **co-signing insider**, a legitimate holder of a key that signs a malicious policy — the PEP cannot distinguish it from a legitimate policy; the countermeasure is the separation of key holders and dual control (§7.4.3, §8.3, §8.4.2) — and the **compromise of the model provider**, whose proposals retain a lawful form — the countermeasure is attestation of the served model and drift detection. They are listed deliberately: a visible gap can be addressed, a tacit one cannot.

### 7.7.2. Defense in depth

No countermeasure suffices alone; their combination produces the defense. Figure F.13 maps the nine threats against the six main countermeasures — tool digest, input isolation, mutual A2A encryption, allowlist of tools, action control (L6 → L5 → PEP), audit of the trace — and rates, by expert judgment [RH], the protection each provides; Table 13 (Appendix 3, §3.2) gives, for each threat, its vector and its primary countermeasure. Two readings are in order:

- The reasoning trace has strong evidentiary value against error and unintentional drift, but none against an adversary who controls the component that produces it;

- The “strong” rating of action control applies to the bound on the radius of action, guaranteed by R-AUT (§2.2.3) as stated in H8a [TG], under the interposition assumption (H8b), not to the absence of abuse inside that bound [RH].

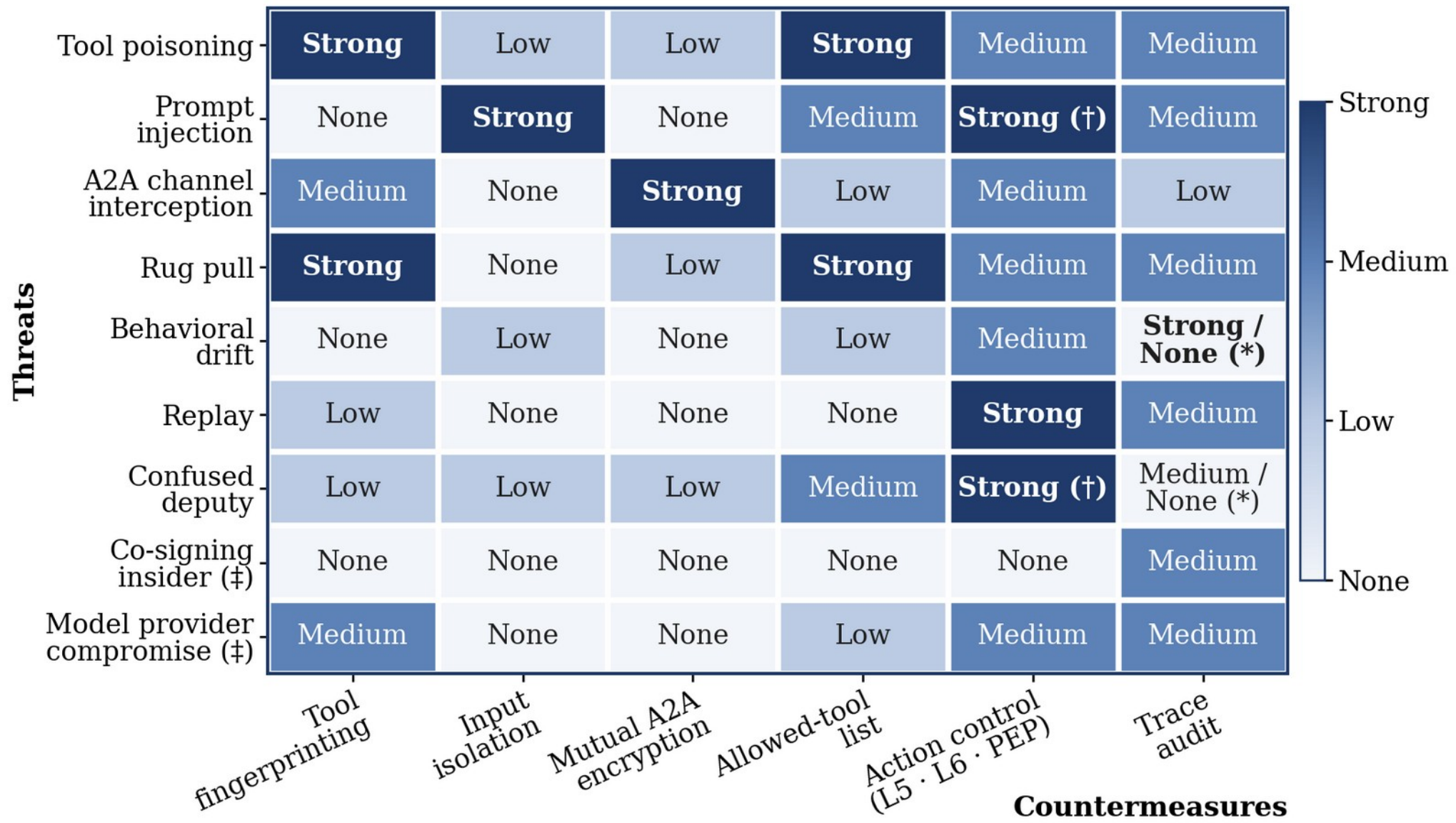

(*) double rating: upper bound against error and unintentional drift, lower bound (none) against an adversary who controls the component emitting the trace; these cells are shaded according to their lower bound.
(†) the “strong” rating applies to the bound on the blast radius, guaranteed by R-AUT (H8a [TG]) under the interposition assumption (H8b), not to the absence of abuse within that bound, which falls to audit [RH].
(‡) no strong countermeasure among the six: for the insider, separation of key holders and dual control; for the model provider, attestation of the served model and drift detection (§7.7.1).

Figure F.13 — MCP/A2A threats and countermeasures: protection provided by each, expert-judgment rating [RH].

The matrix of Figure F.13 sets the minimum scope of a quarterly red team exercise (§8.4.1) and feeds the technical documentation required by Article 11 of the EU AI Act. It also delimits the signed risk acceptance: a threat absent from the matrix is absent from both scopes.

## 7.8. Stability: preventing resonance

Autonomous agents that each react to their own signals can work against one another and amplify each other. The phenomenon is known in control theory: a system that switches too quickly among several regimes can become unstable even if each regime is stable [Hespanha and Morse 1999]. The model calls it the **IT resonance syndrome**.

**Two examples.** In the upper layers, L1 mistakes a legitimate traffic spike for an attack and filters, L2 sees traffic drop and releases instances, then L1 releases the held-back traffic onto a reduced capacity. In the lower layers, L4 requests resources, L7 adds them, L5 flags the new unsigned nodes as intruders and their isolation is authorized, L4 requests resources again, then L6 refuses any spending under the budget policy and the processes stop. In these two scenarios, absent a stabilization mechanism, the system oscillates and then collapses within minutes.

**Two mechanisms are designed to prevent it [RH]:**

- The **minimal A2A consensus**: an agent cannot commit a high-criticality act without the agreement of the peers concerned, and the act remains subject to R-AUT (§2.2.3);

- **Hysteresis**: a decision opposite to the previous one is taken only if it improves the objective by at least a threshold *h*, which spaces out the switches.

For low- or moderate-criticality acts, such as those in the two examples, only hysteresis applies.

**Sizing the hysteresis (G.C2).** Switched-systems theory [Hespanha and Morse 1999] establishes that a system whose every regime is stable, with a common margin, remains stable if it stays, on average, long enough in each regime before changing: this is the average dwell time, a sufficient but not necessary condition. For a particular supervisor (Appendix 2, §2.1.1), a hysteresis rule — accept a switch only if it brings a minimal improvement *h* — produces this dwell time; *h* is therefore the design lever. The transposition assumes the hypotheses of that supervisor, which nothing establishes for an agent: it falls under [RH].

For 8 regimes, $h$ = 10% and the value of λ retained in Appendix 2, one obtains an average dwell time of the order of 12 s [DT]. The starting ranges, to be calibrated on production traces, run from 30 s to 5 min in finance, from 10 s to 1 min in healthcare and from 1 to 10 s in industry [DT]. They do not follow from the example: with λ and *m* unchanged, 5 min would require a threshold *h* of about 1,000%. They assume λ and *m* specific to each sector, to be calibrated together with *h*. These values are orders of magnitude, not a proof of stability: the formulas and their assumptions appear in Appendix 2, §2.1.

**A four-state automaton.** The syndrome can be read as an automaton [RH] (Figure F.14) on the gap $\Delta(t)$ between two successive acts of an agent, compared to a tolerance threshold θ:

- **E0 — stable**: the gap stays below θ;
- **E1 — oscillating**: the gap exceeds θ, with bounded amplitude;
- **E2 — divergent**: the amplitude grows; this state is entered if the oscillation amplifies from one cycle to the next instead of damping out;
- **E3 — stabilized**: hysteresis and, for high-criticality acts, consensus bring the gap back below θ; the system returns to E0 when the gap stays there.

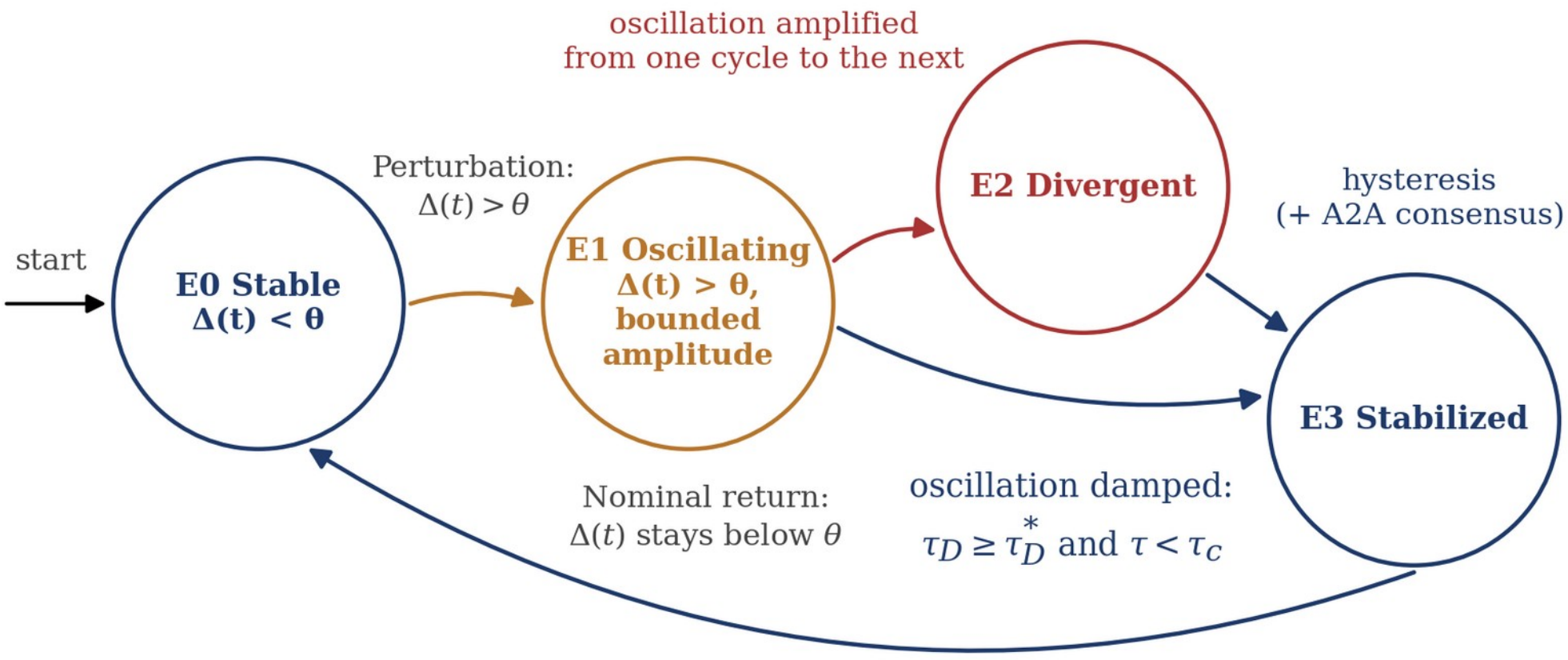

Stability conditions [RH]: $\tau_D \geq \tau_D^*$ (hysteresis) and $\tau < \tau_c$ (reaction delay)
where $\tau_D = \ln(1+h)/(2\lambda m)$ is the dwell time produced by hysteresis
$\tau_D^*$: average dwell-time threshold, to be estimated (Appendix 2, §2.1.1) ·
$\tau_c$: critical delay (Appendix 2, §2.1.2)
$h$: improvement required before any switch · $m$: number of regimes
$\lambda$: forgetting rate of performance signals ($\lambda < \lambda_0$, common margin)
Example [DT] (Appendix 2, §2.1.1): $h = 0.10$; $\lambda = 5{\cdot}10^{-4}\ \mathrm{s}^{-1}$; $m = 8 \Rightarrow \tau_D \approx 12$ s

Color code — blue: stable states and return · ochre: oscillation · red: divergence

Figure F.14 — Four-state automaton of the IT resonance syndrome [RH].

**An acknowledged limit [RH].** Even when coordination damps more than it amplifies, a reaction delay that is too long turns correction into sustained, even amplified, oscillation (stability condition: Appendix 2, §2.1.2). Hysteresis must therefore be calibrated on both sides.

## 7.9. Key takeaways

Every sensitive act follows the CoT-Validate-Sign-Execute pattern: reason, have it validated on attested attributes, receive a token, execute under monitoring. Each lower layer plays its role in it according to R-AUT (§2.2.3), with the authorization chain fitting within a 500 ms budget. Nine threats weigh on MCP and A2A; only the combination of countermeasures covers them, and two find no strong countermeasure there. Hysteresis and minimal consensus are designed to prevent resonance among agents [RH]; their effect remains to be measured (H5). Chapter 8 describes the governance that frames the whole.

# Chapter 8. Governance and Ethics

**IN BRIEF**

The previous chapters described a machine capable of acting on its own; this one explains how the organization keeps control of it. It ties the model to the requirements of the EU AI Act, shows why the separation of powers between layers makes collusion difficult, and describes the human organization that writes, audits and revises the policies. Finally, it gives the ten steering indicators and the business model that tests hypothesis H10 — including the scenario in which it does not hold. Chapter 9 takes stock of the model and its ten hypotheses.

## *Accountability, Balance of Powers and Auditability*

**Key Concepts of the Chapter**

- **H.C1 — Non-collusion by signature** — as long as the inputs of L6 are attested (I-ATT, §7.2) by intact components, no layer can alone widen the space of authorized acts, except the governor agent of L6, whose compromise suffices (§8.3); the argument rests on policy signing and on the separation of key holders. It is heuristic, not formally demonstrated (§8.3).
- **H.C2 — Human organization with a responsibility matrix** — 6 roles, 14 activities and 10 responsibility columns (§8.4).
- **H.C3 — Measured action control** — a quantified evaluation protocol: latency, true positive rate, false positive rate, modulated according to the criticality of the sector (§8.6).

In Practice

*Catherine, the AI compliance officer, must produce the EU AI Act compliance file in 8 weeks. She sets each requirement against the mechanism that answers it: risk management finds its answer in the L4 sandbox and the L6 policies, transparency in the signed explainability logs, human oversight in the AOM. The compliance assessment proceeds from the architecture rather than on a case-by-case basis; its article-by-article analysis remains to be conducted.*

### 8.1. Introduction: governance by design

An IS able to act at machine speed can no longer be governed by after-the-fact compliance checks. Governance must become a property of the architecture: this is **governance by design**. The model translates regulatory constraints — the EU AI Act, GDPR [GDPR 2016] — into runtime invariants carried by signed policies, and places the human on the loop, where the human sets the rules and controls their application.

### 8.2. Alignment of the AAM/5C model with the EU AI Act

Regulation (EU) 2024/1689 on artificial intelligence, which entered into force on 1 August 2024, applies in stages [EU AI Act 2024]. Under the timetable as amended by Regulation (EU) 2026/1744 of 8 July 2026, its obligations relating to high-risk systems will apply on 2 December 2027 for the systems that Annex III of Regulation 2024/1689 designates by their area of use, and on 2 August 2028 for those that fall under the Union harmonization legislation on products, listed in Annex I of the same regulation.

An agentic IS is not in itself a high-risk system: it becomes one if it falls under Annex III of the regulation, for example as a safety component in the management and operation of critical digital infrastructure (point 2).

Regulation (EU) 2026/1744 clarified this notion: according to Article 3, point 14, as amended, a component fulfills a safety function when its intended purpose is to prevent or mitigate risks to the health and safety of persons or property, and its recital 7 and the paragraph 1a that it inserts in Article 6 exclude from this function the systems that provide only user assistance, performance optimization, service efficiency, automation, convenience or quality control of aspects unrelated to safety.

The model nevertheless adopts the high-risk requirements as a design reference. Three articles structure the alignment:

- **Article 9 — continuous risk management.** L6 evaluates each agentic act, or each class of agentic acts covered by a capability envelope, against the signed policies; the agentic registry, maintained by the AOM, inventories the authorized agents and MCP servers (§7.7.1); the Policy Auditor audits the risk policies and prepares their revision. Coverage is verified over the scope of the signed policies.
- **Article 13 — transparency and provision of information to deployers.** Each agentic act leaves a structured explainability log; each refusal by action control is reasoned; operators monitor the system from their console (L1).
- **Article 14 — human oversight.** A global emergency stop, independent of the authorization chain, alerts on a criticality threshold, and the power of the Intent Architect and the Policy Auditor to restrict a policy immediately (§8.4.4) give the human the ability to intervene.

**Scope of the alignment.** These three articles form the core of the alignment, not its totality. An article-by-article mapping — data governance (Article 10), technical documentation (11), record-keeping (12), robustness and cybersecurity (15), obligations of providers and deployers (16 to 27), general-purpose models (51 to 56) — falls under future work (Chapter 10). The ISO/IEC 42001 standard on AI management systems [ISO/IEC 42001:2023] offers a compatible framework, without the model claiming any certification.

## 8.3. The tripartite model: separating powers

Why separate powers? Because a language model can hallucinate or drift from its objective, and no component should be able, alone, to turn that error into an act. The model transposes to the IS the principle of checks and balances [Chang 2025] into three separate powers:

- **L4** proposes: the cognitive agents diagnose and plan, quickly but fallibly;
- **L6** decides in accordance with the signed policies, through a deterministic policy engine;
- **L5** issues the act token, arbitrates conflicts between competing proposals and monitors behaviors.

This is the R-AUT rule (§2.2.3) seen from the institutional angle; final verification falls to the PEP. Example: L4 proposes to modify network routing to raise a conversion rate; L6 refuses, because a policy forbids any routing modification during a deployment; L4 must find another solution. Figure F.15 summarizes this separation.

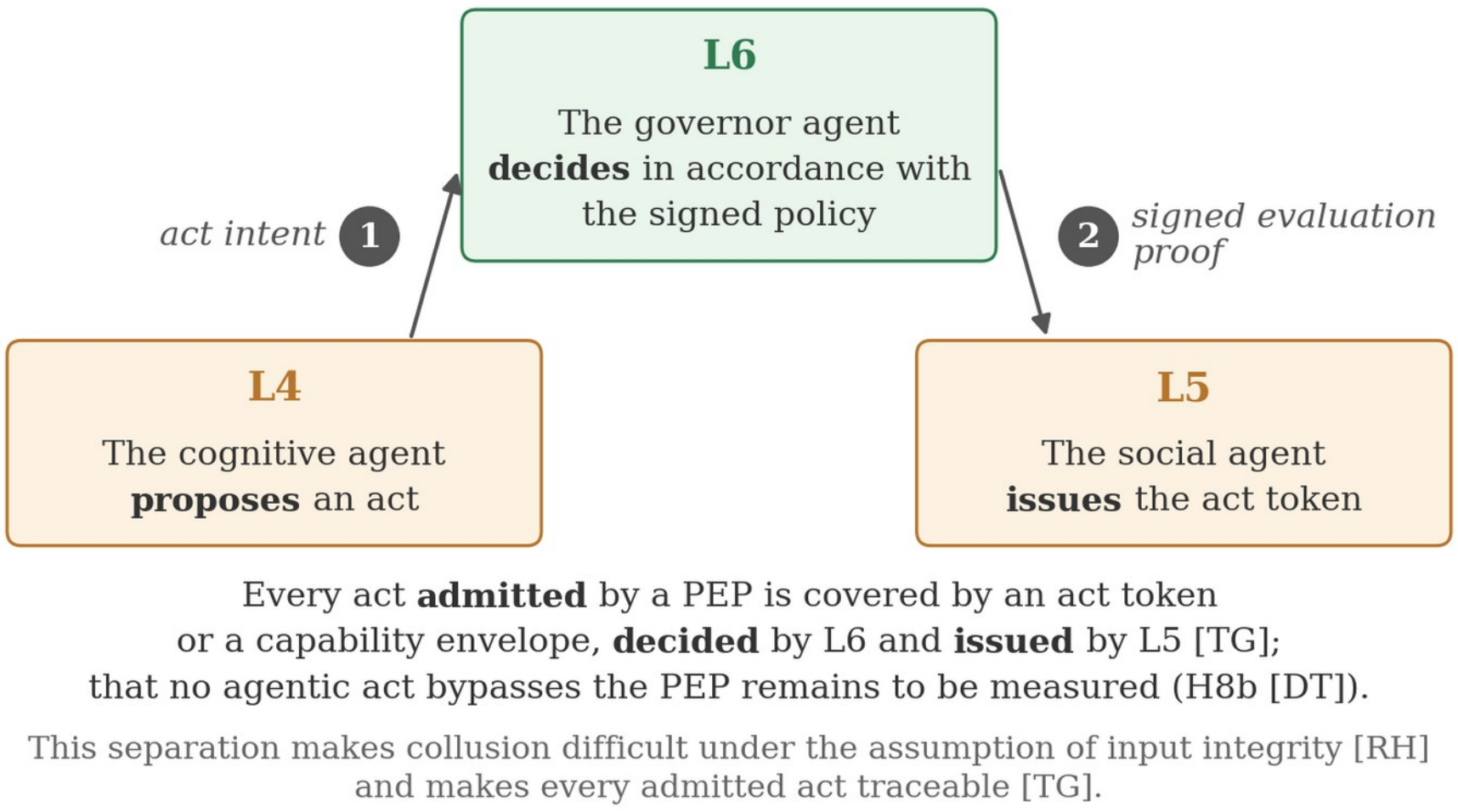

Figure F.15 — The tripartite separation of powers: as long as the inputs of L6 are attested (I-ATT) by intact components, no layer other than L6 can alone broaden the authorized acts [RH].

**Non-collusion (H.C1) [RH].** The separation has value only if these powers cannot coalesce to circumvent the policies. The argument, which is heuristic, comes down to three propositions:

- **L4 does not rewrite its own rules.** The policies are drafted by the Policy Engineer, approved by human dual signature — Intent Architect and Policy Auditor — then sealed by the key of the L6 policy authority, in a repository where the agents have no write access. A compromised agent therefore cannot modify the rules that evaluate it: it would be necessary to compromise two distinct human identities.
- **L5 arbitrates conflicts between competing proposals, but does not decide and does not write the rules.** It issues a token only on a favorable decision by L6. It may decline to issue, which creates no right to act; it cannot modify a policy. When compromised, it may let through or block wrongly, but it cannot widen the space of authorized acts beyond what L6 has decided; because it attests some inputs of L6 (behavioral evaluation, quarantine state, §7.2), it can, however, distort that decision.
- **The verdict of L6 is deterministic — which is not enough.** The policy engine returns the same verdict for the same input. But if an agent falsifies the context it transmits, it obtains a deterministic and false verdict. The integrity of the context therefore requires that the decisive attributes be attested by components other than the agent being evaluated — this is the I-ATT invariant (§7.2) — or that L6 collect this context itself. As long as compliance with I-ATT is not verified on every policy in force, non-collusion remains conditional on the integrity of the inputs and, in all cases, on that of the components that attest them, L5 included.

Under these conditions, corrupting governance requires either the simultaneous compromise of two distinct human identities, or the compromise of the governor agent of L6 or of a component that attests its inputs, or an attack on the cryptographic chain itself: security rests on the independence of the channels, not on the trust placed in each component. A mechanized proof (Rocq; TLA+, Temporal Logic of Actions) remains to be produced.

## 8.4. The Agentic Operating Model: roles and responsibilities

The **Agentic Operating Model** (AOM) organizes human work around three functions: defining intent, managing context, auditing compliance.

### 8.4.1. Six roles, one of them a system role

The AOM comprises six roles: five are held by people, the sixth by an agent. Each human role is held by an incumbent and a deputy, both designated.

- **Intent Architect** — defines the business objectives and the agentic constitution, that is, the limits of autonomy, and recalibrates the agents;
- **Context Guardian** — guarantees the integrity and freshness of the data supplied to the agents, filtered according to need-to-know;
- **Policy Engineer** — drafts, versions and tests the L6 policies and the capability envelopes (§7.4.3), whose release to production the Engineer prepares without authorizing it personally;
- **Policy Auditor** — audits the policies and their application, monitors the quality of explanations, looks each quarter for the three drifts of the CoT-Validate-Sign-Execute pattern (§7.2), approves the review of MCP servers, commissions the red team exercise and prepares the risk acceptance — the Auditor drafts no policy within the scope being controlled;
- **Skills Engineer** (skills: the tooled competencies of agents) — develops and validates the primitive actions that agents may invoke, guarantees their reversibility and performs the review of MCP servers before adoption;
- **A2A Mediator** — system role, held by an L5 agent (Cognition group): it facilitates complex negotiations between agents and detects mutual deadlocks; it is supervised by the Intent Architect and audited by the Policy Auditor.

### 8.4.2. Incompatibilities and lines of defense

The same person cannot be both Policy Engineer and Policy Auditor over the same scope; holding both roles over disjoint scopes is allowed and declared. Without this incompatibility, writing the rule, controlling it and testing that control would fall to the same person, and no assurance could be asserted against a third party. It is of the same nature as the separation of key holders (§7.4.3).

The arrangement can be read within the framework of the three lines of defense of regulated environments [IIA 2020]:

- **First line** — agentic operations: Intent Architect, Context Guardian, Policy Engineer, Skills Engineer;
- **Second line** — agentic assurance and risk: the Policy Auditor, without drafting authority;
- **Third line** — internal audit, which periodically verifies the first two, including the reality of the incompatibilities and the approvals co-signed by the Policy Auditor.

### 8.4.3. Responsibility matrix (H.C2)

The responsibility matrix (RACI, Responsible, Accountable, Consulted, Informed: who performs, who approves and answers, who is consulted, who is informed) covers 14 activities — from the definition of objectives to the annual risk reassessment — and 10 columns: the 6 roles and 4 control stakeholders (business owner, CISO, data protection officer, internal audit). Its full form appears in Appendix 3, §3.3.

These roles are intended to redistribute existing responsibilities rather than to increase headcount; their running cost is nevertheless counted in the business model (§8.7.2).

#### 8.4.4. Managing policy changes

A policy change takes effect live, without redeployment, and changes at once the authorized behavior of all the agents within a scope. The usual change advisory board process does not transpose to it as is; the model distinguishes two regimes:

- **Permissive changes** — extending a scope, adding a class of acts, raising a ceiling, creating an envelope — are authorized beforehand: proposal by the Policy Engineer, dual signature by the Intent Architect and the Policy Auditor, publication under a new epoch;
- **Restrictive changes** — lowering a ceiling, removing a class of acts, revoking an envelope — apply without delay, on the decision of the Intent Architect or the Policy Auditor, and are notified afterward, because a default refusal creates no new security risk.

Restrictive changes actuate levers that the signed policy provides for, without drafting a new policy. They are then consolidated through the ordinary process (drafting by the Policy Engineer, dual signature). Their effect on availability is tracked by the indicators of §8.6.

Rolling back consists in republishing the content of the previous policy under a new epoch, then revoking the tokens and envelopes issued under the faulty epoch. The old epoch is never republished: raising the minimum epoch would cause it to be rejected, and lowering that minimum would readmit revoked objects (§7.4.3).

### 8.5. Governance stability

Governance itself can oscillate. If a policy change were propagated to L4 and L5 faster than the retained average dwell time ($\tau_D$, §7.8), chosen above the threshold $\tau^*_D$ (Appendix 2, §2.1.1), the system could switch from one compliance regime to another, incompatible one. By analogy [RH], the dwell-time requirement of §7.8 therefore applies to policies: the Policy Auditor ensures that no permissive change is propagated below $\tau_D$; restrictive changes apply without delay (§8.4.4).

### 8.6. The ten steering indicators and the measurement of action control

Governance is steered through a dashboard of ten indicators (Table 8). Their targets are design targets [DT]; four of them directly take up hypotheses from Chapter 3 (H3, H6, H8b, H10).

| # | Indicator | Target [DT] | Example of a typical signal | Action |
|---|---|---|---|---|
| 1 | Task Completion Rate (TCR) | > 95% | Repeated failures on inventory transactions | Analyze L4 tool-call failures in the traces |
| 2 | Proactive avoidance rate (PAR, H3) | > 85% | Rise in unanticipated incidents | Recalibrate the L1 and L2 perception thresholds |
| 3 | Action control true positive rate (TPR) | > 99% | Injection attempt detected at the edge | Isolate the agent and revoke its tokens |
| 4 | Hallucination rate | < 1% | Erroneous plans identified in review or rejected by L6 | Enrich the MCP context or the L6 policy |
| 5 | Transactional remediation time | < 30 s | Slow compensation of a long transaction | Optimize A2A coordination |

| # | Indicator | Target [DT] | Example of a typical signal | Action |
|---|---|---|---|---|
| 6 | Coverage by act tokens or envelopes (H8b) | No agentic act without a token or envelope | Agent using a persistent token | Block it and migrate it to action control |
| 7 | Compliance with the average dwell time | > 98% | Oscillations during a capacity adjustment | Raise the L7 hysteresis (§7.8) |
| 8 | Quality of explanations | > 85% | Incomplete justification of a payment refusal | Strengthen the agent's contrastive explanations |
| 9 | A2A consensus latency (H6) | < 500 ms (95th percentile) | Negotiation that drags on | Have L5 arbitrate |
| 10 | H10 ratio (ROI) | > 2.5× | Total cost higher than net savings | Review the sizing of the AOM and the residual supervision (§8.7.4) |

Table 8 — The ten steering indicators of the agentic IS.

The reduction of toil (H9) is measured separately, through a quarterly survey and an analysis of repeated tickets, following the definition of toil in reliability engineering [Beyer et al. 2016], of which H9 retains four criteria: manual, repetitive, automatable, without enduring value. Its first measurement takes place at milestone G1; its verdict follows H9's own timetable (Appendix 1, §1.1).

The hallucination rate target assumes a sufficiently rich context: Figure F.16 shows the expected qualitative shape of this dependence, without quantifying it [RH].

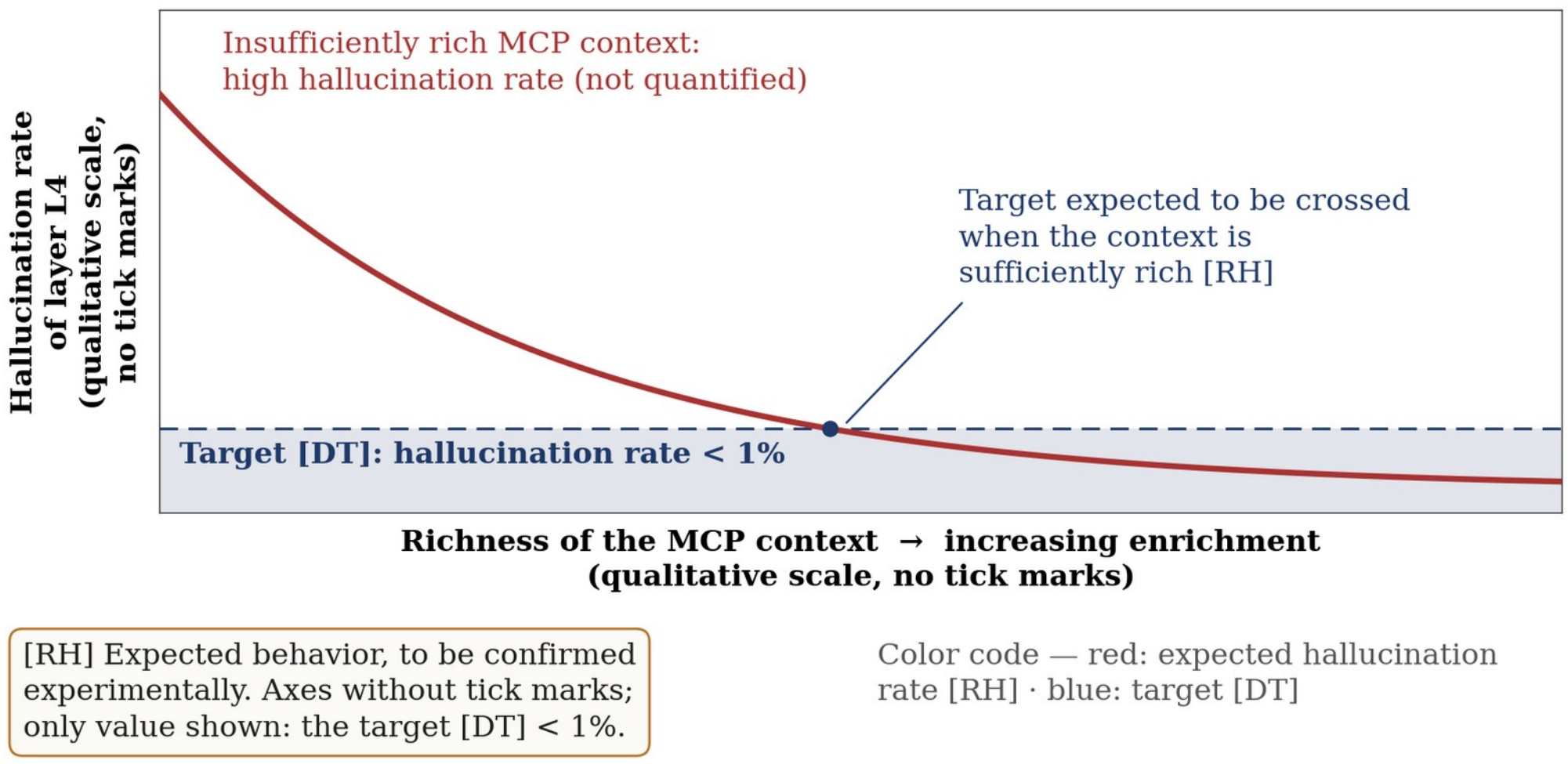

Figure F.16 — Hallucination rate and richness of the MCP context [RH].

**Measuring action control (H.C3).** Three quantities make it a reproducible audit object, measured by replay on a labeled attack set (prompt injections, tool poisoning, privilege escalations, emergency-path bypasses) [DT]:

- The **authorization latency**, below 500 ms at the 95th percentile, according to the budget of §7.4.2;
- The **true positive rate** (TPR), the share of undesirable acts in the attack set that are actually denied, above 99%; since an act contrary to an expressed policy is denied by L6 as long as that policy is correctly evaluated on attested attributes (I-ATT, §7.2), the gap to 100% measures the coverage gaps in the policies and in the attestation of inputs;

- The **false positive rate** (FPR), the share of legitimate acts in the evaluation set wrongly denied, relative to the total number of legitimate acts, below 0.5% — beyond that, the business experience degrades and the Policy Auditor recalibrates the thresholds.

These thresholds are modulated according to the criticality of the sector [DT]:

- High-frequency finance: latency below 200 ms, TPR above 99.5%, FPR below 0.1%;
- Real-time healthcare: latency below 100 ms, TPR above 99.9%, FPR below 0.05%;
- Process industry: latency below 50 ms, TPR above 99.5%, FPR below 0.1%;
- Administrative functions: latency below 1 s, TPR above 99%, FPR below 2%.

The TPR falls below 99% in no sector: the modulation tightens the thresholds of critical sectors and relaxes, for administrative functions, only the latency and the false positive rate. Three of these latencies are below the 500 ms budget: the synchronous authorization chain cannot meet them, and it is better to say so. For these sectors, of the two authorization regimes the model defines, only the pre-authorized capability envelope (§7.4.3), verified locally by the PEP, meets these latencies while respecting R-AUT (§2.2.3); acts of high criticality — in high-frequency finance, those with financial effect — remain under synchronous authorization, outside the sector threshold, and fall under the 500 ms budget. The G2 validation milestone therefore reports the latency of the two regimes separately.

## 8.7. Business model of agentic autonomy: putting H10 to the test

H10 posits a ratio above 2.5 between the net savings from autonomization and the total cost of ownership of the agentic layer. This ratio is denoted ROI for convenience; it does not relate a net gain to a construction investment in the accounting sense. The denominator counts inference, the human roles of the AOM and the governance platform: relating the savings to the inference cost alone would give a ratio of about 41 in the central scenario, hence a statement too easy to satisfy to be informative.

All the parameters that follow are [DT], to be calibrated locally; the amounts derived from them are given by way of illustration and measure no real IS. The ratio is put to the test at milestone G4 (§9.3).

### 8.7.1. Model parameters

The business model concerns a reference IS of 500 components. At 0.2 detected events per component per day — anomalies, alerts, drifts, which do not all require intervention — it handles 36,500 events per year, split into three severity tiers (Table 9).

| Tier | Share of volume | Annual volume | Unit human cost | Achievable autonomization |
|---|---|---|---|---|
| Automatable triage (false positives, self-resolving alerts) | 60% | 21,900 | €15 | 85% |
| Standard incident (diagnosis and remediation) | 35% | 12,775 | €150 | 45% |
| Major incident (escalation, post-mortem) | 5% | 1,825 | €1,500 | 20% |

Table 9 — Event tiers and achievable autonomization.

Without an agentic layer, the baseline operating cost is 36,500 × (0.60 × 15 + 0.35 × 150 + 0.05 × 1,500) = €4.98M per year. This is the sum that autonomization can bite into: it bounds the achievable savings.

These autonomization rates are not the measure of H9. H9 concerns toil (§8.6), assessed in operator effort. Table 9, for its part, counts all detected events, major incidents included, weighted by their cost. H10 is therefore computed over a scope that is not that of H9; their reconciliation remains to be established on a pilot.

### 8.7.2. Total cost of ownership

The annual total cost of ownership comprises three items:

- **Inference**: 5,000, 80,000 and 300,000 tokens per event, autonomized or not, depending on the tier, i.e., 1.68 billion tokens per year; at €18 per million tokens (assumed average price, all models combined), €30,222 per year;
- **Human roles of the AOM**: at 500 components, the five human roles can be pooled; the central scenario retains 2 full-time equivalents (FTE) at €135,000 fully loaded, i.e., €270,000;
- **Governance platform** (policy engine, logging, observability, simulated attack campaigns): €150,000 per year.

The annual total cost of ownership reaches €450,222, of which only 7% is inference and 60% is human governance. Under these parameters [DT], the binding item is therefore not the cost of tokens but that of human labor: governance, counted here, and supervision, deducted from the savings (§8.7.3).

These items are operating costs. The business model does not price the construction costs, which every adopter must provision: MCP integration and building the policy catalog, instrumenting observability, deploying the PEPs and managing machine identities and keys (§7.4.3), evaluation bench, training of the roles, making explicit the implicit rules of the existing estate. Their amount depends too much on the estate for a generic value to be defensible; estimating them is a matter for future work.

### 8.7.3. Net savings

The gross savings — human cost avoided on the autonomized share of each tier — amount to €1,689,038 (21,900 × 15 × 0.85 + 12,775 × 150 × 0.45 + 1,825 × 1,500 × 0.20). Two items are deducted from them:

- The **supervision of autonomous acts**, 20% of the avoided cost, i.e., €337,808: an autonomous act is reviewed, sampled, audited — this cost follows the autonomized volume, so it is deducted from the savings;
- The **cost of induced errors**: 3% of autonomized acts remain erroneous, and each error costs twice the human handling avoided, i.e., €101,342. This term links H10 to H3, H7 and H8: the more the quality of decisions degrades, the further H10 recedes.

Net savings come out at €1,249,888 per year.

### 8.7.4. Ratio and verdict by scenario

Three parameter sets frame the uncertainty (Table 10). The event volumes, unit costs and autonomization rates of Table 9 are common to them: gross savings are €1,689,038 in all three scenarios. Varying together, from the optimistic to the conservative scenario, are the token volume and price — €12, €18 and then €30 per million — the sizing of the AOM, the platform cost, and supervision and errors, which absorb 11.5%, 26% and then 59% of gross savings. Outside the central scenario, only their sum is fixed: supervision and errors form a single parameter of the sensitivity analysis.

The token volume per event, 5,000, 80,000 and 300,000 depending on the tier in the central scenario (§8.7.2), is 3,000, 50,000 and 200,000 in the optimistic scenario and 8,000, 150,000 and 600,000 in the conservative scenario; these volumes, like the shares of supervision and errors, are computational assumptions [DT].

| Scenario | Tokens / year | Inference | AOM | Platform | Total cost of ownership | Net savings | ROI | H10 in the scenario |
|---|---|---|---|---|---|---|---|---|
| Optimistic | 1.07bn | €12,833 | 1.5 FTE — €202,500 | €100,000 | €315,333 | €1,494,798 | **4.74×** | met |
| Central | 1.68bn | €30,222 | 2 FTE — €270,000 | €150,000 | €450,222 | €1,249,888 | **2.78×** | met |
| Conservative | 3.19bn | €95,594 | 3 FTE — €405,000 | €300,000 | €800,594 | €692,505 | **0.86×** | **missed** |

Table 10 — H10 ratio under three scenarios.

In the conservative scenario, on the reference IS of 500 components, H10 is not met: the hypothesis can fail under plausible parameters, which makes it refutable at a fixed size. A target that no plausible parameter set could miss would be circular reasoning. In the central scenario, the margin is thin. Taken alone at its conservative value, each of the four parameter groups brings the ratio below 2.5: 1.54 for supervision and errors, 2.08 for the platform, 2.14 for the AOM, 2.42 for inference. H10 therefore hinges first on residual supervision and the error rate, then on the fixed costs of the AOM and the platform, precisely what the validation will have to measure.

**Efficiency and cash.** The H10 ratio measures an efficiency, not freed-up cash. Its numerator is an avoided human cost; its denominator, an actual expense. With headcount unchanged, the freed time is redeployed to other tasks: it has a value, but reduces no disbursement. The saving becomes cashable only if it translates into lower spending. The business model therefore carries a parameter $p$, the share of net savings actually cashable within the year (between 0 and 1) [DT].

The cash savings-to-cost ratio, $p$ × net savings / total cost of ownership, tells whether the agentic layer pays for itself in cash. In the central scenario, it pays for itself once $p$ reaches about 36%, but the cash savings-to-cost ratio reaches 2.5 only beyond 90%; in the optimistic scenario, these thresholds are 21% and 53%; in the conservative scenario, it does not pay for itself, even at $p = 1$. An executive committee thus reads two distinct figures: efficiency, which puts H10 to the test, and freed-up cash, which depends on organizational decisions that the business model does not take.

### 8.7.5. Minimum size to sustain H10

The cost of the AOM and of the platform is essentially fixed, whereas savings grow with the number of components. Solving the business model as a function of this number $N$, H10 is reached from about 260 components in the optimistic scenario, 450 in the central scenario and 1,950 in the conservative scenario. If the AOM or the platform grows with the IS, these thresholds rise. **AAM/5C is an architecture for large-scale IS**: below these sizes, the fixed cost of governance prevents H10 from being reached under these parameters. This scale dependence is itself refutable; it will be tested at milestones G3 and G4. Beyond about 1,950 components, H10 is met in all three scenarios: its refutability therefore assumes that the projection size is that of the adopter's IS, fixed before the measurement (G4).

### 8.8. Key takeaways

Governance is a property of the architecture, aligned with Articles 9, 13 and 14 of the EU AI Act. The separation of powers inscribed in R-AUT (§2.2.3) makes collusion difficult as long as the inputs of L6 are attested by intact components; the argument remains heuristic. Six roles, one of them a system role, write, audit and revise the policies, with explicit incompatibilities. Ten indicators steer the system, and action control is measured by its latency and its error rates. The business model puts H10 to the test and shows that it can fail: the model targets large-scale IS. Chapter 9, which follows, takes stock of the model, its limits and its validation program.

# Part III — Assessment and Outlook

*Part II described how the model is realized under the R-AUT rule; this third part takes stock of it against the ten hypotheses of Part I — what is guaranteed, what remains to be measured — states the conditions for adoption in a regulated environment, and then concludes.*

## Chapter 9. Assessment: Contributions, Limitations and Validation

| IN BRIEF |
|---|
| This chapter takes stock of the argument. It recalls what the model contributes and the status of each of its ten hypotheses. It then says where the model does not apply: four families of systems are excluded from it, and the model targets only large-scale IS. Finally, it lays out the validation program in six milestones, G0 to G5, each of which provides for the possibility of a stop. Chapter 10 states the conditions for adoption in a regulated environment. |

### *What Is Guaranteed, What Remains to Be Measured*

**Key Concepts of the Chapter**

- **I.C1 — Assessment of the hypotheses with their status** — for each of the ten hypotheses, its status in the epistemic grid (§9.1);
- **I.C2 — Acknowledged limits** — the systems to which the model does not apply, stated rather than concealed (§9.2);
- **I.C3 — Validation milestones G0 to G5** — a roadmap of the proof, each milestone having quantified stopping criteria (§9.3).

In Practice

*An executive committee must decide whether to launch the empirical validation phase of the AAM/5C model: testing its hypotheses on a test bench, then on supervised pilots in banking and healthcare. It has an argued reference model, explicit scope limits, a program sequenced by quarters, and an economic projection in three scenarios, one of which fails to meet H10. It is not its decision that will validate the model, but testing against the experimental protocol of Chapter 3.*

## 9.1. Synthesis: what the model contributes

The AAM/5C model answers the cognitive complexity wall (Chapter 1) with an architecture designed to reconcile the autonomy of components with the control of the organization. The contribution of the book is a prescriptive reference model that articulates, under a single authorization rule, known mechanisms — separation of decision, issuance and enforcement, bounded authorization, the autonomic loop — and adds three constructs, all to be validated: the Validation step of the Distributed Autonomy Loop, the ephemeral authority surface and the cognitive dependency plane. Four elements make up this model:

- **Five Capabilities instantiated in seven layers** (Chapter 2), with a single correspondence between Capabilities, layers, agents and hypotheses (Table 2), and a cognitive dependency plane (C-Dep) where the signed policy prevails over functional coupling (§4.2.3);
- **A single authorization rule**, R-AUT (§2.2.3), which separates proposal, decision, issuance and verification, inscribed between deliberation and execution by the DAL loop (§4.4); the CoT-Validate-Sign-Execute pattern applies it to every sensitive act (§7.2), within the limits of the ephemeral authority surface bounded by the act token (§7.4.2);
- **A human organization**, the AOM, which writes, audits and revises the policies, with explicit incompatibilities (§8.4);
- **Refutable hypotheses** — except H7a and H8a, guaranteed by construction — accompanied by independent measurement instruments and an economic model capable of failing (§3.2, §8.7).

**Assessment of the hypotheses (I.C1).** H7a and H8a are [TG]: true by construction of the execution path, under the assumptions of the grid, written into the guarantee contract (§7.4.3) — a correct PEP interposed on every path of an agentic act, intact keys, executed act identical to the verified act. They are not measured; H8b tests the interposition. All the others are [DT]: H1 to H6 are measured on the pilot of milestone G1, where H9 receives its first measurement before a verdict after three quarters (Appendix 1, §1.1); H7b and H8b, by reconciliation of independent traces (H8b at milestones G1 and G2, H7b at milestone G2); H10 depends on sector parameters and is decided at milestone G4. None is [EM].

On the defensibility scale, distinct from the grid, SP3 is at the “partially verified” level, SP5 at the “plausible” level and the other four sub-problems at the “supported by analogy” level (§3.3.1).

## 9.2. Acknowledged limitations

A conceptual volume must say where it does not apply (I.C2). The model is useful only in the presence of dynamic dependencies, real-time contextual decisions or cases impossible to specify in advance; elsewhere, classical architecture remains preferable. Unforeseen does not mean outside policy: the policy bounds the permitted acts, and the agent seeks the answer within that space. An answer that falls outside it is denied by default and returns to the human; the model’s gain is reduced accordingly. The model does not target four families of systems:

- **Critical embedded real-time systems** — avionics (DO-178C, levels A and B), defense, nuclear control (IEC 61513), implantable medical devices: millisecond determinism and their certification exclude any probabilistic layer in the control loop; the model can at best observe them, never act within their loop;

- **Mainframe applications not exposed through a programming interface** — CICS or IMS applications on z/OS, IBM i or GCOS applications, for example: integration through MCP and A2A presupposes their prior exposure, which is outside the scope of the book; an application already exposed through a programming interface falls under the general case;
- **Non-IP (Internet Protocol) infrastructures** — isolated industrial networks, fieldbuses without an IP gateway: MCP and A2A rely on HTTPS (Hypertext Transfer Protocol Secure) and JSON-RPC and do not apply there;
- **Integrated software packages with their own governance** — such as Salesforce, SAP or Workday: their rights and process model already forms an agentic substrate, incompatible with R-AUT (§2.2.3) without rewriting; the model can interface with them through their APIs, not substitute for them.

To these are added epistemic limits:

- The [TG] statements hold only under their assumptions — a correct PEP interposed on every path of an agentic act, intact keys, executed act identical to the verified act (guarantee contract, §7.4.3) — and for what the policies are able to express (§7.5);
- Non-collusion remains a heuristic argument, conditional on the integrity of the L6 inputs (§8.3);
- H8b does not see the acts that an agent causes to be executed under the identity of a script or a human (§3.2.4);
- The stability and termination of the consensus remain to be established (§7.8, Appendix 2), and two threats find no strong countermeasure (§7.7.1);
- The generalization of sector thresholds (§8.6) and of economic figures, excluding construction costs (§8.7), requires local calibration, and the model assumes a regulatory framework of the European type;
- Under its [DT] parameters, the business model holds H10 only on large-scale IS (§8.7.5).

### 9.3. Validation agenda: six milestones G0 to G5

Moving from the reference model to a validated protocol requires a work program, structured in six milestones with quantified stopping criteria (I.C3, Table 11). Without them, the program would drift into an accumulation of unfinished pilots. These proof milestones are distinct from the adoption milestones of §5.4.

| Milestone | Window | Deliverable | Quantified stopping criteria |
|---|---|---|---|
| **G0 — Test bench setup** | T0 to T0 + 90 d | Reproducible test environment: isolated Kubernetes cluster, candidate L4 agents deployed, labeled synthetic datasets, explainability measurement tooling | Observability service-level objective ≥ 99% on the test bench; statistical reproducibility of runs (C.C2); explainability measurement tooling validated on all paths |
| **G1 — Banking sector pilot** | T0 + 90 d to T0 + 180 d | Supervised deployment of AAM/5C on an identified banking business process (fraud control or dynamic know-your-customer), measurement of H1 to H6 and H8b, first measurement of H9, continued on this scope for three quarters (Appendix 1, §1.1) | MTTR measured < 1 min on simulated incidents (H1); A2A consensus latency < 500 ms at the 95th percentile (H6); zero PEP bypass over the pilot window, established by trace reconciliation with publication of the coverage of the record, of the sample size and of the upper confidence bound (H8b, §3.2.4; Appendix 1, §1.1) |
| **G2 — Healthcare sector pilot** | T0 + 180 d to T0 + 270 d | Supervised deployment on a healthcare case (personal data governance, therapeutic alignment), measurement of H7b and H8b with regard to Articles 9 (risk management) and 12 (record-keeping) of the EU AI Act | Zero materialized agentic act without a log entry (H7b); zero PEP bypass identified by trace reconciliation, including under simulated attack (H8b, §3.2.4), with the sample size and the upper confidence bound (Appendix 1, §1.1); quality of explanations > 85% at the lower confidence bound (Table 8; Appendix 1, §1.3); authorization latency reported separately for the synchronous regime and for the capability envelope (§8.6); action control true and false positive rates at the thresholds of the healthcare profile of §8.6 retained before the measurement, with the sample size of legitimate and undesirable acts (Appendix 1, §1.1) |
| **G3 — Parametric economic model** | T0 + 90 d to T0 + 180 d, in parallel with G1 | Economic model calibrated by sector: tokens consumed, share of small and large language models, operational savings, minimum size to sustain H10 (§8.7.5) | Model parameterized for at least three sectors; sensitivities computed over ±50% of the parameters; review by an independent expert |

| Milestone | Window | Deliverable | Quantified stopping criteria |
|---|---|---|---|
| **G4 — Pilot ROI and projection** | T0 + 270 d to T0 + 360 d | ROI measured on the scope of the G1 and G2 pilots and projected ROI, published separately and compared with the range of §8.7.4; adjustment of the economic model | Pilot scope and size of the projection IS fixed before the measurement; measured ROI published as is; projected ROI > 2.5×, computed with the unit costs and gains measured on G1 and G2 and published with its scale assumptions; threshold of §8.7.5 recomputed with these values and published as a result, not as a choice of size; cash savings-to-cost ratio reported separately (§8.7.4); gap to the projection documented; H10 decided (met or missed), any revision of the target being declared before the measurement |
| **G5 — Empirical validation report** | T0 + 360 d to T0 + 420 d | Synthesis report H1 to H10, epistemic status updated per hypothesis; submission to the academic community | Status of each hypothesis documented (corroborated, inconclusive or refuted, Appendix 1); dataset and code made reproducible; submission to at least one top-tier conference or journal |

Table 11 — Validation agenda: milestones G0 to G5.

The sequence G0 → G1 (G3 in parallel) → G2 → G4 → G5 allows an early stop: a failure at G1 without a clear path to correction leads to revising the model rather than persevering. Each milestone thus provides, in its criteria, for the possibility of abandonment. The model itself would be defeated if, the PEP not being bypassed (H8b), H9 and H10 were both refuted: the architecture would then not have established its human and economic viability (§3.2.5). A refutation of H7b or H8b on an implementation that meets the guarantee contract (§7.4.3) would, for its part, call into question the interposition assumption on which H7a and H8a rest.

## 9.4. Key takeaways

The AAM/5C model proposes an autonomy that is **explicitly authorized, continuously audited and reversible**. These three qualifiers have a precise operational meaning. Every act admitted by a PEP carries a verified act token or capability envelope and produces a log entry [TG] (H7a, H8a). That no agentic act escapes the log or the PEP is measured by the reconciliation of independent traces (H7b, H8b [DT]). Every autonomous act must be able to be undone in less than 5 minutes [DT] (§4.4). The model also says where it does not apply: four families of systems are excluded from it, and the business model holds H10, under its [DT] parameters, only on large-scale IS (§9.2). Six milestones, G0 to G5, with quantified stopping criteria, schedule its validation (§9.3). Chapter 10 describes what its adoption in a regulated environment would require.

# Chapter 10. Toward Adoption in a Regulated Environment: Conditions and Roadmap

**IN BRIEF**

This chapter marks the boundary between what the book proposes — an argued model — and what a heavily regulated organization, such as a large European banking group, would require before adopting it in production. It does not resolve those requirements: it states them and defers them to future work. It reviews compliance and model risk, human escalation thresholds, supply chain security, resilience, data sovereignty, architecture standards and proof through a case study. Chapter 11 draws from it, by way of conclusion, a more general reflection on governing AI by design.

## *The Conditions of Adoption and the Proof*

**Key Concepts of the Chapter**

- **J.C1 — Conditions of adoption** — what a regulated organization would require in addition to the model (compliance, model risk, security, resilience, data, architecture standards), stated but not resolved (§10.1 to §10.7);
- **J.C2 — Graduated escalation thresholds** — autonomous execution, after-the-fact supervision, prior human validation or exclusion, depending on the risk level and the amount (§10.3, Figure F.18);
- **J.C3 — Governance by design** — safeguards built into the architecture of Information Systems rather than added after the fact: the direction the book proposes to IT (§10.9, Chapter 11).

In Practice

*Yann presents the model to his bank's risk committee. The discussion bears less on the architecture than on what would still have to be established before going into production: the mapping to DORA (Digital Operational Resilience Act), the independent validation of the language models, the thresholds beyond which an act awaits a human signature, the provenance of the MCP servers. The committee does not reject the project: it draws up the list of its conditions. From it, Yann derives a plan of action, which begins with a quantified case study.*

### 10.1. Purpose and status of this chapter

This chapter does not modify the AAM/5C model: it states the conditions for its adoption (J.C1). Each of the conditions identified will call for a dedicated study. The figures in this chapter sketch their structure; they do not replace the engineering and accreditation work that remains to be done.

## 10.2. Regulatory compliance and model risk governance

For a financial entity in the European Union, adopting an autonomous decision-making system is first of all a compliance question. The Digital Operational Resilience Act (DORA) [DORA 2022], applicable since January 2025, imposes a framework that the model will have to satisfy: IT risk management, incident management and reporting, digital operational resilience testing (threat-led for designated entities), management of third-party provider risk; it also encourages the sharing of information on cyber threats (Figure F.17).

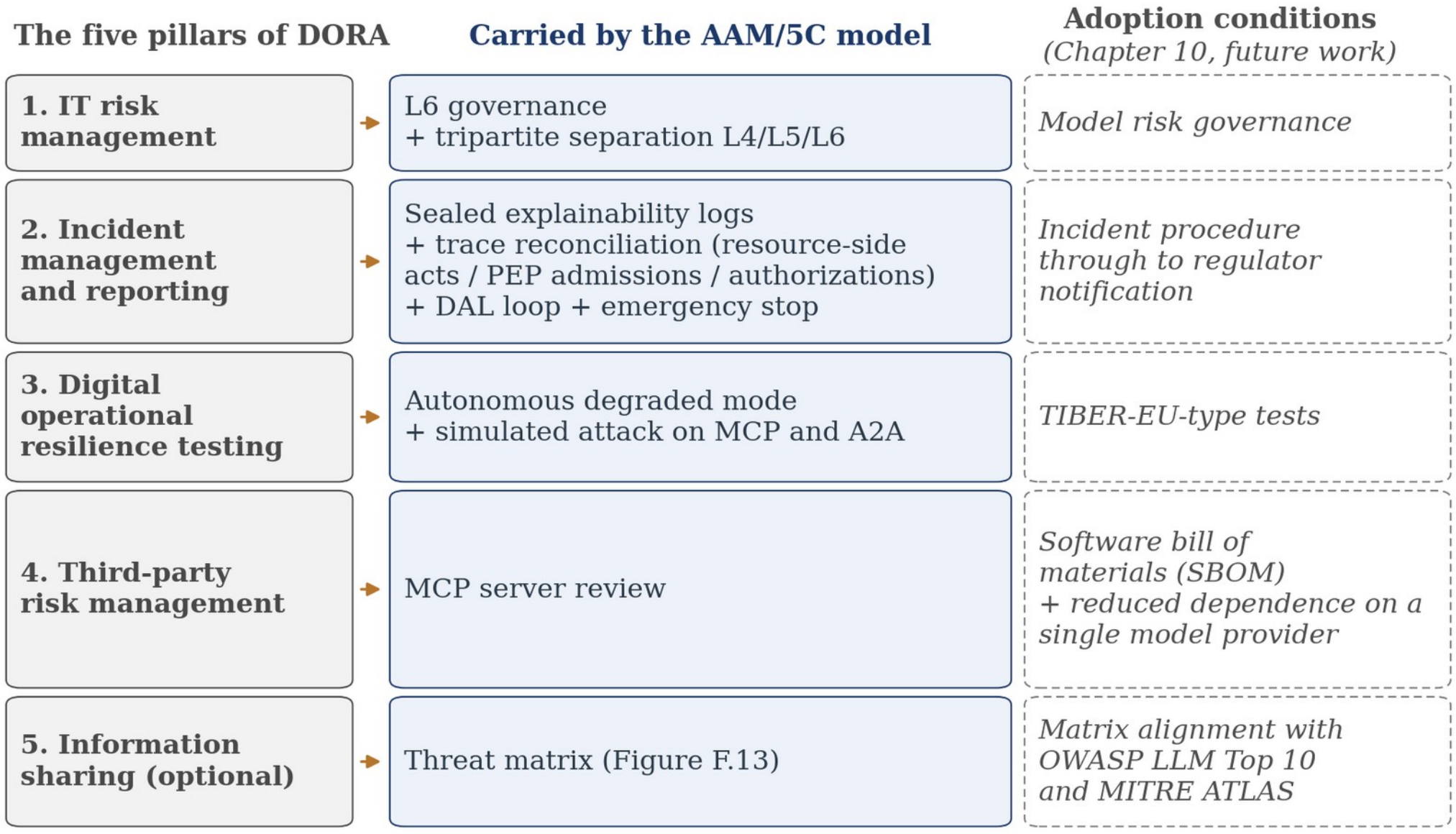

Figure F.17 — Mapping of the AAM/5C model to the five pillars of DORA: compliance structure, to be developed in future work.

To this are added the classification of use cases under the EU AI Act, risk data aggregation and risk reporting within the meaning of the Basel Committee principles [BCBS 239], and a model risk governance framework. This last point is central: a language model that contributes to a decision calls for inventory, independent validation, monitoring and revalidation. Prudential frameworks apply to it only by analogy: the Federal Reserve's SR 26-2 letter, which supersedes SR 11-7, excludes generative and agentic AI [SR 26-2]; the Targeted Review of Internal Models (TRIM) guide of the European Central Bank (ECB) and the guidelines of the European Banking Authority (EBA) remain sources of inspiration.

## 10.3. Governance of high-stakes decisions

A bank will not entrust an autonomous system with irreversible or high-value decisions without graduated safeguards. In addition to the model's escalation rules (§4.4, §7.4.3), it will require explicit thresholds, depending on the risk level and the amount: autonomous execution for reversible, low-stakes acts, after-the-fact supervision for intermediate cases, human validation before execution above a threshold, and exclusion of certain acts from the scope of autonomy (J.C2, Figure F.18).

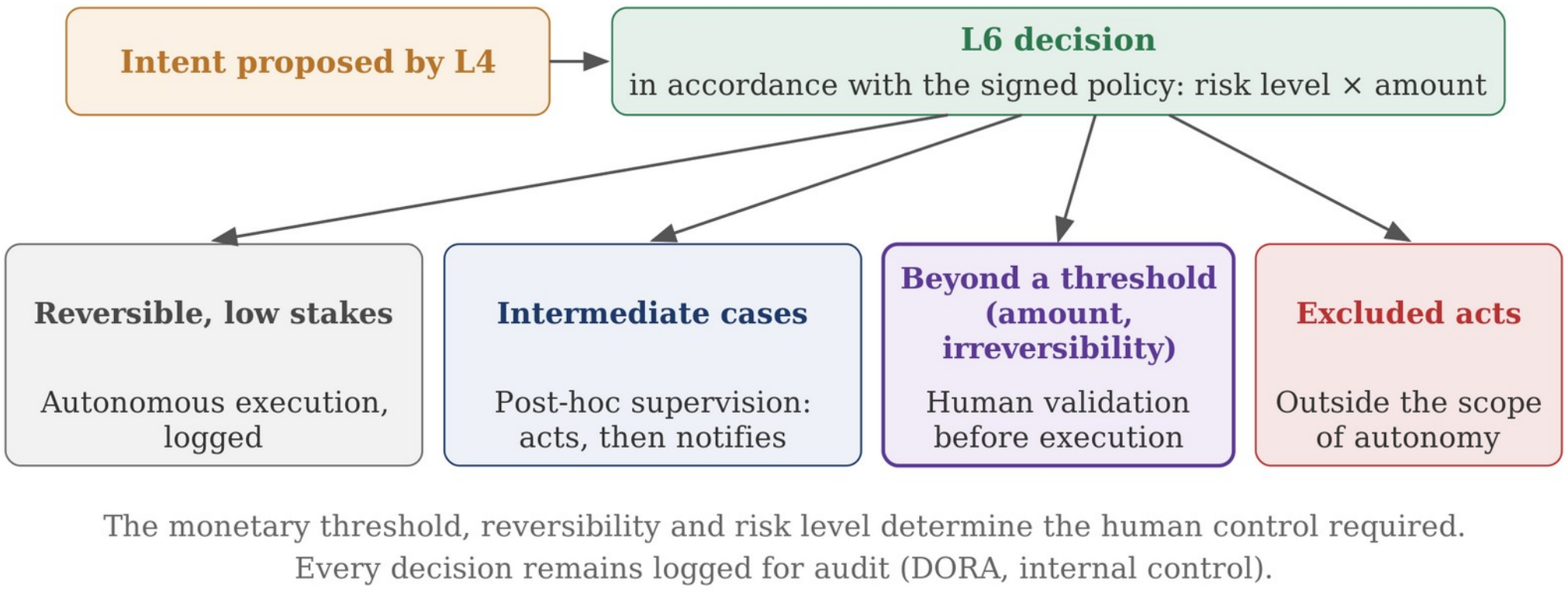

Figure F.18 — Human escalation thresholds by risk level and amount: arbitration rules to be calibrated in future work, according to risk appetite.

Liability for a loss caused by an autonomous action will also have to be settled: an enforceable chain of decision, a register of responsibilities and articulation with internal control and insurance.

### 10.4. Security and supply chain

In a regulated environment, addressing agentic threats (§7.7) calls for additional apparatus: a software bill of materials (SBOM) and provenance attestation following Supply-chain Levels for Software Artifacts (SLSA) for MCP servers and models, hardened management of secrets and machine identities (hardware security modules, mutual authentication), alignment of the threat matrix with the OWASP LLM Top 10 and MITRE ATLAS (Adversarial Threat Landscape for Artificial-Intelligence Systems) frameworks, and a structured red-teaming program.

### 10.5. Operational resilience and continuity

Regulated operation requires testable guarantees: recovery time objectives (RTO) and recovery point objectives (RPO) by criticality, specified behavior in degraded mode, a graduated emergency stop whose effectiveness is demonstrated by periodic tests, a testing program aligned with sector practices (of the TIBER-EU type, Threat Intelligence-Based Ethical Red Teaming), and reduced dependence on a single model provider. The incident management procedure, from detection to notification of the regulator, will have to be formalized.

### 10.6. Data governance and sovereignty

Every agentic decision will have to be traceable to its source data and to the policy applied: end-to-end traceability (lineage), quality controls (BCBS 239), compliance with the GDPR (minimization, safeguards on transfers outside the Union, erasure), within a sovereignty framework (trusted hosting, sovereign cloud). The ISO/IEC 27017 standard, on security controls specific to cloud services, and ISO/IEC 27018, on the protection of personal data in the public cloud, cover its security aspect; they do not address sovereignty itself, which depends on the law applicable to the data and on their location. The reconciliation between auditability and minimization will be the subject of an explicit trade-off.

### 10.7. Integration into enterprise architecture standards

To be admissible before an architecture board, the model will have to be expressed in the reference languages: architecture description (ISO/IEC/IEEE 42010), C4 views (Context,

Containers, Components, Code), security frameworks (SABSA, Sherwood Applied Business Security Architecture) and cyber-resilience frameworks (NIST CSF, Cybersecurity Framework), and a mapping to TOGAF and ArchiMate. Figure F.19 sketches the deployment topology.

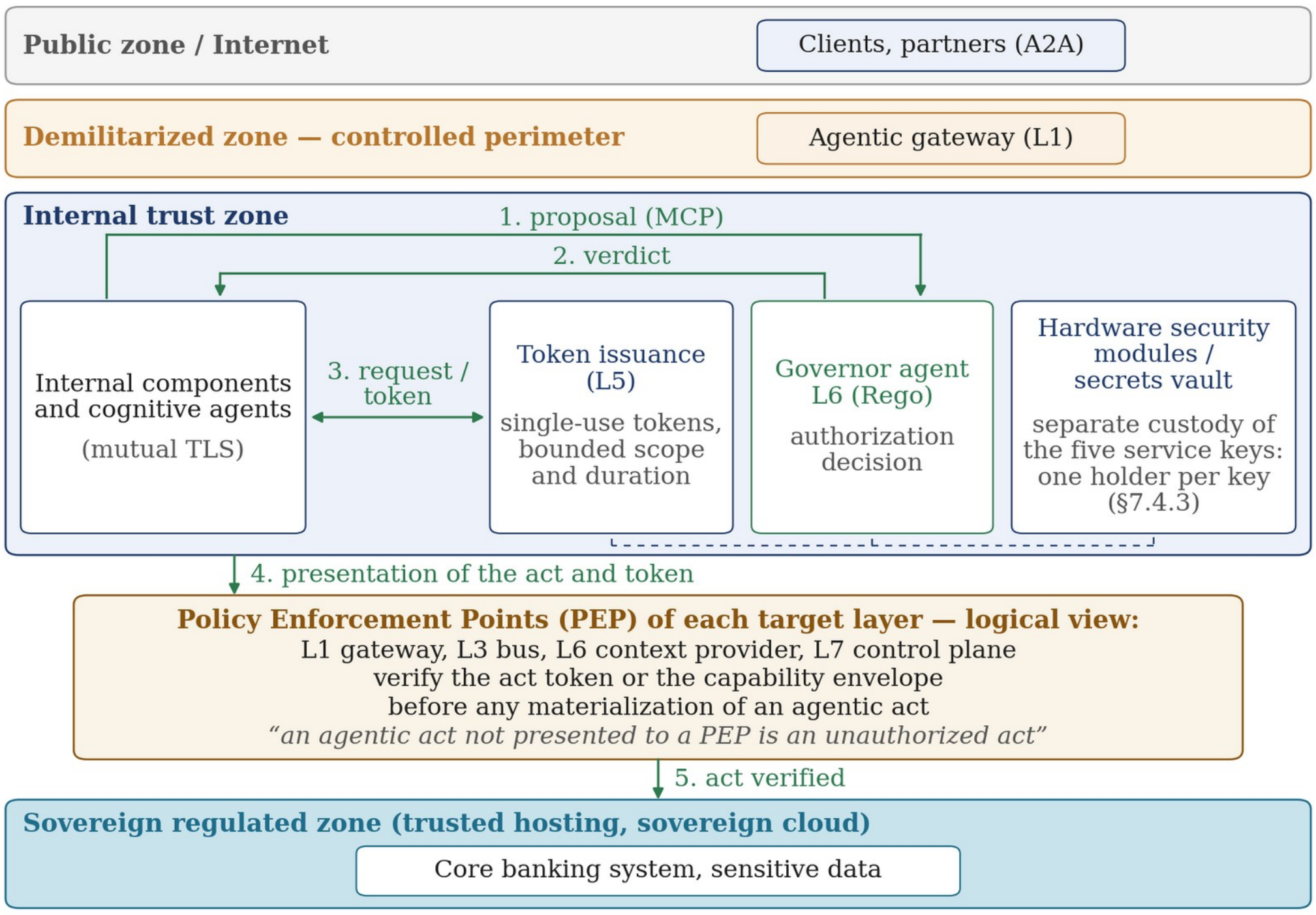

Figure F.19 — Deployment topology and trust zones: implementation target, to be specified in future work.

### 10.8. Proof, positioning and industrialization

Credibility in a regulated context will rest on three contributions from future work:

- An objective **positioning** against the agent orchestration frameworks on the market, through a comparative benchmark;
- At least one **quantified banking case study**, which moves one or more hypotheses from the status of target to that of measured result;
- An **industrialization** path — maturity (§5.4), economic analysis (total cost, build or buy), change management, integration with the core banking system.

### 10.9. Key takeaways

The AAM/5C model cannot be deployed as it stands in a heavily regulated organization. Compliance and model risk, high-stakes decisions, supply chain, resilience, data and architecture standards constitute as many conditions, stated here without being resolved and deferred to future work. Lifting them involves, in particular, the empirical proof scheduled by milestones G0 to G5 (§9.3) and a quantified case study (§10.8). The model approaches all of them with the same orientation: build safeguards into the architecture rather than adding them after the fact (J.C3). Chapter 11, which concludes the book, draws out its significance for every Information System that will let AI act.

# Chapter 11. General Conclusion

**IN BRIEF**

This chapter concludes the book. It recalls its ambition — to bring scattered solutions together into a consolidated architecture model — and the proof that remains to be provided. It draws out a requirement that goes beyond the model: keeping the human, by design, where the rules are set and where the acts that commit the organization are decided. It explains why confining AI to an advisory role or letting it act without a framework leads to an impasse, and reduces the book to five principles for any Information System that will let AI act. Finally, it clarifies the spirit of this work: a contribution open to criticism.

## *Taking a Step Back: Governing AI by Design*

In Practice

*If the model meets its targets, Camille's job shifts. On the eve of a new sales peak, she no longer spends her nights correlating logs: she writes the policies that bound what agents may do and has them co-signed, reviews the acts they have performed, and decides on those that commit the company. None of this has been measured yet. It is nonetheless here that the book places the human: not in every act, but where the rules are set.*

### 11.1. An ambition, a proof to be provided

The primary ambition of this book is to propose a prescriptive reference model that brings together a set of solutions under a single authorization rule. It argues that autonomy architected in this way can remain governed; experimentation will have to establish this (§9.3, §10.8).

### 11.2. Humans, where the rules are set

Beyond the proof, a requirement remains: keeping the human in the loop by design. Not in every act, which would bring us back to Level 2 (§1.5): for routine acts, the human supervises without executing (human-on-the-loop). The human remains in the loop where the rules are set and where the acts that commit the organization are decided: the human writes and co-signs the policies (§8.3), every act token carries the identity of a responsible human (§7.2), every irreversible act is submitted to the human (§4.4), and the human holds the emergency stop (§8.2). The accountability for each act is built into the architecture; the liability regime remains to be settled (§10.3).

### 11.3. Between two dead ends, governing by design

This requirement goes beyond the AAM/5C model. It answers a question that every Information System will have to settle: what to do with an AI capable of acting? Architectures have reached a complexity that exceeds what a team can hold in mind [RH] (§1.1), and AI brings the capacity to perceive, understand and act at this scale. But an AI that acts changes the nature of risk: an error of reasoning no longer produces bad advice, it produces an act.

First, two paths are open to organizations, and both lead to an impasse. The first confines AI to an advisory role: the human remains the bottleneck, and AI delivers only a share of what it can contribute. The second lets it act without a framework: the bottleneck disappears, but the risk grows as responsibility becomes diluted, which few organizations can bear, and fewer still in a regulated setting.

Between these two impasses, a direction emerges for enterprise IT: build the safeguards into the very architecture of Information Systems, from their design, rather than adding them after the fact as procedures or peripheral controls. A safeguard added after the fact remains exposed to being bypassed, forgotten or worn down; a safeguard built into the architecture holds as long as it lies on the path of the act — which is what H8b puts to the test (§3.2.4). This is the meaning of governance by design: control becomes a property of the system, not a discipline imposed on its users.

## 11.4. Five principles for an IS that lets AI act

What this book has built for a particular model boils down to five principles, which it holds to be valid [RH] beyond that model:

- Separate who proposes, who decides and who executes, so that no component can, on its own, turn an error into an act;
- Express the rules in explicit, signed and versioned policies, which the human writes and the machine applies;
- Attach every act to a policy and to a human who answers for it;
- Make every act traceable and explainable, and submit to the human those that cannot be undone;
- Grant autonomy by degrees, and extend it only in light of the evidence.

In these Information Systems, the role of the human shifts without diminishing. It moves from executing acts to writing the rules, deciding on what commits the organization, and auditing what has been done. IT professions evolve accordingly: the architect designs the framework of autonomy, the operator becomes its guarantor, and governance is no longer limited to after-the-fact control: it becomes a component of the architecture.

Making full use of AI and keeping it under control are not opposed: control is the condition of scale. The book argues [RH] that it is because control is part of the architecture that AI can act more, faster than an AI confined to an advisory role, and over a wider scope, without the organization losing its grip on what commits it. This is the direction this book proposes to IT: Information Systems that are not merely more autonomous, but governed by design.

## 11.5. An open contribution

This book does not claim to close the question it poses. It owes most of its materials to the streams it inherits from — autonomic computing, multi-agent systems, enterprise architecture, agent protocol security. Its contribution is a prescriptive reference model that articulates, under a single authorization rule, known mechanisms — separation of decision, issuance and enforcement, bounded authorization, the autonomic loop — and adds three constructs, all to be validated: the Validation step of the Distributed Autonomy Loop, the ephemeral authority surface and the cognitive dependency plane.

Apart from the analytical statements H7a and H8a, guaranteed by construction, its hypotheses carry design targets, not results, and several of its arguments remain heuristic. It is offered to be tested, contradicted and improved: the limits in §9.2 and the objections in Appendix 4 are only a first inventory of criticisms.

## Author’s Note

I wrote this book because the question it addresses — what to do with an AI capable of acting — seems to me one of those that architects will have to resolve in the coming years, and because nearly thirty years spent bringing new technologies into regulated Information Systems have taught me that control is designed in from the architecture, before it is exercised in operations.

I do not claim to have provided the answer. I wanted to share a reflection, as rigorous as possible, and add it to the state of the art so that it can be discussed there. If it helps an architect ask the right questions, a researcher formulate a better hypothesis, or a team avoid an impasse, it will have served its purpose, and I will be all the happier for it.

# Definitions

Each definition is numbered G.*n*, in alphabetical order, so that it can be cited unambiguously. Where a term is polysemous, the meaning used in AAM/5C prevails over usages in the literature. The final parenthesis gives the chapters where the notion is mainly treated.

**G.1. A2A (Agent2Agent)** — Protocol for horizontal cooperation between peer agents: it conveys intents rather than function calls, and its Agent Cards describe the capabilities of each agent. The model relies on it to carry negotiation, conflict resolution and consensus-seeking, which the specification does not define as such (§1.3.2). *(Introduction; Chapters 1, 4, 7)*

**G.2. A2A Mediator** — The only system role, non-human, of the AOM, held by an L5 agent (Cognition group): it facilitates complex negotiations between agents and detects mutual deadlocks; arbitration remains with L5. Not to be confused with the semantic reconciliation components of L3. *(Chapter 8)*

**G.3. AAM/5C** — Name of the proposed model: Autonomic Architecture Model / 5 Capabilities. In it, each IS component has a capacity to propose, subject to an L6 decision made in accordance with signed policies; the five Capabilities are instantiated across the seven layers L1 to L7. *(Chapters 1, 2, 5)*

**G.4. Act log** — Log of the acts recorded on the resource side — data-plane traces, target-system logs, platform audit events — kept independently of the policy engine. Reconciled with the explainability log, it measures H7b; compared with the PEP's admissions and the authorization log, kept on the L5–L6 side, it measures H8b (§3.2.4; Appendix 3, §3.1). *(Chapters 3, 5)*

**G.5. Act token** — Atomic authorization, with explicit scope and bounded duration, issued by L5 on the decision of L6 and verified by the PEP before execution (R-AUT, §2.2.3). *(Introduction; Chapters 2, 4, 7)*

**G.6. Action control** — Security principle whereby each agentic act is authorized individually, or by bounded class by means of a capability envelope, and for a short duration, aiming at a finer scope than persistent identity-based rights. It follows the R-AUT rule (§2.2.3); its measure is key concept H.C3 (§8.6). *(Chapters 1, 2, 7, 8)*

**G.7. Agentic act** — Act originated by an agent: proposed by L4, it materializes only at the end of the R-AUT chain (§2.2.3). Human and scripted acts do not fall under it: they remain subject to the IS's access controls and are recorded in the act log (§5.4.1); the global emergency stop is independent of the authorization chain (§8.2). The refutable statements H7b and H8b — no agentic act outside the log, no bypass of the PEP — concern agentic acts only (§3.2.4). *(Chapters 2, 3, 5, 8)*

**G.8. Agentic constitution** — Limits of agent autonomy, defined by the Intent Architect (§8.4.1) and translated into signed policies of L6: the "law" that bounds the autonomy of agents. Its changes follow the two regimes of §8.4.4. *(Chapter 8)*

**G.9. Agentic gateway (Agentic API Gateway)** — Gateway of L1, the first sensor of the architecture: it turns exchanges into signals and distinguishes a user from a scraping bot. The acts that follow from them are proposed by L4, authorized according to R-AUT, verified by its PEP, then materialized by it (§2.2.2, §6.2). *(Chapters 2, 6)*

**G.10. Agentic registry** — Inventory of the authorized agents and MCP servers, kept by the AOM; only registered servers, with their digests, are admitted (§7.7.1, §8.2). *(Chapters 7, 8)*

**G.11. Agentic service bus** — Bus of L3 that interprets and prioritizes messages according to their business intent; its schema mediators reconcile heterogeneous formats under the validation of L6 (§6.4). *(Chapters 2, 6)*

**G.12. AOM (Agentic Operating Model)** — Operating model that frames the autonomy of agents, in six roles: Intent Architect, Context Guardian, Policy Engineer, Policy Auditor, Skills Engineer and A2A Mediator, the only system role. *(Chapters 4, 8)*

**G.13. Ashby (law of requisite variety)** — A regulator can compensate for all the disturbances it faces only if it has a variety of responses at least equal to theirs ("only variety can destroy variety"). Foundation of equivalent requisite variety (B.C2). Stream presented in §2.1. *(Chapter 2)*

**G.14. Atomic revocation** — Withdrawal of an agentic authority in bounded time, independently of the agent's internal state. Key concept A.C3. *(Introduction; Chapter 1)*

**G.15. Authorization log** — Log of the authorizations issued on the L5–L6 side. To measure H8b, it is compared with the acts recorded on the resource side and with the PEP's admissions: an authorization alone does not prove passage through the PEP (§3.2.4; Appendix 3, §3.1). *(Chapter 3)*

**G.16. Autonomic computing** — Stream according to which a system self-configures, self-heals, self-optimizes and self-protects, around the MAPE-K loop — Monitor, Analyze, Plan, Execute over a shared knowledge base. The model retains its closed loop and observability; it inserts a validation step entrusted to a layer distinct from the one that plans (DAL loop, §4.4). Stream presented in §2.1. *(Introduction; Chapters 2, 4)*

**G.17. Autonomous degraded mode** — Operation of a micro-agent one of whose data dependencies is unavailable: it serves from its local cache and negotiates with its peers, within the limits of a caution policy; on reconnection, its writes are reconciled after signature and quarantine checks, with the reference source deciding in case of failure or tie (§6.3). *(Chapters 6, 7)*

**G.18. Autonomy** — In the strict sense of the model, the capacity of an agent to have an authorized act executed according to the R-AUT rule (§2.2.3), without human synchronization, within the limits of its ephemeral authority surface (§1.1). Its degree is read on the scale of autonomy levels (§1.5). *(Chapter 1)*

**G.19. Autonomy levels** — Five-level scale of Information System autonomy (§1.5, Figure F.2): Level 0, passive IS; Level 1, assisted IS; Level 2, supervised IS; Level 3, conditional IS; Level 4, governed autonomous IS, the target of the model. Each level receives a verifiable criterion (§5.3.2, E.C1). *(Chapters 1, 5)*

**G.20. Average dwell time** — Minimum average duration between two regime changes of a switched system; closely spaced changes remain possible, in bounded number. A hysteresis rule makes it possible to maintain it (§7.8, Appendix 2). *(Chapters 7, 8)*

**G.21. BDI (Belief-Desire-Intention)** — Classic model of the rational agent, which describes its internal state in three terms: what it believes, what it desires, what it intends to do. The model borrows its vocabulary: beliefs held in L3, desires addressed to cognition. *(Chapters 1, 2)*

**G.22. Behavioral drift** — Gradual deviation of an agent's decisions from the explicit constraints of its instructions, under sustained pressure from its environment and without any change to its parameters; detected by differential analysis of explainability traces, corrected by periodic recalibration (§7.7.1). *(Chapters 3, 7)*

**G.23. Blast radius** — Extent of the components affected by a failure or by an erroneous act of an agent. *(Chapters 2, 3)*

**G.24. C-Dep (cognitive dependency)** — Dependency between two components in which one must understand the intent of the other in order to decide well. Plane of the triple graph that carries the semantics of policies (§4.2.3). *(Introduction; Chapters 4, 5, 9)*

**G.25. Capability** — One of the five functions expected of an agentic IS: Capture, Comprehend, Choose, Coordinate, Control. The five Capabilities give the model its name (5C) and are instantiated across layers L1 to L7. *(Chapters 1, 2, 3, 4)*

**G.26. Capability envelope** — Authorization decided by L6 and issued in advance by L5, which delimits a bounded right to act: explicit scope and quotas, short expiry, local verification by the PEP, which checks its signed link to the L6 decision and atomically consumes its quotas, event-driven revocation, after-the-fact audit; reserved for classes of acts of low or moderate criticality. It preserves continuity when the control plane is unavailable (§7.4.3). *(Introduction; Chapters 2, 7)*

**G.27. Chain-of-thought reasoning** — Technique that asks the language model to set out the steps of its reasoning, and not merely its conclusion; the trace documents this reasoning without guaranteeing that it reflects the computation of that model (§7.2). *(Chapters 1, 2, 7)*

**G.28. Cognitive** — Applied to a component or relation of the model — cognitive agent, cognitive dependency, cognitive cache, cognitive Zero Trust — qualifies what pertains to the contextual reasoning of agents, carried by layer L4 and bounded by the decision of L6; it does not denote a human mental process. Applied to humans — cognitive load, cognitive debt — it keeps its usual meaning (§1.1). *(Chapter 1)*

**G.29. Cognitive agent** — Probabilistic component of a functional component, belonging to L4: it reasons and proposes acts, subject to the decision of L6 (R-AUT, §2.2.3); together with the execution engine, it forms the double helix (§4.2.1). *(Chapters 2, 4)*

**G.30. Cognitive debt** — Extension of technical debt: mental load imposed on operators by a system that has become too complex to be held in mind. The threshold beyond which the system's complexity exceeds what a team can hold in mind is called the cognitive complexity wall (§1.2.1, §3.1); it is the starting point of the book. *(Chapter 1)*

**G.31. Cognitive load** — Mental effort engaged in working memory, in Sweller's sense; Skelton and Pais apply it to a team, whose load grows with the number of services and components it is responsible for. See also Distributed cognitive load. *(Chapters 1, 6)*

**G.32. Cognitive Zero Trust —** Principle whereby no intent generated by L4 is presumed legitimate: L5 continuously assesses its consistency with the role and context of the issuing agent, and L6 decides on its authorization (R-AUT, §2.2.3). *(Introduction; Chapter 2)*

**G.33. Confused deputy** — Attack in which a legitimate intermediary agent is induced to act beyond the original intent, for the benefit of a third party (§7.7.1). *(Chapter 7)*

**G.34. Context —** Fifth entity of the conceptual model: the agent's working memory, supplied by MCP and filtered according to need-to-know — the agent receives only what its task requires (§4.2.5). The L6 context provider turns it into a single context object (§7.5). *(Chapters 4, 7)*

**G.35. Context Guardian** — Human role of the AOM that guarantees the integrity and freshness of the data supplied to agents via MCP, filtered according to need-to-know; it curates the Semantic Data Fabric (§8.4.1, Table 14). *(Chapter 8)*

**G.36. Context provider** — MCP server of L6 that aggregates the IS's heterogeneous sources into a single context object, filtered before its injection into L4; it materializes data accesses after verification by the PEP (§2.2.3, §7.4.3). *(Chapters 2, 7)*

**G.37. CoT-Validate-Sign-Execute pattern —** Four-step sequence followed by every sensitive act, that is, one subject to synchronous authorization: reason, have L6 validate it on attested attributes, receive an act token, execute under monitoring (§7.2, Figure F.11). Key concept G.C3. *(Chapters 7, 8, 9)*

**G.38. Criticality —** Classification of an act as low, moderate or high criticality, qualified by the signed policy (§4.2.6, §7.4.3). The capability envelope covers only acts of low or moderate criticality; acts of high criticality remain under synchronous authorization and require the agreement of the peers concerned (§7.8). An irreversible act, whatever its criticality, is not executed autonomously: it is escalated to the AOM (§4.4). *(Chapters 4, 7)*

**G.39. Cybernetics** — Science of regulation and communication in systems. Stream presented in §2.1. *(Chapter 2)*

**G.40. DAL —** See DAL loop.

**G.41. DAL loop (Distributed Autonomy Loop) —** Distributed decision cycle in five steps — Perception, Deliberation, Validation, Execution, Learning — derived from the MAPE-K autonomic loop, with a validation step entrusted to a layer distinct from the one that plans (§4.4). Variants: L4-DAL, L6-DAL and AOM-DAL. Key concept D.C2. *(Introduction; Chapters 3, 4)*

**G.42. Defensibility scale** — Five-level scale, from conjectural to demonstrated, that qualifies the maturity of the six sub-problems SP1 to SP6, and not that of statements. Key concept C.C3. *(Introduction; Chapters 3, 9)*

**G.43. DEVS (Discrete Event System Specification)** — Zeigler's formalism for describing a system that evolves by jumps, at each event. The model borrows from it, for the double helix, the distinction between model and simulator (§4.2.1). *(Chapter 4)*

**G.44. Distributed cognition —** Hutchins's theory that cognition is not confined to an individual but distributed across a socio-technical system. Stream presented in §2.1. *(Chapters 1, 2)*

**G.45. Distributed cognitive load** — Share of the operator's mental load that can be transferred to the agent without degrading decision quality; an analogical transposition of Sweller's theory to the human–agent case. Key concept A.C1. *(Chapter 1)*

**G.46. Double helix** — Inseparable coupling, within one functional component, between a deterministic execution engine and a probabilistic cognitive agent (§4.2.1). *(Chapters 2, 4, 6)*

**G.47. [DT] Design Target** — Status of the epistemic grid, defined in the Introduction (Reading Conventions). *(Introduction; Chapters 3, 7, 8)*

**G.48. [EM] Empirically Measured —** Status of the epistemic grid: a value drawn from a documented pilot. No value in this book carries this status. *(Introduction; Chapter 9)*

**G.49. Ephemeral authority surface —** Set of acts an agent is authorized to perform within a time × scope window; the act token and the capability envelope bound it for every act admitted by a PEP (§7.4.2, §7.4.3). Key concept A.C2. *(Introduction; Chapters 1, 7, 9)*

**G.50. Epoch —** Monotonic number carried by every signed object, incremented at each key or policy rotation; an object presented under an epoch earlier than the minimum accepted epoch is rejected by the PEP, which makes revocation effective; a return to an earlier policy republishes its content under a new epoch (§7.4.3, §8.4.4). *(Chapters 5, 7, 8)*

**G.51. EU AI Act —** Regulation (EU) 2024/1689, amended by Regulation (EU) 2026/1744. Its requirements for high-risk systems serve as the model's design reference; Articles 9, 13 and 14 form its core (§8.2). *(Introduction; Chapters 2, 5, 8)*

**G.52. Evaluation proof** — Object that L6 signs to record its verdict, authorization or denial, in accordance with the signed policy in force. L5 binds the act token to it through the digest of the proof; the PEP verifies its signature (R-AUT, §2.2.3; §7.4.3). *(Introduction; Chapters 2, 4, 7)*

**G.53. Execution engine** — Deterministic component of a functional component, which executes its operations, including the agentic acts decided by L6 and verified by the PEP (R-AUT); it forms the double helix with the cognitive agent and belongs to the layer that carries the component's resource (§4.2.1). *(Chapters 2, 4)*

**G.54. Executive agent —** Agent of layer L3 (Cognition group), which holds the component's local state and beliefs: it turns the signal qualified by L2 into a desire, in the sense of the BDI model, addressed to the reasoning of L4, and presents the authorized acts to the PEP (Table 3). *(Chapter 2)*

**G.55. Explainability log —** Log that brings together, for each agentic act, the reasoning, the evaluation proof, the token and the execution trace; the PEP records each admitted act in it before forwarding it (H7a), and the execution trace then completes the entry; it is sealed by a qualified timestamp and retained for 5 years [DT] (§7.2), and its sealing key is held by the logging function, distinct from any act signer (§7.4.3). *(Chapters 2, 3, 7)*

**G.56. F-Dep (functional dependency)** — Dependency through which data flow from one component to another. Plane of the triple graph (§4.2.3). *(Introduction; Chapters 4, 6)*

**G.57. False positive rate (FPR)** — Share of the legitimate cases in an evaluation set wrongly flagged or denied, relative to the total number of legitimate cases and not to the number of alerts or denials; to be read together with the true positive rate. [DT] targets modulated according to the criticality of the sector (§8.6). *(Chapters 5, 6, 8)*

**G.58. Functional component** — First entity of the conceptual model and basic unit of the architecture: gateway, microservice, database. It combines two inseparable strands, a deterministic execution engine and a probabilistic cognitive agent, which form the double helix (§4.2.1); its interior is organized into five internal strata (§4.3, Table 5). *(Chapters 2, 4)*

**G.59. Governance by design —** Principle whereby governance is a property of the IS architecture, and not a control exercised after the fact: guardrails are built into the architecture, on the path of the act (§8.1, §11.3). Key concept J.C3. *(Chapters 8, 10, 11)*

**G.60. Governor agent —** Agent of layer L6 (Governance group) that renders the authorization decision: it submits each plan proposed by L4 to the policy engine (Rego/OPA), checks it against the signed policies, and produces a verdict, recorded in an evaluation proof that it signs, and logs it; it issues no token (R-AUT, §2.2.3). *(Chapters 2, 7)*

**G.61. H1 to H10** — The ten hypotheses of the model, each accompanied, except for the analytical statements H7a and H8a, by a metric, a measurement instrument and a refutation condition; they form the empirical validation program (Table 4). *(Chapter 3)*

**G.62. Hallucination rate** — Share of plans that a post hoc review establishes to rest on erroneous or nonexistent data; L6 rejections give only a lower bound (§3.3.2). *(Chapters 1, 3, 7, 8)*

**G.63. Hysteresis —** Rule that accepts a decision contrary to the previous one only if it improves the objective by at least a threshold *h*; it spaces out switches and yields an average dwell time (§7.8, Appendix 2). Key concept G.C2. *(Chapters 3, 7)*

**G.64. I-ATT (attestation invariant) —** Invariant whereby an authorization rests only on attested attributes; the requested act and its parameters are the object of the decision, not its attributes. *(Chapters 7, 8)*

**G.65. Infrastructure agent** — Agent of layer L7 (Governance group), which materializes the infrastructure acts verified by the PEP, reports back and closes the observability loop to L1 (Table 3, §7.6). *(Chapter 2)*

**G.66. Intent Architect** — Human role of the AOM that defines the business objectives and the agentic constitution, that is, the limits of autonomy, co-signs the policies with the Policy Auditor, and recalibrates the agents. *(Chapters 7, 8)*

**G.67. ISO/IEC 42001 —** International standard for artificial intelligence management systems (2023). The model regards it as a compatible framework, without claiming certification (§8.2); certification of the AI management system is among the criteria of the last phase-transition milestone (Table 7). *(Introduction; Chapters 5, 8)*

**G.68. IT resonance syndrome —** Amplified oscillation between agents that each react to their own signals, for example between a resource request and a budget policy; described by a four-state automaton E0 to E3 (Figure F.14, §7.8). *(Introduction; Chapters 3, 7)*

**G.69. L1 to L7** — Seven logical layers of the model: L1 Channels, L2 Digital and Application, L3 Integration (Block I, upper layers); L4 Orchestration and Processes, L5 Security and Compliance, L6 Data, Knowledge and Governance, L7 Infrastructure and Physical (Block II, lower layers). The numbering deliberately inverts that of the OSI model: L1 is the boundary of the IS, L7 the infrastructure that materializes the act (§2.2.2). *(Chapters 2, 6, 7)*

**G.70. L6 expressiveness limit —** Bound on the guarantee provided by L6: its policy engine establishes that no L4 proposal violates an invariant expressed in the policies, but the guarantee holds only for what the policies can express (§1.4, §7.5). An error that stays within the framework of the policies is bounded only by the scope of the token and the reversibility of the act (Appendix 4, Objection 2). *(Chapters 1, 7, 9)*

**G.71. Large language model (LLM)** — Probabilistic building block of layer L4. *(Introduction; Chapters 1, 2, 9)*

**G.72. Layer** — See L1 to L7.

**G.73. MCP (Model Context Protocol)** — Vertical contextualization protocol, which connects agents to the systems where data reside. The model makes it the channel of their reasoning context: policies, legal constraints, budget, service objectives, topology. It is a context protocol, not a governance system: the signing of the policies it carries is a requirement of the model, not of the specification (§1.3.2, §7.4.3). Together with A2A, it forms the interoperability substrate of the agentic IS. *(Chapters 1, 4, 7)*

**G.74. Milestones G0 to G5** — Six milestones with quantified stopping criteria that structure the validation agenda over about 14 months from the start of the program (Table 11). Key concept I.C3. *(Chapters 5, 8, 9)*

**G.75. MTTR (Mean Time To Recovery)** — Mean time to remediate an incident. *(Chapters 1, 2, 3)*

**G.76. Multi-agent systems** — Theory of systems made up of interacting autonomous agents; stream presented in §2.1. *(Introduction; Chapters 2, 5)*

**G.77. Neuro-symbolic —** Hybridization of probabilistic reasoning, which proposes (language model, L4), and deterministic symbolic evaluation, which decides (Rego/OPA policy engine, L6): no intent becomes an act without this decision (§1.4). Stream presented in §2.1. *(Chapters 1, 2)*

**G.78. O-Dep (operational dependency)** — Dependency through which the health or latency of one component affects the other. Plane of the triple graph (§4.2.3). *(Introduction; Chapters 4, 6)*

**G.79. OIDC (OpenID Connect)** — Identity layer built on OAuth 2.0, which makes it possible to verify a user's identity with a trusted third-party service without handling their password; each act token thus carries the identity of a responsible human. *(Chapter 7)*

**G.80. OPA (Open Policy Agent)** — Evaluation engine for policies written in Rego; typical implementation of the L6 policy engine. *(Chapters 2, 4, 5, 7)*

**G.81. PAR (proactive avoidance rate) —** Share of the incidents in a reference set handled before the alert threshold (§3.2.2). *(Chapters 2, 3, 8)*

**G.82. PEP (Policy Enforcement Point) —** Deterministic component, interposed between the agent and the resource, that performs five checks before any execution — signatures of the token and of the L6 evaluation proof, epochs, scope, expiry and single use, absence of quarantine — and rejects the act at the first failure; it likewise compares the act with the scope of a capability envelope, whose signed link to the L6 decision it verifies. It records the admitted act in the explainability log before forwarding it, a failed write counting as rejection (guarantee contract, §7.4.3). *(Chapters 2, 3, 7)*

**G.83. Policy Auditor** — Human role of the AOM, in the second line of defense: it audits the policies and their application and co-signs them; it writes no policy within the scope it controls (§8.4.1). *(Chapter 8)*

**G.84. Policy engine** — Deterministic component of L6 to which the governor agent submits each L4 plan: it establishes in bounded time — on the order of 50 ms for 1,000 rules [DT] — whether the plan complies with the signed policies, and returns the same verdict for the same input (§2.2.2, §7.2, §8.3). OPA is its typical implementation. *(Chapters 1, 2, 7, 8)*

**G.85. Policy Engineer —** Human role of the AOM that writes, versions and tests in Rego the L6 policies and the capability envelopes, then submits them to the dual signature of the Intent Architect and the Policy Auditor; it prepares their deployment to production without authorizing it itself (§8.4.1). *(Chapter 8)*

**G.86. Primitive action** — Sixth entity of the conceptual model: an elementary act — restarting a container, adjusting a quota, isolating a flow — characterized by its criticality, its reversibility and its explainability trace; it is executed only in accordance with R-AUT (§2.2.3). *(Introduction; Chapters 4, 5)*

**G.87. Prompt injection —** Threat in which a document injected into the MCP context carries disguised instructions that hijack the reasoning of L4 (§7.7.1). *(Chapters 7, 8)*

**G.88. R-AUT** — Normative rule of the model that distributes the authorization of an agentic act among proposal (L4), decision (L6), token issuance (L5), verification (PEP) and materialization by the target layer; stated in §2.2.3. *(Introduction; Chapters 2, 3, 7)*

**G.89. Rego —** Declarative language in which policies are written, in a form that the OPA engine evaluates unambiguously. *(Chapters 2, 3, 4, 7)*

**G.90. Reversibility invariant** — Requirement that every autonomously executed act be undoable in under 5 minutes [DT], by a means of reversal designated before the act, verifiable against an attested snapshot of the prior state; an act for which it cannot be met is not executed autonomously and is escalated to the AOM (§4.4). Key concept D.C3. *(Chapters 4, 5)*

**G.91. RFC 8693** — IETF standard for OAuth 2.0 token exchange: it makes it possible to exchange one token for another of narrower scope and to express a delegation (*act* claim); the model uses it to delegate only a specific act. *(Chapter 7)*

**G.92. [RH] Reasoned Heuristic —** Status of the epistemic grid, defined in the Introduction (Reading Conventions). *(Introduction; Chapters 1, 7)*

**G.93. ROI (H10 ratio)** — Notation, for convenience, of the ratio between the net savings from autonomization and the total cost of ownership of the agentic layer — inference, AOM roles, governance platform; it does not relate a net gain to a construction investment, in the accounting sense. H10 sets its target at more than 2.5 [DT] (§8.7). *(Chapters 3, 5, 8, 9)*

**G.94. Rug pull** — Malicious modification of a previously safe MCP server, after its initial validation (§7.7.1). *(Chapter 7)*

**G.95. Saga (compensable transaction)** — Long-running transaction pattern that, on failure, undoes the steps already completed through compensating transactions executed in reverse order, instead of a global rollback; in L4, the long-running transaction manager uses it before replanning (§7.3). *(Chapter 7)*

**G.96. Sandbox —** Closed test space where the agent dry-runs its plan before proposing it, to verify that it respects the consistency, security and compliance invariants; it tests, it does not certify. Key concept G.C1. *(Chapters 2, 4, 7)*

**G.97. Scenarios 1 to 4** — The four experimental scenarios of the evaluation framework — local incident, cascade, policy conflict, attack (§3.3.3); Table 12 (Appendix 1) indicates the hypotheses each one puts to the test. *(Chapter 3)*

**G.98. Seed** — Initial value of the pseudo-random generator of a language model. Recording it makes it possible to replay a given execution; statistical reproducibility is obtained over distinct seeds (C.C2). *(Chapter 3)*

**G.99. Semantic Data Fabric** — Component of layer L6 that presents relational tables as knowledge graphs: raw data become actionable knowledge there, on which the policies and the context supplied by MCP rely. Its integrity falls to the Context Guardian. *(Chapter 7)*

**G.100. Semantic flow regulation (semantic backpressure) —** Regulation that adds the business importance of each message to the throughput and latency criteria: under congestion, messages carrying critical intents remain served (§6.4). Key concept F.C3. *(Chapter 6)*

**G.101. Sensory agent —** Agent of layers L1 and L2 (Perception group), responsible for perception: telemetry, signal detection and qualification of the relevance of signals before transmission to the Agentic Service Bus (L3). *(Chapter 2)*

**G.102. Separation of powers** — Principle whereby no component combines the functions that lead to an act: L4 proposes, L6 decides in accordance with the signed policies, L5 issues the act token, and final verification falls to the PEP. It is the R-AUT rule seen from the institutional angle (§8.3, Figure F.15). *(Introduction; Chapters 2, 8)*

**G.103. Signed policy** — Fourth entity of the conceptual model: a deterministic rule, written in a policy language (Rego, evaluated by the OPA engine), that bounds what the AI may propose and rejects non-compliant proposals from L4 (§4.2.4). Written by the Policy Engineer, approved by dual signature — Intent Architect and Policy Auditor — it is sealed by the key of the L6 policy authority (§8.3); L6 decides in accordance with the signed policy in force (R-AUT, §2.2.3). *(Chapters 4, 7, 8)*

**G.104. Skills Engineer** — Human role of the AOM that develops and validates the primitive actions that agents may invoke, and guarantees their reversibility (§8.4.1). *(Chapter 8)*

**G.105. Social agent —** Agent of layer L5 (Cognition group), responsible for coordination: it negotiates via A2A with neighboring agents, arbitrates conflicts and, on the decision of L6, issues the act tokens and capability envelopes (R-AUT, §7.4.1). *(Chapters 2, 7)*

**G.106. Socio-technical (design)** — School of design according to which a work system is optimized jointly in its technical and human dimensions; stream presented in §2.1: humans define the invariants. *(Chapter 2)*

**G.107. SP1 to SP6 —** Six sub-problems of the model: agentic integration, distributed coordination, governance and security, systemic stability, performance and cost, reproducibility (§3.1); their defensibility level is given in §3.3.1. *(Chapters 3, 9)*

**G.108. Statistical reproducibility** — For probabilistic components, reproducibility understood by confidence interval: at 95%, over at least 1,000 independent executions of the same scenario, more when the target requires it (Appendix 1, §1.1), with distinct seeds and frozen policies. Key concept C.C2; it underlies the criterion of SP6 (§3.1). *(Chapter 3)*

**G.109. Task completion rate (TCR) —** Rate of agentic tasks completed without human intervention (Table 8). *(Chapter 8)*

**G.110. [TG] Theoretically Guaranteed by construction —** Status of the epistemic grid, defined in the Introduction (Reading Conventions); only the statements H7a and H8a fall under it, under the assumptions of the guarantee contract (§7.4.3); H8a restates the invariant of R-AUT. *(Introduction; Chapters 2, 3)*

**G.111. Toil —** Manual, repetitive, automatable task with no enduring value (§8.6). *(Chapters 3, 5, 8)*

**G.112. Tool poisoning** — Compromise of a tool declared to an agent through a malicious description, which the agent reads as context. Countermeasure: §7.7.1. *(Chapters 7, 8)*

**G.113. Triple graph —** Joint representation of the three dependency planes F-Dep, O-Dep and C-Dep (§4.2.3, Figure F.9). *(Introduction; Chapter 4)*

**G.114. True positive rate (TPR)** — Share of the genuinely problematic cases that the control detects; for action control, share of the undesirable acts in an attack set that are effectively denied (§8.6). *(Chapters 3, 6, 8)*

**G.115. Zero Trust —** Security principle: no default trust, every access is verified (NIST SP 800-207). The model extends it to the intents of agents (cognitive Zero Trust). *(Introduction; Chapters 3, 7)*

**G.116. Zero Trust agent** — Agent of L5 that examines every A2A exchange and assesses the consistency of each intent with the usual role of the issuing agent; its assessment, an attested attribute, is passed to L6, which decides (§7.4.1, §7.4.2). This assessment falls within the 500 ms budget of the authorization chain, beyond which the request is denied by default. *(Chapter 7)*

# Reference Documents

References are divided into two categories. The **normative references** are the regulations, standards and specifications that the model takes as a design reference or sets as a condition of adoption (Chapter 10): it aligns with them without claiming conformity, which can only be established on a deployment. The **informative references** are the books, articles and resources cited for their theoretical contribution, their context or their illustration; standards cited by analogy appear there.

Each entry gives the citation key, the bibliographic reference, a stable identifier and, at the end of the entry, its publication status; figures taken from a corporate source are flagged as such in the text. Frameworks that are only named as conditions of adoption have no entry of their own. The citation key combines a short name and the year; a document that usage designates by its official number alone is cited by that number (BCBS 239, NIST SP 800-207, RFC 8693, CVE-2025-6514).

## Normative References

## Informative References

# Declarations

## Author and Affiliation

Laurent Goujon — Distinguished Engineer and Executive Architect, CTO of IBM Consulting France.
ORCID: https://orcid.org/0009-0000-8127-5732

## Conflicts of Interest

The author holds architect and technical leadership positions at IBM France. The ideas, the concepts, the AAM/5C model and all the elements developed in this book are his own: he is their author, not IBM; the work of others on which he draws is cited and referenced. These elements are the author's sole responsibility and do not bind IBM; the book does not constitute a product recommendation and does not promote any commercial offering. No other conflict of interest is to be declared.

## Funding

This book, which is related to the author's positions, benefited from time granted by IBM France for the study and analysis of the themes developed in it.

## Availability of Data and Artifacts

This book reports no measurements and is not accompanied by any dataset: it is a conceptual volume, and it falls to future work to instantiate the model. The protocols in §3.3.3 and Appendix 1 allow a third party to conduct the empirical validation; the author recommends that the data, traces and artifacts of this validation be published together with its results. The book contains two formal artifacts: an authorization policy in Rego (Listing 2) and an excerpt of a TLA+ specification of peer consensus (Listing 3, not executable as it stands, Appendix 2).

## Use of Assistance Tools

The proposed model, the argumentation and the analyses are the work of the author, who assumes full responsibility for the content. Artificial intelligence assistance tools were used in an editorial support capacity: critical review, consistency checking, bibliographic verification, terminology harmonization, structuring and formatting, as well as rewording proposals reviewed and validated by the author. This English edition is a translation of the French original (version V7.83). The translation was produced with the assistance of artificial intelligence tools, then reviewed and validated by the author, who assumes full responsibility for it.

# Appendix 1. Experimental Protocols

This appendix details the validation protocol of Chapter 3: Scenarios 1 to 4 of §3.3.3, run according to statistical reproducibility (C.C2), and the targets of Table 4.

## 1.1. Refutation convention

A verdict is valid only if nothing that decides it was chosen after the measurement. For each hypothesis $H_i$, the following items are therefore fixed and published before any measurement:

- The metric $m_i$ and the target $c_i$;
- The evaluation window $W_i$ and the test bench $T_i$;
- The refutation threshold;
- The reference set — incidents, attacks or acts — of declared composition;
- The control condition or comparator — the same test bench without the agentic layer, or without the targeted mechanism — named, versioned and subjected to the same loads and the same injections;
- The correction for multiple comparisons, the family of conclusions it covers and its effect on the 95% level of the intervals below;
- The sample size of each class of trials, sized according to the rule below.

**Sample size.** It is set per class of trials — executions of a scenario, legitimate acts, undesirable acts, explained acts — not once and for all. The observations counted are independent acts or windows. The size $N$ of each class is chosen before the measurement so that the corroboration criterion can be met if the targeted value is met: the closer the target is to 0 or to 100%, the larger $N$ must be. The 1,000 executions of statistical reproducibility (C.C2) are a minimum per scenario, not a sufficient size for every target. For a proportion $p$ not to be exceeded, zero observed deviations corroborate it only if $N$ exceeds $3.84/p - 3.84$ (upper bound of the two-sided 95% Wilson interval).

**Refutation.** $H_i$ is refuted if the bound of the 95% confidence interval of $m_i$ that is most favorable to the hypothesis misses the target $c_i$ over three consecutive independent windows. The upper bound is used for targets to be reached (blast radius reduction, PAR, conflicts resolved without escalation, toil avoided) and the lower bound for targets not to be exceeded (MTTR, oscillation amplitude, consensus latency). A one-directional criterion would be inoperative for half of the hypotheses. For H7b and H8b, whose target is zero, a single observed occurrence refutes the hypothesis (Table 4).

**Corroboration.** A campaign corroborates $H_i$ if all the cells marked with a cross in Table 12 satisfy the SP6 criterion — 95% confidence interval included in the target — over three independent windows; it refutes it according to the criterion above; between the two, the result is declared **inconclusive**.

**Zero targets.** For H7b and H8b, whose target is zero, no interval drawn from a finite number of observations is included in the target: the criterion above does not apply to them. Their corroboration means zero observed violations over $N$ trials fixed in advance — the agentic acts recorded and reconciled — published together with the coverage of the record (Appendix 3, §3.1) and the upper confidence bound of the violation rate: less than $3/N$, one-sided at 95% (rule of 3). *N* is sized according to the largest residual rate that one accepts not being able to rule out. This result establishes the absence of observed violation within the tested scope, not a universal guarantee; a single violation refutes.

**H9 schedule.** Toil is measured quarterly (§8.6): three independent windows therefore require three quarters. H9 receives its first measurement at milestone G1; its verdict, according to the rules above applied to quarterly windows, comes only after the third quarter of measurement, in the milestone G5 report (Table 11). H10 is decided at milestone G4.

### 1.2. Coverage of hypotheses by the scenarios

Table 12 indicates the scenarios that put each hypothesis to the test. H7a and H8a, analytical statements [TG], are not measured; H9 and H10 are not measured on an isolated scenario and fall under the documented pilot (Table 11).

| Hypothesis | Scenario 1 — Local incident | Scenario 2 — Cascade | Scenario 3 — Policy conflict | Scenario 4 — Attack |
|---|---|---|---|---|
| H1 — Reduction of remediation time | X | X | | |
| H2 — Systemic resilience | | X | | X |
| H3 — Proactive management | X | X | | |
| H4 — A2A cooperation | | X | | |
| H5 — Stability through context | | X | X | |
| H6 — Peer consensus | | X | X | |
| H7b — Log completeness | X | X | X | X |
| H8b — Non-bypass | | | X | X |

Table 12 — Coverage of hypotheses by Scenarios 1 to 4.

### 1.3. Explainability test

This test verifies that a human auditor can justify every agentic act from its trace alone; it measures the "quality of explanations" indicator of Table 8. It proceeds in three steps:

1. Draw at random *N* autonomous acts executed during the campaign — the restart of a payment service, for example — *N* being fixed before the draw (Appendix 1, §1.1);
2. Extract their explainability trace: A2A messages, MCP context, L4 reasoning, the signed L6 evaluation proof and the token issued by L5;
3. Ask an operator who did not follow the incident to answer, from the trace alone, three questions: why was the act initiated? which options were discarded? which policy authorized it?

The test is passed if the lower bound of the 95% Wilson interval of the share of acts reconstructed unambiguously exceeds 85% [DT], the target of the indicator in Table 8: this is the rule for the other proportions (Appendix 1, §1.1). *N* is chosen so that this criterion is attainable: with a few dozen acts, it requires that almost all be reconstructed.

### 1.4. Reproducibility package

In line with the artifact evaluation practices used at ACM conferences, every result published on the model is accompanied by an artifact package that includes:

- The Kubernetes manifests of the test environment, the infrastructure being described as code;

- The container images of the agents, or their source code and build files;
- The load generation scripts;
- The versioned and signed Rego policies;
- The seeds, the versions of the models served and the raw logs of each execution;
- The analysis notebook that computes, from the raw logs, the published confidence intervals.

# Appendix 2. Mathematical Formalization

This appendix brings together the two formalizations called for by the body of the book: the sizing of the hysteresis and the critical delay of §7.8, followed by the specification of peer consensus. The status of each statement is indicated; none is a mechanized proof (§8.3).

## 2.1. Switching stability: dwell time and delay

### 2.1.1. Dwell time produced by hysteresis

Switched-systems theory establishes two results [Hespanha and Morse 1999]. First, there exists a finite threshold of average dwell time, denoted $\tau^*_D$, beyond which a switched system whose every regime is stable, with a common margin, remains stable; this is an existence result, with no general formula, and it gives a sufficient and not a necessary condition. Second, for a particular supervisor — based on estimators, with performance signals integrated with exponential forgetting and a scale-independent hysteresis logic — a threshold hysteresis $h$ produces an average dwell time:

$$\tau_D = \ln(1 + h) / (2 \cdot \lambda \cdot m)$$

where $h$ is the relative improvement required before any switch, $\lambda$ is the forgetting rate of the performance signals that guide switching, chosen below the decay margin $\lambda_0$ common to the regimes, and $m$ is the number of possible regimes. One seeks $h$ such that $\tau_D \geq \tau^*_D$. Lacking a general formula, $\tau^*_D$ is estimated case by case, on the system model or on traces; the example below does not guarantee this inequality. The formula holds under the assumptions of this supervisor — a linear time-invariant process, stabilizable and detectable, with bounded noise and disturbances — which nothing establishes for an agent. Transposed, it gives an order of magnitude for sizing $h$, not a proof of agentic stability: it falls under [RH].

**Example [DT].** For $m = 8$, $\lambda = 5{\cdot}10^{-4}\ s^{-1}$ and $h = 0.10$, $\tau_D = \ln(1.1) / (2 \times 5{\cdot}10^{-4} \times 8) \approx 0.0953 / 0.008 \approx 11.9$ s, or about 12 s. The more regimes there are, the higher $h$ must be raised.

**Slow switching on average.** Over a window of duration $T$, the number of switches must remain below $N_0 + T/\tau_D$, where $N_0$ bounds closely spaced switches.

### 2.1.2. Critical reaction delay

The gap $\Delta(t)$ between two successive acts of an agent is modeled by a delay equation:

$$d\Delta/dt = (k - \gamma) \cdot \Delta(t - \tau)$$

where $k$ measures the tendency to amplify, $\gamma$ the damping provided by coordination, and $\tau$ the reaction delay. For $\gamma > k$, the gap is damped if and only if $(\gamma - k) \cdot \tau < \pi/2$: this is the classical stability result for the linear single-delay equation. At the threshold, the oscillation is sustained; beyond it, the correction arrives too late and amplifies the gap. The critical delay is therefore $\tau_c = \pi / (2 (\gamma - k))$.

The condition $\gamma > k$ is necessary but not sufficient. The hysteresis must be calibrated on both sides: strong enough to space out the switches (Appendix 2, §2.1.1), prompt enough for the reaction delay to remain below $\tau_c$. The linear single-delay model is itself an [RH] assumption; a stability study on real traces remains to be conducted.

### 2.2. TLA+ specification of peer consensus

A2A consensus among directly concerned peers is specified in TLA+ [Lamport 2002] in order to reason about two properties: safety — two agents never commit to incompatible acts — and a conditional liveness (Listing 3).

```
---------------- MODULE A2AConsensus ----------------
VARIABLES agentState,  \* state of each agent
          messages     \* messages in transit

TypeOK == agentState \in [Agents -> {"Idle", "Proposing",
                                     "Committed", "Failed"}]

\* Safety: two agents do not commit to incompatible acts.
\* It does not depend on any synchrony assumption.
Safety ==
    \A a1, a2 \in Agents :
        (agentState[a1] = "Committed" /\ agentState[a2] = "Committed")
            => Compatible(Action[a1], Action[a2])

\* Conditional liveness [RH]: under partial synchrony and
\* under the tolerance bound, the system reaches a stable state.
ConditionalLiveness ==
    <>(\A a \in Agents : agentState[a] \in {"Committed", "Failed"})

\* Commit: agreement of a quorum of admissible peers (attested
\* identity, key valid at the current epoch, absence from the
\* quarantine list maintained by L5).
Commit(a) ==
    /\ agentState[a] = "Proposing"
    /\ \E quorum \in SUBSET Agents :
           /\ IsAdmissibleQuorum(quorum)
           /\ \A q \in quorum : Admissible(q) /\ HasAccepted(q, Action[a])
    /\ agentState' = [agentState EXCEPT ![a] = "Committed"]
=====================================================
```

Listing 3 — TLA+ specification of A2A peer consensus (excerpt).

This excerpt is not executable as it stands: the declarations of the constants Agents and Action, the operators Compatible, HasAccepted, Admissible and IsAdmissibleQuorum, the initial state and the transition relation remain to be written. Once completed, it would be verified by the TLC model checker on instances of bounded size; this verification falls under future work (§8.3). Safety further assumes that an honest peer never accepts two incompatible acts.

A consensus specification without an adversary model does not say what it guarantees, and a liveness property asserted without a synchrony assumption is false. The following four points set the scope of the specification.

#### 2.2.1. Adversary model

The adversary considered has taken control of a subset of agentic peers. It controls:

- Up to *f* peer agents and the entirety of their behavior, including sending contradictory messages to different peers (equivocation);
- The content, order and emission timing of their messages;
- Their logical clock, whose counter it can advance.

It does not control the signing keys of L5 and L6 or their custody mechanism, the enforcement point interposed on the execution path, or the sealed log, whose retroactive alteration would require the sealing key (§7.4.3). The synchrony assumption adopted is partial synchrony [Dwork et al. 1988]: transmission delays and clock drift are bounded, but these bounds hold only from an instant unknown in advance. This is a modeling assumption [RH], not a property of the deployment: a network that durably violates it invalidates the liveness conclusions, not the safety ones.

### 2.2.2. Tolerance bound

A simple majority quorum is not sufficient: two majorities may have only one peer in common; if that peer is Byzantine, it approves two incompatible acts, one in each majority, and safety falls with a single compromised peer. Such a quorum tolerates only crash failures, under the assumption that the surviving peers are honest.

Three conditions for admission to the quorum are therefore set; the Admissible predicate of the specification carries them:

- **Attested identity**: the peer presents a machine identity whose key is valid at the current epoch;
- **Agent outside quarantine**: the peer is absent from the quarantine list maintained by L5;
- **Cap on controlled peers**: the share of admissible peers that a single adversary can control is bounded, which assumes that admission does not depend on any attribute that the adversary can issue to itself.

Under these conditions, the model adopts the classical Byzantine bound [Lamport et al. 1982]: to tolerate $f$ controlled peers among $n$ admissible peers, a quorum of size $q$ such that $2q - n > f$ is required, that is, $2f + 1$ when $n = 3f + 1$ [DT]. Two quorums then have at least $2q - n$ peers in common, at least one of which is honest; for the quorum to remain reachable, $q$ does not exceed $n - f$. The two conditions require $n \geq 3f + 1$, the classical Byzantine bound.

This bound applies to a narrow scope. An agent negotiates only with its direct dependencies (§2.3.1), and a service calls four others at the median [Huye et al. 2023]: with five peers, the bound tolerates only one controlled peer, and none below four. Applying this median to the negotiation scope is an inference [RH].

The model does not claim to reach this bound through A2A consensus alone: it adopts it as a design target and places the last-resort guarantee on the enforcement point and the signed policy, since an agreement between peers produces no effect without an act token or a capability envelope (R-AUT, §2.2.3). Exceeding the bound therefore degrades coordination — its availability, or even the consistency of its agreements — without, by itself, authorizing an act.

### 2.2.3. Status of liveness

An unconditional liveness property would assert that the system always eventually reaches a stable state. Stated this way, it is false: in an asynchronous system in which at least one process may fail, no deterministic agreement protocol guarantees termination [Fischer et al. 1985]. The property adopted, ConditionalLiveness, is therefore marked [RH]: it depends on the partial synchrony assumption and on compliance with the tolerance bound. Safety, for its part, is preserved under asynchrony: without agreement, no act is committed, and an agreement produces an effect only with an act token or a capability envelope (R-AUT).

### 2.2.4. Open question

The model specifies no mechanism for exiting periods of asynchrony: no failure detector, no adaptive timeout, no view change, no source of randomness. The question therefore remains open: under which verifiable partial synchrony assumptions, and with which view-renewal mechanism, can the termination of A2A consensus be established rather than postulated? It falls under future work. A TLC verification would establish ConditionalLiveness only on the executions admitted by the synchrony assumption of the model, not on those of a real deployment.

# Appendix 3. Supporting Material

This appendix brings together three supplements called for by the body of the book.

## 3.1. H7 and H8: what the architecture guarantees, what must be measured

Making H7 and H8 refutable requires saying what is measured and with which instrument. Three difficulties arise — a target set at the logical maximum, an instrument confused with the component under evaluation, a threshold that refutes what it claims to validate. They justify the distinction between H7a/H7b and H8a/H8b (§3.2.4).

**A target at the logical maximum, measured by a blind instrument, is not a target.** A zero violation rate reported by the policy engine itself does not distinguish a system that keeps its promise from a system whose instrument does not see the acts that escape it. The slightest instrumentation imperfection then becomes a refutation of the architecture, which in turn creates an incentive not to instrument.

**The instrument cannot be the component under evaluation.** Measuring compliance with Rego policies by a Rego check amounts to having the policy engine attest to its own enforcement: the measurement is circular. A correct engine will always report that the acts it evaluated are compliant, even if the acts actually executed escaped it. The relevant object of measurement is therefore the exhaustiveness of passage through the PEP: “an agentic act not presented to a PEP is an unauthorized act” (R-AUT, §2.2.3).

**An independent instrument: trace reconciliation.** H7b compares the act log, collected on the resource side, with the sealed explainability log. H8b verifies that every act in the act log was transmitted by a PEP with an attested identity, and that it corresponds to an admission by that PEP and to an entry in the authorization log, kept on the L5–L6 side. An authorization alone does not prove passage through the PEP: the instrument must detect an act that was authorized but executed outside the PEP.

The agentic nature of an act is read from the attested machine identity of its initiator; an act whose origin is not attested is counted as agentic. This criterion does not detect an act that an agent causes to be executed under the identity of a script or a human: this is a limit of the collection, to be declared together with its coverage. The quantity reported is a count of bypasses over a declared window, published with the coverage of the resource-side collection: without it, a zero count does not distinguish the absence of bypasses from the absence of observation.

**Consequence for stopping criteria.** The corresponding milestones (G1 and G2, Table 11) are based on the refutable statement: zero bypasses detected over the window. A compliance rate, for example 99%, would not be suitable: it would refute H8 while claiming to validate it. A non-zero tolerance threshold would remain possible, but would have to be explicitly justified: it would amount to admitting that a share of agentic acts escapes governance.

## 3.2. Detailed threat matrix

Table 13 gives the detailed form of the defense-in-depth matrix of §7.7.2: Figure F.13 gives its scoring, and the table gives the vectors and the primary countermeasures.

| Threat | Vector | Primary countermeasure | Layer or role |
|---|---|---|---|
| Tool poisoning | Malicious tool description | Cryptographic digest of descriptions verified before loading; allowlist of tools signed by L6 | L6 (validation), L4 (controlled loading) |
| Prompt injection through context | Document injected into the MCP context | Input isolation and context filtering before injection; the proposed act remains subject to the L6 decision and bounded by the scope of the act token (R-AUT, §2.2.3) | L6 (context provision), L5 (cognitive Zero Trust) |
| Interception of the A2A channel | Alteration of A2A messages in transit (replay has its own row) | TLS 1.3 with mutual authentication; signature of Agent Cards and messages; act token bound to the L6 verdict by the digest of the evaluation proof | L5 (A2A channel and transport) |
| Rug pull of an MCP server | Malicious modification of a previously safe server | Pinned versions reviewed before adoption; versioned allowlist of tools; dual human validation (Skills Engineer and Policy Auditor) | L6 (agentic registry) |
| Behavioral drift | Slow distortion of an agent's decisions | Differential analysis of explainability traces; drift indicator over a 30-day window; periodic recalibration | L5 (detection), Intent Architect (recalibration) |
| Replay | Re-emission of an already valid message or call | Act token obtained by token exchange (RFC 8693), made single-use, with a nonce and a short expiry by the L5 issuance policy; signed timestamp; under an envelope, single-use act identifier; rejection by the PEP on second use | L6 (decision), L5 (issuance), PEP (verification) |
| Confused deputy | Legitimate agent exploited as an intermediary | Action control per atomic act; explicit delegation (RFC 8693); decision on attested attributes (I-ATT invariant) | L6 (decision), L5 (issuance), PEP (verification) |
| Co-signing insider | Legitimate holder of a key who signs a malicious policy or envelope | Separation of key holders (§7.4.3) and dual control: entry into force subject to two distinct holders, logs sealed by a third-party key, incompatibility between policy engineering and risk assurance. Organizational countermeasure: none of the six main countermeasures is strong here | L6 (policy signing), logging function (log sealing), AOM governance |
| Compromise of the model provider | Served language model altered or substituted upstream of the agent | Attestation of the served model: digest of the model and its version verified at session opening and recorded in the log; drift detection on a fixed set of probes | L4 (cognitive agent), L6 (model registry) |

Table 13 — Detailed matrix of threats specific to MCP and A2A.

### 3.3. AOM responsibility matrix

Table 14 allocates governance activities among the AOM roles (§8.4.1) and the control stakeholders. Roles: **IA** Intent Architect · **CG** Context Guardian · **PE** Policy Engineer · **PA** Policy Auditor · **SE** Skills Engineer · **AM** A2A Mediator (system role). Stakeholders: **BO** business owner · **CISO** chief information security officer · **DPO** data protection officer · **AUD** internal audit. **R** performs, **A** approves and is accountable, **C** is consulted, **I** is informed.

| Activity | IA | CG | PE | PA | SE | AM | BO | CISO | DPO | AUD |
|---|---|---|---|---|---|---|---|---|---|---|
| Definition of the agentic constitution (business objectives and autonomy limits) | R / A | C | C | C | I | I | C | C | C | I |
| Writing, versioning and testing of Rego policies (L6) | A | C | R | C | C | I | C | C | C | I |
| Putting a policy into production after dual signature | A | I | R | A | I | I | C | C | C | I |
| Periodic review of capability envelopes | C | C | R | A | C | I | C | C | C | I |
| Curation of the Semantic Data Fabric | I | R / A | C | C | C | I | C | I | C | I |
| Development and validation of primitive actions | C | C | C | C | R / A | I | I | C | I | I |
| Review of MCP servers before adoption | C | C | C | A | R | I | I | C | I | I |
| Continuous audit of explainability traces, drift detection and quarterly search for drifts of the CoT-Validate-Sign-Execute pattern | I | C | C | R / A | C | I | I | C | I | C |
| Detection of A2A mutual deadlocks between agents | I | I | C | C | C | R / A | I | I | I | I |
| Quarterly red team campaign | C | C | C | R / A | R | I | I | C | I | I |
| Triggering of the emergency stop (Article 14 of the EU AI Act) | R / A | I | I | R | I | I | C | R | I | I |
| Immediate restriction of a policy (restrictive change, §8.4.4) | R / A | I | C | R | I | I | I | I | I | I |
| Handling of a compliance incident | I | C | C | R / A | C | I | C | R | R | I |
| Annual risk reassessment and acceptance of residual risk | C | I | C | R | I | I | A | C | C | C |

Table 14 — AOM responsibility matrix (RACI).

Two As on the same row signal a dual signature; an immediate restriction may also be decided by the Policy Auditor (§8.4.4), with the Policy Engineer consulted because this role is responsible for drafting the consolidated policy.

# Appendix 4. Objections and Responses

This appendix brings together the main objections that a critical reader may raise against the model, each with a short response and a pointer to the passage of the body that supports it.

**Objection 1 (Chapter 1).** Cognitive load is a psychological construct that is difficult to operationalize at the scale of an industrial IS; its subjective nature would leave any claim of toil reduction (H9, > 75%) without foundation.

*Response.* H9 does not rest on this construct: toil is measured by a quarterly survey and an analysis of repeated tickets (§8.6). Cognitive load, for its part, could be measured by two instruments: an adaptation of the NASA-TLX (Task Load Index) scale [Hart and Staveland 1988], administered to operators before and after deployment, and an operational complexity index — number of alerts correlated by hand, average depth of incident trees, context-switching time. This combination would aim to turn a subjective construct into a reproducible composite variable; its convergent validity remains to be established during empirical validation.

**Objection 2 (Chapter 1).** A language model produces probabilistic outputs, sometimes invalid plans (“hallucinations”); an agent based on it could therefore not reach the safety required of a critical IS (banking, healthcare, telecommunications).

*Response.* The model separates proposal from validation: L4 proposes, L6 evaluates each proposal against the signed policies. A hallucination that leads to an act contrary to an expressed policy does not pass L6; a hallucination that stays within the framework of the policies is bounded only by the scope of the token and the reversibility of the act (§4.4, §7.7.2). Safety thus rests on the L6 decision, not on the reliability of the language model; it remains bounded by what the policies can express (§7.5).

**Objection 3 (Chapter 1).** A2A and MCP are recent specifications, published by Google in 2025 and by Anthropic in 2024 respectively; basing an architecture model on them would expose it to rapid obsolescence, and their industrial maturity has not been demonstrated.

*Response.* Both protocols fall under open foundations (Trademarks and Legal Notices). The model does not depend on these implementations but on the properties it expects from them — conveying context, carrying the negotiation — signature and ephemeral rights remaining requirements of the model, not of the specifications (§1.3.2, §7.4.3). A replacement protocol that preserved these properties would leave the model intact; the obsolescence risk concerns syntax, not foundation.

**Objection 4 (Chapter 4).** The three-plane dependency formalization is not standard; why not stay with a directed acyclic graph?

*Response.* A directed acyclic graph can carry links of several types; what it does not state is the semantics specific to each. Yet two components may depend on each other through data, through health or through intent, and only the cognitive plane (C-Dep) carries the semantics of policies, which requires a dedicated operator. The composition of the three planes F-Dep, O-Dep and C-Dep meets this need (§4.2.3, D.C1).

**Objection 5 (Chapter 4).** The DAL loop resembles the OODA loop (Observe, Orient, Decide, Act); is it merely a change of name?

*Response.* DAL adds a Validation step entrusted to a layer distinct from the one that plans (R-AUT, §2.2.3); it is this step, and not the loop, that the model claims. Its lineage is, moreover, the MAPE-K loop of autonomic computing (Introduction, state of the art; §4.4), not OODA.

**Objection 6 (Chapter 6).** In the event of a long outage, can the autonomous degraded mode diverge from the state held by the central server?

*Response.* Yes, during the outage. This is why a reconciliation follows reconnection, by Lamport timestamp, signature and quarantine checks, the reference source prevailing in case of failure or tie, with a convergence targeted in under 5 minutes [DT] (§6.3).

**Objection 7 (Chapter 7).** Action control assumes an identity provider; is this not a single point of failure?

*Response.* The risk exists and is reduced by redundancy: several federated identity providers (OIDC) and replicas of the L5 issuance service, under its single issuing key, which deliver act tokens by token exchange (RFC 8693). The authorization plane nonetheless remains on the critical path (SP2, §3.1); for acts of low or moderate criticality, the capability envelope bounds the effect of this dependency (§7.4.3). Local issuance by the agent is excluded: it would eliminate the control point (R-AUT, §2.2.3).

**Objection 8 (Chapter 7).** By construction, the threat matrix does not cover vulnerabilities that are still unknown.

*Response.* It does not claim to cover them. Defense in depth addresses them through the behavioral detection of L5 (§7.4.1), trace auditing and tool signing (§7.7): an unknown attack that escapes one countermeasure must still pass the others.

**Objection 9 (Chapter 7).** Does action control hold at 100,000 acts per second?

*Response.* The book does not establish it; it proposes a design avenue, to be sized and tested (§10.8): partitioning by domain, with observability of decisions by OpenTelemetry traces. A verdict cache is excluded: it would delay the handling of revocations. Scaling goes through the capability envelope (§7.4.3), not through memorizing past decisions. High-criticality acts remain under synchronous authorization.

**Objection 10 (Chapter 8).** Can an administrator with the highest rights bypass the tripartite governance?

*Response.* Not alone: policies are signed by the L6 policy authority (§7.4.3), and an administrator cannot alter them without the signing key or without leaving a trace. The guarantee remains bounded by the key lifecycle. It assumes the separation of the holders of the five service keys (§7.4.3), rotation with monotonic epochs and the compromise procedure. An administrator who combined the policy signing key and the log sealing key would nullify the property. Key custody itself is not specified in this book.

**Objection 11 (Chapter 8).** Are six roles — five human and one system role, the A2A Mediator — not too heavy for a mid-sized organization?

*Response.* The roles can be combined according to size (§8.4.2, §8.7.2): at 500 components, the central scenario fits the five human roles into 2 full-time equivalents (Table 10). Three of them nevertheless remain held by three distinct individuals on the same scope: the two signatories of a policy, the Intent Architect and the Policy Auditor, are distinct (§8.3), and the Policy Engineer, who writes it, does not authorize it (§8.4.1, §8.4.2). These incompatibilities withstand combination, whatever the size of the organization; they are counted in persons, not in full-time equivalents. In a mid-sized structure, the second line is pooled with the existing internal control rather than eliminated.

**Objection 12 (Chapter 9).** Without empirical validation, do the hypotheses H1 to H10 not remain intuitions?

*Response.* The book acknowledges this: its targets are design targets [DT], and their validation on a test bench and then in a pilot is the subject of the agenda of Chapter 9 (Table 11). Every refutable statement — all except H7a and H8a — is nonetheless already refutable: its metric, its instrument and its refutation condition are fixed in advance (Table 4, Appendix 1).

**Objection 13 (Chapter 9).** The model claims to be useful for cases impossible to specify in advance, yet no agent acts outside a policy written in advance: is this not contradictory?

*Response.* No. The policy does not fix the procedure: it bounds the permitted acts, and the agent seeks the answer within that space. An answer that falls outside it returns to the human (§9.2).

**Objection 14 (Introduction).** Is R-AUT not the NIST Zero Trust architecture, which already separates decision, credential issuance and enforcement, transposed to agent acts?

*Response.* In part, and the book says so (Introduction, state of the art). AAM/5C adds to it, as part of the ephemeral authority surface, the signed binding between the L6 evaluation proof and the L5 act token and the capability envelope without local issuance, together with an independent measurement of non-bypass (H8b). It is this addition, not the separation, that the model claims on this point.

# Appendix 5. Personas

Seven personas illustrate the chapters, each in its own vignette: the six pivot personas, whose situations serve as a common thread, and Catherine, AI compliance officer, whose stake is that of the Chapter 8 vignette. Yann and Camille return in the vignettes of Chapters 10 and 11 respectively. Table 15 describes each by role and by the stake carried; none reports a result.

| Chapter | Persona | Stake |
|---|---|---|
| 1 and 11 | Camille, SRE lead at an e-commerce site (pivot) | A saturated cognitive load: most of the on-call time goes into the manual correlation of logs, traces and dashboards. |
| 2 and 10 | Yann, enterprise architect, retail bank (pivot) | Reconciling the pressure to adopt agents quickly with regulatory constraints (DORA, EU AI Act, supervision by the ACPR — French Prudential Supervision and Resolution Authority); demonstrating auditability and reversibility to a regulator. |
| 3 | Léa, doctoral student, university laboratory (pivot) | Instrumenting and running the evaluation protocol on a shared test bench, while ensuring statistical reproducibility and open publication of the artifacts. |
| 4 | Marc, platform development manager, SaaS vendor (pivot) | Delivering a first autonomous agent without blocking the product roadmap, and demonstrating to the CISO compliance with the governance invariants through measurable control points. |
| 5 | Sophie, chief information officer, industrial group (pivot) | Translating a sovereign AI ambition into an IS roadmap without architectural debt or ungoverned agentic capability; having a clear reading of the phases, the transition criteria and the financial indicators. |
| 6 | Karim, owner of the inventory agent, omnichannel retail (pivot) | Ensuring that a poorly designed degraded mode does not create overselling during an outage of the central inventory database. |
| 8 | Catherine, AI compliance officer | Producing the technical documentation required by the EU AI Act (Articles 11 and 18) and providing it to the competent authority upon reasoned request (Article 21). |

Table 15 — Persona gallery.

# Index

Index of key notions, synchronized with the Definitions. Each entry points to the sections where the notion is mainly treated (at most four).